\documentclass[useAMS,usenatbib,usegraphicx]{mn2e}
\usepackage{times,natbib,epsfig,graphicx,rotating,lscape,rotating,amssymb,color,subfigure}
\usepackage[usenames,dvipsnames,svgnames,table]{xcolor}

\def \arcsec{^{\prime\prime}}
\def \arcmin{^{\prime}}
\def\msun{{\rm M}_{\odot}}
\def\rsun{{\rm R}_{\odot}}
\def\rhosun{{\rho}_{\odot}}
\def\mjup{{\rm M}_{\rm J}}
\def\rjup{{\rm R}_{\rm J}}
\def\rhojup{{\rho}_{\rm J}}

\def\teff{{\rm T}_{\rm eff}}
\def\sred{S_{\rm red}}
\def\occfit{{\sc occfit~}}
\def\qprime{Q_{\star}^{\prime}}

\title[WTS-2 b]{WTS-2 b: a hot Jupiter orbiting near its tidal
  destruction radius around a K-dwarf} \author[J. L. Birkby et
al.]{J. L. Birkby$^{1}$\thanks{E-mail: birkby@strw.leidenuniv.nl
    (JLB)}, M.  Cappetta$^{2}$, P. Cruz$^{3}$,
  J. Koppenhoefer$^{2,4}$, O.  Ivanyuk$^{5}$, A. J. Mustill$^{6,7}$,
  \newauthor S. T. Hodgkin$^{8}$, D. J. Pinfield$^{9}$,
  B. Sip\H{o}cz$^{9}$, G.  Kov\'acs$^{8}$, R. Saglia$^{2,4}$,
  Y. Pavlenko$^{5,9}$, \newauthor D. Barrado$^{3}$, A.
  Bayo$^{10,11}$, D. Campbell$^{9}$, S. Catalan$^{9,12}$, L.
  Fossati$^{13}$, \newauthor M.-C. G\'alvez-Ortiz$^{14}$, M.
  Kenworthy$^{1}$, J. Lillo-Box$^{3}$ E. L. Mart\'in$^{14}$,
  D. Mislis$^{8}$, \newauthor E. J. W. de Mooij$^{15}$,
  S. V. Nefs$^{1}$, I. A. G. Snellen$^{1}$, H. Stoev$^{3}$,
  J. Zendejas$^{2}$, \newauthor C. del Burgo$^{16}$, J.  Barnes$^{9}$,
  N. Goulding$^{9}$, C. A. Haswell$^{13}$, M. Kuznetsov$^{5,16}$,
  \newauthor N. Lodieu$^{17,18}$, F. Murgas$^{17}$,
  E. Palle$^{17,18}$, E.  Solano$^{3,19}$, P. Steele$^{2}$, R. Tata$^{17,18}$\\\\
  $^{1}$Leiden Observatory, Leiden University, Niels Bohrweg 2, 2333CA Leiden, The Netherlands\\
  $^{2}$Max-Planck-Institut f\"ur extraterrestrische Physik,  Giessenbachstrasse, D-85741 Garching, Germany\\
  $^{3}$Depto. Astrof\'{\i}sica, Centro de Astrobiolog\'{\i}a (INTA-CSIC), ESAC campus, P.O. Box 78, E-28691 Villanueva de la Ca\~nada, Spain\\
  $^{4}$Universitatssternwarte Scheinerstrasse 1, D-81679 Munchen, Germany\\
  $^{5}$Main Astronomical Observatory of Ukrainian Academy of  Sciences, Golosiiv Woods, Kyiv-127, 03680, Ukraine\\
  $^{6}$Departamento de F\'isica Te\'orica, Facultad de Ciencias,  Universidad Aut\'onoma de Madrid, Cantoblanco, 28049 Madrid, Spain\\
  $^{7}$Lund Observatory, Department of Astronomy and Theoretical Physics, Lund University, Box 43, SE-221 00 Lund, Sweden\\
  $^{8}$Institute of Astronomy, University of Cambridge, Madingley  Road, Cambridge, CB3 0HA, UK\\
  $^{9}$Centre for Astrophysics Research, University of Hertfordshire,  College Lane, Hatfield, Hertfordshire AL10 9AB, UK\\
  $^{10}$European Southern Observatory, Alonso de C\'ordova 3107,  Vitacura, Santiago, Chile\\
  $^{11}$Max Planck Institut f\"ur Astronomie, K\"onigstuhl 17, 69117,  Heidelberg, Germany\\
  $^{12}$Department of Physics, University of Warwick, Gibbet Hill Rd, Coventry CV4 7AL, UK\\
  $^{13}$Department of Physical Sciences, The Open University, Walton Hall, Milton Keynes, MK7 6AA, UK\\
  $^{14}$Centro de Astrobiolog\'ia (CSIC-INTA). Crta, Ajalvir km 4. E-28850, Torrej\'on de Ardoz, Madrid, Spain\\
  $^{15}$Department of Astronomy and Astrophysics, University of Toronto, 50 St. George Street, Toronto, ON M5S 3H4, Canada\\
  $^{16}$Instituto Nacional de Astrof\'\i sica, \'Optica y Electr\'onica, Luis Enrique Erro 1, Sta. Ma. Tonantzintla, Puebla, Mexico\\
  $^{17}$Instituto de Astrof\'isica de Canarias, Calle V\'ia L\'actea s/n, E-38200 La Laguna, Tenerife, Spain\\
  $^{18}$Departamento de Astrof\'isica, Universidad de La Laguna (ULL), E-38205 La Laguna, Tenerife, Spain\\
  $^{19}$Spanish Virtual Observatory Thematic Network, Spain}
\begin{document}
\pagerange{\pageref{firstpage}--\pageref{lastpage}}
\pubyear{2014}
\maketitle
\label{firstpage}

\begin{abstract}
  We report the discovery of WTS-2 b, an unusually close-in 1.02-day
  hot Jupiter ($M_{P}=1.12\mjup$, $R_{P}=1.363\rjup$) orbiting a K2V
  star, which has a possible gravitationally-bound M-dwarf companion
  at $0.6$ arcsec separation contributing $\sim20$ percent of the
  total flux in the observed $J$-band light curve.  The planet is only
  1.5 times the separation from its host star at which it would be
  destroyed by Roche lobe overflow, and has a predicted remaining
  lifetime of just $\sim40$ Myr, assuming a tidal dissipation quality
  factor of $\qprime=10^{6}$. $\qprime$ is a key factor in determining
  how frictional processes within a host star affect the orbital
  evolution of its companion giant planets, but it is currently poorly
  constrained by observations. We calculate that the orbital decay of
  WTS-2 b would correspond to a shift in its transit arrival time of
  $T_{\rm shift}\sim17$ seconds after $15$ years assuming
  $\qprime=10^{6}$. A shift less than this would place a direct
  observational constraint on the lower limit of $\qprime$ in this
  system. We also report a correction to the previously published
  expected $T_{\rm shift}$ for WASP-18 b, finding that $T_{\rm
    shift}=356$ seconds after $10$ years for $\qprime=10^{6}$, which
  is much larger than the estimated $28$ seconds quoted in WASP-18 b
  discovery paper. We attempted to constrain $\qprime$ via a study of
  the entire population of known transiting hot Jupiters, but our
  results were inconclusive, requiring a more detailed treatment of
  transit survey sensitivities at long periods. We conclude that the
  most informative and straight-forward constraints on $\qprime$ will
  be obtained by direct observational measurements of the shift in
  transit arrival times in individual hot Jupiter systems. We show
  that this is achievable across the mass spectrum of exoplanet host
  stars within a decade, and will directly probe the effects of
  stellar interior structure on tidal dissipation.
\end{abstract}


\section{Introduction}
The orbital period distribution of gas giants is a fundamental
property of planetary systems and places constraints on their
formation processes, migration mechanisms, and future evolution. The
observed period distribution is not smooth. The majority of hot
Jupiters, i.e. those with a semi-major axis $a<0.1$ AU, are found in a
`pile-up' at periods of $P\sim3-4$ days ($\sim0.035-0.045$ AU),
whereas only four hot Jupiters are found in very close in orbits
($\lesssim0.02$ AU, $\lesssim1$ days), namely WASP-18 b, WASP-19 b,
WASP-43 b, and WASP-103 b (\citealt{Hel09,Heb10,Hel11,Gil14},
respectively). The sharp decline of hot Jupiters in orbital periods
less than two days is a genuine feature of the exoplanet period
distribution, confirmed by both ground-based and space-based planet
searches, e.g.  \textit{Kepler} \citep{How11}, and SuperWASP
\citep{Hel12}.  This suggests that very close-in hot Jupiters are
relatively rare, else current instrumentation would easily detect them
on account of their very frequent and deep transits, and large RV
variations in comparison to longer period, smaller planets.

As a result, \citet{Hel11} argue that extreme systems like WASP-19 b
are approximately one hundred times less common than those hot
Jupiters in the pile-up, indicating that it is either difficult to get
gas giants into very close orbits, or that they are quickly destroyed
by strong tidal forces once they arrive. The latter would imply that
very close-in hot Jupiters with old host stars are in the last few
percent of their lifetimes, which raises a further question of how
likely it is to have observed these systems in a transient phase of
their orbital evolution. Despite extensive theoretical work, our
understanding of how tidal forces influence the orbital evolution of
giant planets is poorly constrained by observation. The efficiency of
the dissipation of the orbital energy due to frictional processes in
the star is usually parameterised by a stellar tidal quality factor
$\qprime$. Studies of binary star systems estimate its value to be
$\qprime\sim10^{6}$ (see e.g. \citealt{Mei05}) and analysis of the
tidal evolution of a small sample of exoplanets has found some
evidence for consistency with this value ($10^{6}<\qprime<10^{9}$,
\citealt{Jack08}). On the other hand, a recent exoplanet population
study which tuned the value of $\qprime$ until the distribution of
remaining planet lifetimes was statistically likely, found
$\qprime\gtrsim10^{7}$ at the $99\%$ confidence level \citep{Pen12}
for its specific set of initial conditions. However, direct
observational measurements of $\qprime$ in individual systems,
i.e. the observation of a decaying orbital period, do not currently
exist.  $\qprime$ is the dominant factor in setting the pace of the
orbital evolution for very close in hot Jupiters and the unusually
short predicted remaining lifetimes for planets such as WASP-18 b and
WASP-19 b have lead to a number of suggested modifications to the
theory of stellar tides that reduce the efficiency of the
dissipation. For example, \citet{Win10} speculate that the observed
increase in misalignment between the planetary orbit and stellar spin
axes for hot Jupiters orbiting hotter stars depends on the depth of
the convective zone in the host star.  Here, cooler stars with deeper
convective envelopes dissipate the orbital energy more efficiently
resulting in a faster alignment of the stellar obliquity, in keeping
with theoretical studies (e.g.  \citealt{Bark09,Bark10,Pen11}). Others
suggest that there is a complicated dependency on the planetary mean
motion that results in zones of inefficient tidal dissipation during
inspiral \citep{Ogi07}, or even possible mass loss effects that act to
slow the orbital evolution of the planet \citep{Li10}.

In this paper, we present the discovery and characterisation of the
hot Jupiter WTS-2 b. It is the second planet to be detected in the
infrared light curves of the WFCAM Transit Survey (WTS)
\citep{Cap12,Kov12}, and orbits a mid-K dwarf star at just 1.5 times
the separation at which it would be destroyed by tidal forces, making
it a useful benchmark in constraining the theory of stellar tides. The
remainder of this paper is organised as follows: in
Section~\ref{sec:WTS}, we briefly summarise the goals of the WTS and
its atypical observing strategy, the reduction procedure used to
generate the infrared light curves, and the processes used to identify
WTS-2 b as a transiting candidate and the checks performed before
proceeding with its follow-up observations. Section~\ref{sec:obs}
describes all of the follow-up data we obtained for WTS-2 b and their
data reduction. We characterise the WTS-2 host star in
Section~\ref{sec:char_star}, and derive the corresponding properties
of its planetary companion WTS-2 b in Section~\ref{sec:char_planet}.
Section~\ref{sec:falsepos} summarises our investigation into possible
blending scenarios. In Section~\ref{sec:discussion}, we calculate and
discuss the tidal evolution and remaining lifetime of WTS-2 b. We
calculate the expected shift in its transit arrival time after 10
years, assuming that its orbit is decaying under tidal forces with
$\qprime=10^{6}$, and we give a correction to the previously published
expected shift in the transit arrival time for WASP-18 b in
Section~\ref{sec:wasp-18}. In Section~\ref{sec:pop}, we also attempt
to constrain $\qprime$ using the known population of hot Jupiters.
Finally, we assess the potential for characterising the atmosphere of
WTS-2 b using ground-based telescopes in Section~\ref{sec:followup}.
Our conclusions are summarised in Section~\ref{sec:conclusion}.

\section{The WFCAM Transit Survey}\label{sec:WTS}
The WTS was a photometric monitoring campaign that covered $\sim6$
sq. degrees of the sky. It used the $3.8$ m United Kingdom Infrared
Telescope (UKIRT) on Mauna Kea, Hawaii, in conjunction with the
Wide-Field Camera (WFCAM), to observe at infrared wavelengths
($J$-band, $1.25\mu$m). The survey began on $5^{\rm th}$ August
2007. A detailed description of the WTS and its goals can be found in
\citet{Kov12}, \citet{Bir12}, and \citet{Zen13}, but its main features
are recounted here briefly for reference. The WTS light curves were
observed at infrared wavelengths in order to maximise sensitivity to
photons from M-dwarfs. However, for the earlier-type stars in the WTS
fields, infrared observations had the added advantage of being less
sensitive to low-level star spot modulation, thus providing more
stable light curves in which to hunt for planets \citep{Goul12}. The
WTS covered four fields distributed in RA so that at least one field
was always visible within 15 degrees of zenith from Mauna Kea. This
was key to the survey's observing strategy as it operated as a back-up
program in the highly efficiently queue-scheduled operational mode of
UKIRT, observing in sky conditions that the UKIRT large programs, such
as UKIDSS \citep{Lawr07}, could not use. Consequently, the majority of
the WTS observations were taken in the first hour of the night when
the atmosphere is still cooling and settling; however, the back-up
nature of the program served to randomise the observing pattern. The
exact field locations were chosen to minimise giant contamination,
while maximising the number of early M-dwarfs and maintaining
$E(B-V)<0.1$ mag, which kept the fields at $b>5$ degrees. WTS-2 b was
found in the `19 hr field', which was centred at RA$=19^{h}$,
Dec$=+36^{d}$ and contained $\sim 65,000$ stellar sources at $J\leq17$
mag. Note that this field is very close to, but does not overlap with
the Kepler field-of-view \citep{Bat06}, which has been shown to have a
low fraction of late-K and early-M giants at optical magnitudes
comparable to the WTS \citep{Mann12}.

\subsection{Observation and reduction of the UKIRT/WFCAM $J$-band
  time-series photometry}\label{sec:jband}
The infrared light curves of the WTS were generated from time-series
photometry taken with the WFCAM imager \citep{Cas07} mounted at the
prime focus of UKIRT. WFCAM consists of four $2048\times2048$
$18~\mu$m pixel HgCdTe Rockwell Hawaii-II, non-buttable, infrared
arrays. The arrays each cover $13.65\arcmin\times13.65\arcmin$
($0.4\arcsec$/pixel) and are arranged in a square paw-print pattern,
separated by $94$ per cent of an array width. The four WTS field cover
$1.5$ sq. deg. each, which requires $8$ pointings of the WFCAM
paw-print, tiled together to give uniform coverage. The WTS observed a
$9$-point jitter pattern of 10 second exposures at each pointing,
resulting in a cadence of one data point per 15 minutes in any given
one hour observing block ($9 \times 10$ s $\times 8$ + overheads).

The 2-D image processing of the WFCAM images and the generation of the
WTS light curves is described in detail by \citet{Kov12} and closely
follows the methods of \citet{Irw07}. In summary, we remove the dark
current and reset anomaly from the raw images, apply a flat-field
correction using twilight flats, then decurtain and sky
subtract. Astrometric and photometric calibration was achieved using
2MASS stars in the field-of-view \citep{Hod09}. To generate the light
curves, we made a master catalogue of source positions using a stacked
image of the $20$ best frames and used it to perform list-driven,
co-located, variable aperture photometry. For WTS-2, the best aperture
radius (i.e. the one that gave the smallest RMS) was equal to
$\sqrt{2}$ times the typical FWHM of the stellar images across all
frames i.e. 3.5 pixels ($1.98^{\prime\prime}$). In an attempt to
remove systematic trends in the light curves, e.g. those caused by
flat-fielding inaccuracies or varying differential atmospheric
extinction across the wide field-of-view, we fit a 2-D quadratic
polynomial to the flux residuals in each light curve as a function of
the source position on the detector. This step can significantly
reduce the RMS of the brightest objects in wide-field surveys
\citep{Irw07}. Finally, we removed residual seeing-correlated effects
by fitting a quadratic polynomial to the flux residuals in each light
curve as a function of the stellar image FWHM on the corresponding
frame.

\begin{figure*}
\centering
\includegraphics[width=\textwidth]{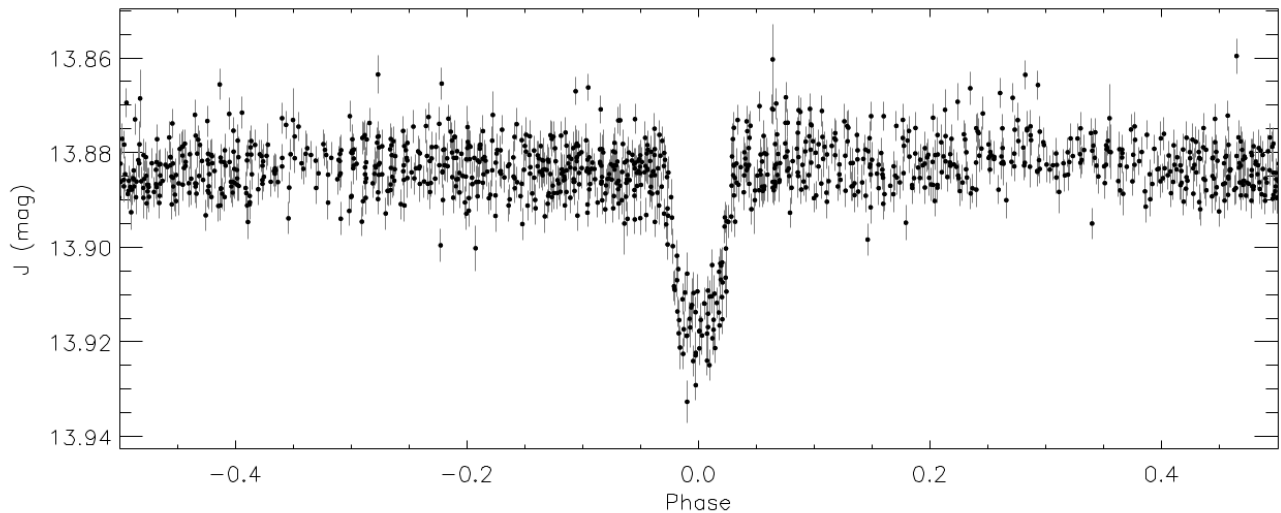}
\caption{The full phase-folded discovery light curve of WTS-2 b from
  WTS $J$-band observations. The data are not binned. The
  out-of-eclipse RMS is 5.3 mmag. The data for this figure are given
  in Table~\ref{tab:lc_J}. Note that this light curve has not been
  corrected for dilution by the additional faint light source also
  within the aperture (see Section~\ref{sec:lucky}).}
\label{fig:lc_J}
\end{figure*}

The resulting $J$-band light curves for the 19hr field have a median
RMS of $\sim1$ per cent ($\sim10$ mmag) or better for $J\leq16$ mag, with
a per data point precision of $\sim3-5$ mmag for the brightest targets
(saturation occurs at $J\sim13$ mag)\footnote{The RMS is calculated
  using the robust median of absolute deviations (MAD) estimator,
  scaled to the equivalent Gaussian standard deviation
  (i.e.. RMS$\sim1.48\times$MAD).}. The out-of-eclipse data in the
light curve of WTS-2 ($J_{\rm WFCAM}=13.88$ mag) has a per data point
precision of $5.3$ mmag. The full, phase-folded, unbinned $J$-band
light curve of WTS-2 b is shown in Figure~\ref{fig:lc_J}, and the data
are given in Table~\ref{tab:lc_J}.

\begin{table}
  \centering
  \begin{tabular}{@{\extracolsep{\fill}}ccc}
    \hline
    \hline
    HJD&$J_{\rm WFCAM}$&$\sigma_{J_{\rm WFCAM}}$\\
    &(mag)&(mag)\\
    \hline
    2454317.810999&13.9219&0.0033\\
    2454317.823059&13.9245&0.0032\\
    ...&...&...\\
    \hline
  \end{tabular}
  \caption{The observed WFCAM
    $J$-band light curve data for WTS-2 b without correction for dilution. Magnitudes
    are given in the WFCAM system. \citet{Hod09} provide conversions
    for other systems. The errors, $\sigma_{J}$, are estimated using a
    standard noise model, including contributions from Poisson noise
    within the object aperture, sky noise, readout noise and errors in the
    sky background estimation. (This table is published in full in the
    online journal and is shown partially here for guidance regarding
    its form and content.)}
  \label{tab:lc_J}
\end{table}

\subsection{Detection and prioritisation of WTS transit candidates}\label{sec:detect}
The vast sample of stars in the WTS and its randomised observing
strategy do not permit a straight-forward eyeball search for transits
in the light curves, so we undertook several steps to reduce the
enormity of this task. All stellar sources in the 19hr field with $J<
17$ mag were first passed through the box-least-squares transit
detection algorithm {\sc occfit}, which is described in detail by
\citet{Aig04}. Like all ground-based transit surveys, the processed
WTS light curves suffer from residual correlated red noise, which can
mimic transit events. We therefore adjusted the detection significance
statistic, $S$, calculated by \occfit to account for the presence of
red noise following the model of \citet{Pon06b} to give $S_{\rm red}$.
In order to qualify as a WTS transit candidate, a detection must have
$\sred\geq5$. We also rejected transit detections in the period range
$0.99<P<1.005$ days, as the majority of these were found to be aliases
caused by the observing window function of our ground-based survey.

In the final step before eyeballing the remaining light curves, we
used $ZYJHK$ single epoch photometry from WFCAM (see
Section~\ref{sec:broadband}), plus complementary $griz$ photometry
from SDSS DR7 \citep{Yor00} to create a spectral energy distribution
(SED) for each object and estimate its effective temperature (see
\citealt{Bir12} for details). The effective temperature, $\teff$, was
converted to an approximate stellar radius for each source, using the
stellar evolution models of \citet{Bar98}, at an age of 1 Gyr with a
mixing length equal to the scale height. Assuming a maximum planetary
radius of $2\rjup$, we defined an envelope of transit depths as a
function of stellar radius that were consistent with planetary transit
events. Only detections with changes in flux ($\Delta F$)
corresponding to $R_{\star}\sqrt{(\Delta F)} \leq 2\rjup$ were allowed
through to the eyeballing stage. It is important to note firstly that
\occfit tends to under-estimate transit depths because it does not
allow for the trapezoidal shape of a transit, nor does it account for
limb-darkening effects. Secondly, the models we use to estimate the
stellar radii systematically under-estimate the temperature of
solar-like stars \citep{Bar98}, making our first estimates of $\teff$
too cool for stars of earlier type than M-dwarfs and hence initial
radius estimates that are too small. Both of these factors combined
make it unlikely that genuine hot-Jupiter transit events are rejected
by this final selection criterion.

The $\sim3500$ candidates that survived to the eyeball stage were
mostly false-positives arising from nights of bad data or singular bad
frames that we do not filter from the data. We also removed binary
systems that were detected on half their true orbital period (as is
favoured by the detection statistic). Overall, with this method we
detected $40$ good transiting candidates, including WTS-2 b, which has
$\sred=23$, an {\sc occfit}-detected period of $1.0187$ days, an
initial estimated stellar effective temperature $\teff=4777$ K, and an
{\sc occfit}-detected transit depth of $\Delta F=0.031$, corresponding
to an estimated planet radius of
$\sqrt{0.031}\times0.82\rsun\sim1.4\rjup$.

Before proceeding with follow-up observations, we checked that the
stellar density calculated from the phase-folded light curve of WTS-2
b matched the estimated stellar type from the initial SED model fit,
using the method described by \citet{Sea03}. A large discrepancy would
suggest a blended or grazing binary system. We found a light curve
stellar density of $\rho_{\star}\sim1.49\rhosun$, which is within
$\sim0.05\rhosun$ of the model density for a $4777$ K star at 1 Gyr in
the \citet{Bar98} models. The close agreement between the stellar
densities from SED modelling and the phase-folded light curve
triggered the follow-up observations to characterise WTS-2 b.

\section{Follow-up observations and data reduction}\label{sec:obs}
\subsection{Multi-wavelength single epoch broadband
  photometry}\label{sec:broadband}
In order to measure the photometric colours and estimate the spectral
type of WTS-2 (and all the other sources in the WTS), we used WFCAM to
observe single, deep exposures of the four WTS fields in five filters
($ZYJHK$), with exposure times 180, 90, 90, $4\times90$, and
$4\times90$ seconds, respectively. The 2-D image processing for these
data are the same as described in Section~\ref{sec:jband}. For WTS-2,
we also obtained Johnson $B$-, $V$- and $R$-band single epoch
photometry on the nights of $8^{\rm th}$ and $22^{\rm nd}$ March 2012
at the University of Hertfordshire's Bayfordbury Observatory
(latitude$=51.8$ degrees North, longitude$=0.1$ degrees West). We used
a Meade LX200GPS $16$-inch $f/10$ telescope fitted with an SBIG
STL-6303E CCD camera, and integration times of 300 seconds per band.
Images were bias, dark, and flat-field corrected, and the extracted
aperture photometry was calibrated using three bright reference stars
within the image. The quoted photometric uncertainties for this data
combine the contribution from the signal-to-noise of the source
(typically $\sim20$) with the scatter in the zero-point from the
calibration stars.

\begin{figure}
\centering
\includegraphics[width=0.5\textwidth]{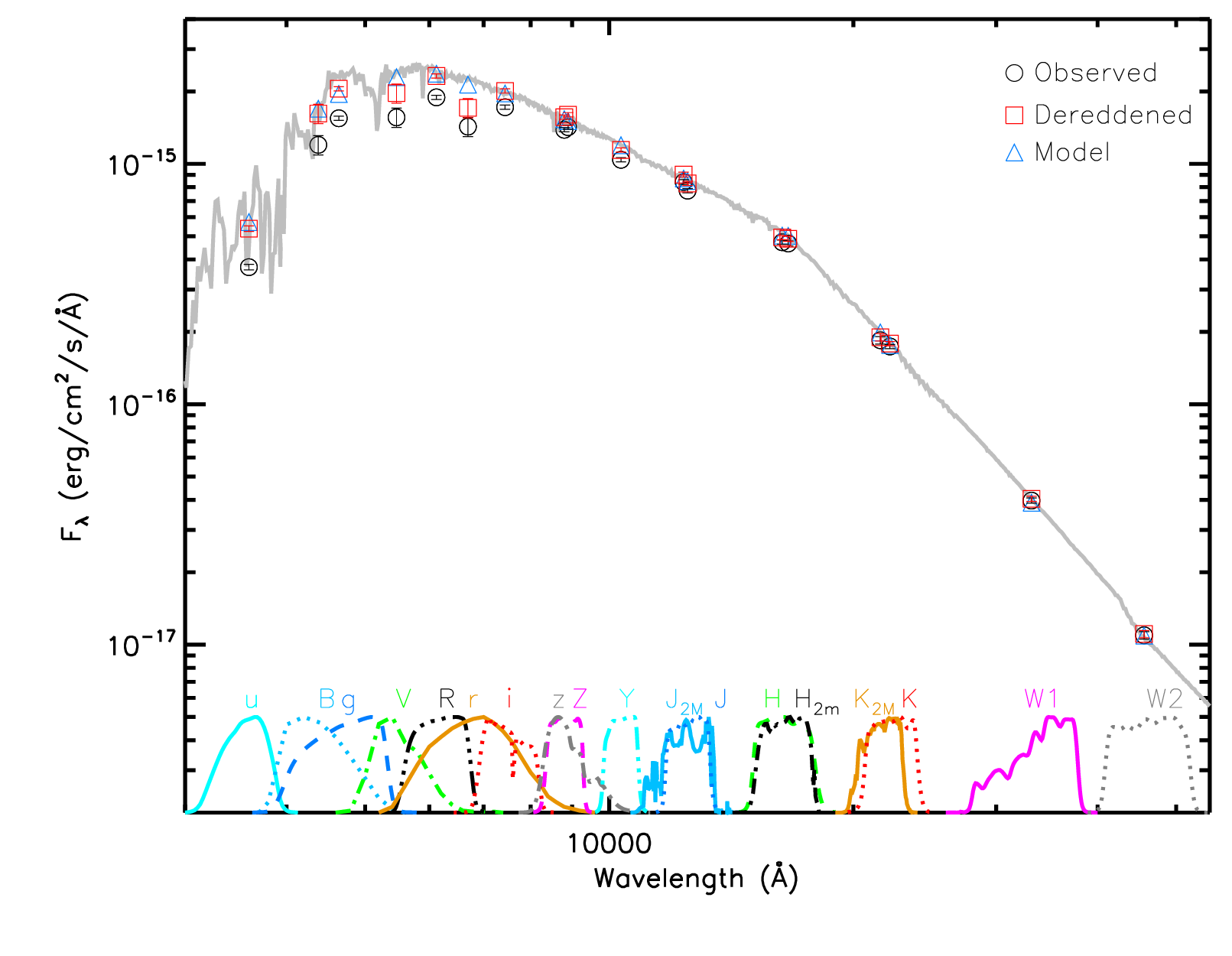}
\caption{The spectral energy distribution of WTS-2. The best-fitting
  Kurucz model spectrum from a $\chi^{2}$ analysis (see
  Section~\ref{sec:SED}) is overlaid in grey ($\teff=5000$ K), while
  the synthetic photometry for the corresponding observed bandpasses
  are shown by the blue open triangles. The observed data are shown by
  the black open circles and the dereddened photometry is shown by the
  red open squares. Note that the errors on the photometry include the
  photon error listed in Table~\ref{tab:broadband}, plus a $2\%$
  uncertainty added in quadrature to the WFCAM and SDSS bandpasses to
  allow for calibration between the different surveys. However, the
  magnitudes have not been adjusted for contamination by a faint red
  source of third light within 0.6 arcsec of the host star (see
  Section~\ref{sec:lucky}) The symbols are generally bigger than the
  errors. The filter transmission profiles for our observed bandpasses
  (see Table~\ref{tab:broadband}) are shown by the lines at the
  bottom of the plot.}
\label{fig:SED}
\end{figure}

A further nine photometric data points at optical and infrared
wavelengths were gathered for WTS-2 using the publicly available
Virtual Observatory SED
Analyzer\footnote{http://svo2.cab.inta-csic.es/svo/theory/vosa} ({\sc
  vosa}, \citealt{Bay08,Bay13}), including $ugriz$ from the Sloan
Digital Sky Survey data release 7 (SDSS DR7) \citep{Yor00}, $JHK$ from
the Two Micron All Sky Survey (2MASS, \citealt{Skr06}), and $W1W2$
from the Wide-field Infrared Survey Explorer (WISE,
\citealt{Wrig10}). We do not give the $W3$ and $W4$ bandpasses as they
fall below the WISE $5\sigma$ point source sensitivity for
detection. We also note that the $u$-band for SDSS photometry is
affected by a known red leak in the filter and has been assigned an
accordingly larger error\footnote{See
  http://www.sdss3.org/dr8/imaging/caveats.php}. All of the available
single epoch broadband photometry for WTS-2 is reported in
Table~\ref{tab:broadband} and plotted in Figure~\ref{fig:SED}. The
data are used in Section~\ref{sec:char_star} to determine the
best-fitting SED for WTS-2.

\begin{table}
  \centering
  \begin{tabular}{lrrr}
    \hline
    \hline
    Filter&$\lambda_{\rm eff}$ (\AA)&EW (\AA)&Magnitude\\
    \hline
    SDSS-$u$&3546&558&$18.361\pm0.039$\\
    Johnson-$B$&4378&1158&$16.8\pm0.1$\\
    SDSS-$g$&4670&1158&$16.283\pm0.004$\\
    Johnson-$V$&5466&890&$15.9\pm0.1$\\
    SDSS-$r$&6156&1111&$15.464\pm0.003$\\
    Johnson-$R$&6696&2070&$15.3\pm0.1$\\
    SDSS-$i$&7471&1045&$15.146\pm0.003$\\
    WFCAM-$Z$&8802&927&$14.501\pm0.003$\\
    SDSS-$z$&8918&1124&$14.959\pm0.005$\\
    WFCAM-$Y$&10339&999&$14.352\pm0.004$\\
    2MASS-$J$&12350&1624&$13.928\pm0.025$\\
    WFCAM-$J$&12490&1513&$13.963\pm0.003$\\
    WFCAM-$H$&16338&2810&$13.470\pm0.002$\\
    2MASS-$H$&16620&2509&$13.464\pm0.026$\\
    2MASS-$K_{s}$&21590&2619&$13.414\pm0.039$\\
    WFCAM-$K$&22185&3251&$13.360\pm0.003$\\
    WISE-$W1$&34002&6626&$13.292\pm0.027$\\
    WISE-$W2$&46520&10422&$13.368\pm0.038$\\
    \hline
  \end{tabular}
  \caption{Broadband photometry for WTS-2. All reported magnitudes are
    in the Vega system except the SDSS photometry, which is in the AB
    magnitude system. These magnitudes have not been corrected for
    reddening, nor for the dilution by the
    faint red source within 0.6 arcsec of the host star (see
    Section~\ref{sec:lucky}). $\lambda_{\rm eff}$ is the effective
    wavelength defined as the mean wavelength weighted by the
    transmission function of the filter, and EW is the equivalent
    width of the bandpass.}
\label{tab:broadband}
\end{table}

\subsection{INT/WFC $i$-band time-series photometry}
In order to confirm the transit of WTS-2 b and to help constrain the
transit model, on $18^{\rm th}$ July 2010 we obtained further
time-series photometry in the Sloan $i$-band using the Wide Field
camera (WFC) on the $2.5$ m Isaac Newton Telescope (INT) at Roque de
Los Muchachos, La Palma. A total of $67$ frames covering the full
transit with some out-of-transit baseline were obtained with exposures
times of $90$ seconds, at a cadence of $1$ data point every $2.45$
minutes (the overheads include the CCD read-out time plus time allowed
for the auto-guider to place the star back onto the exact same pixel
after every exposure).

\begin{figure}
\centering
\includegraphics[width=0.49\textwidth]{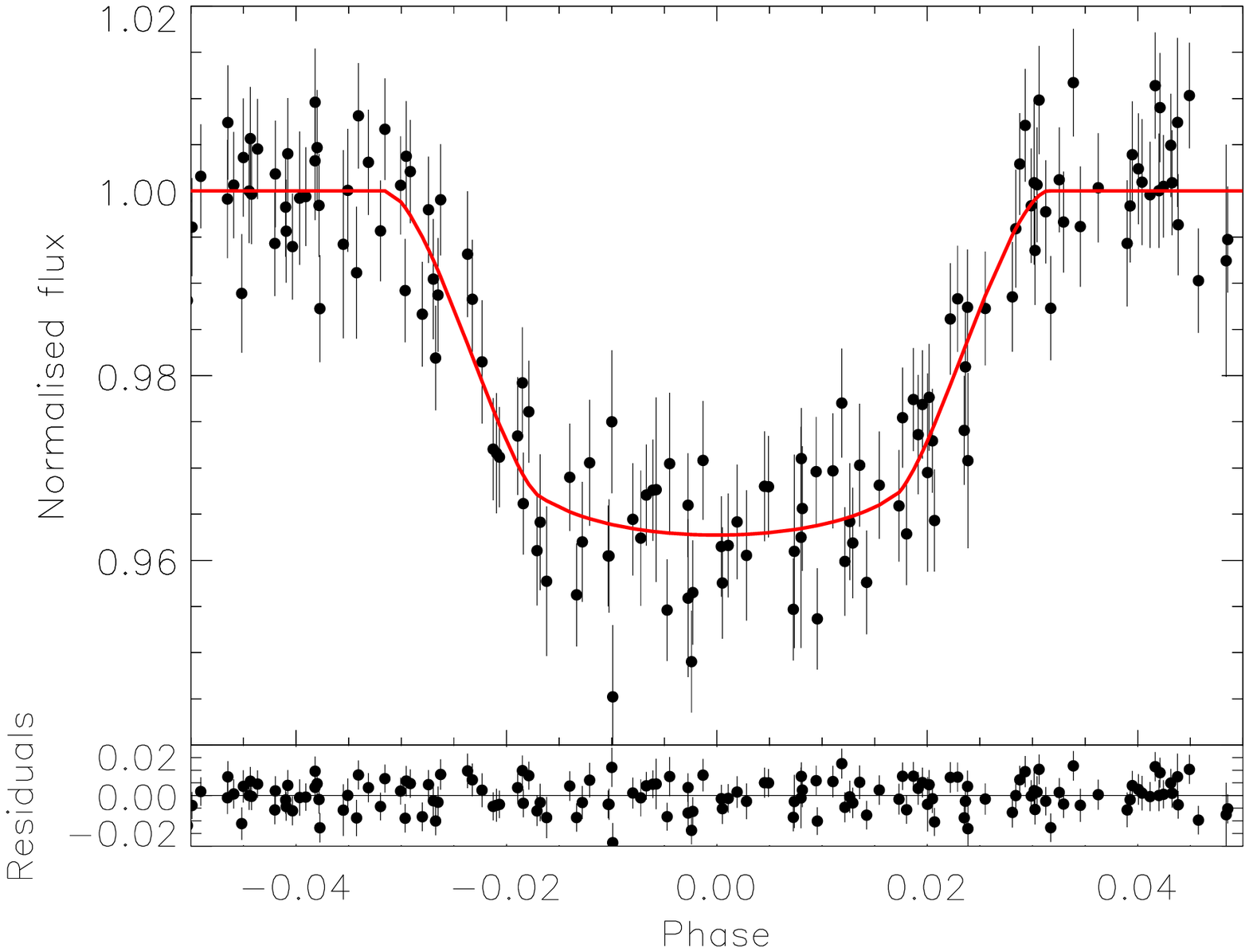}\\
\includegraphics[width=0.49\textwidth]{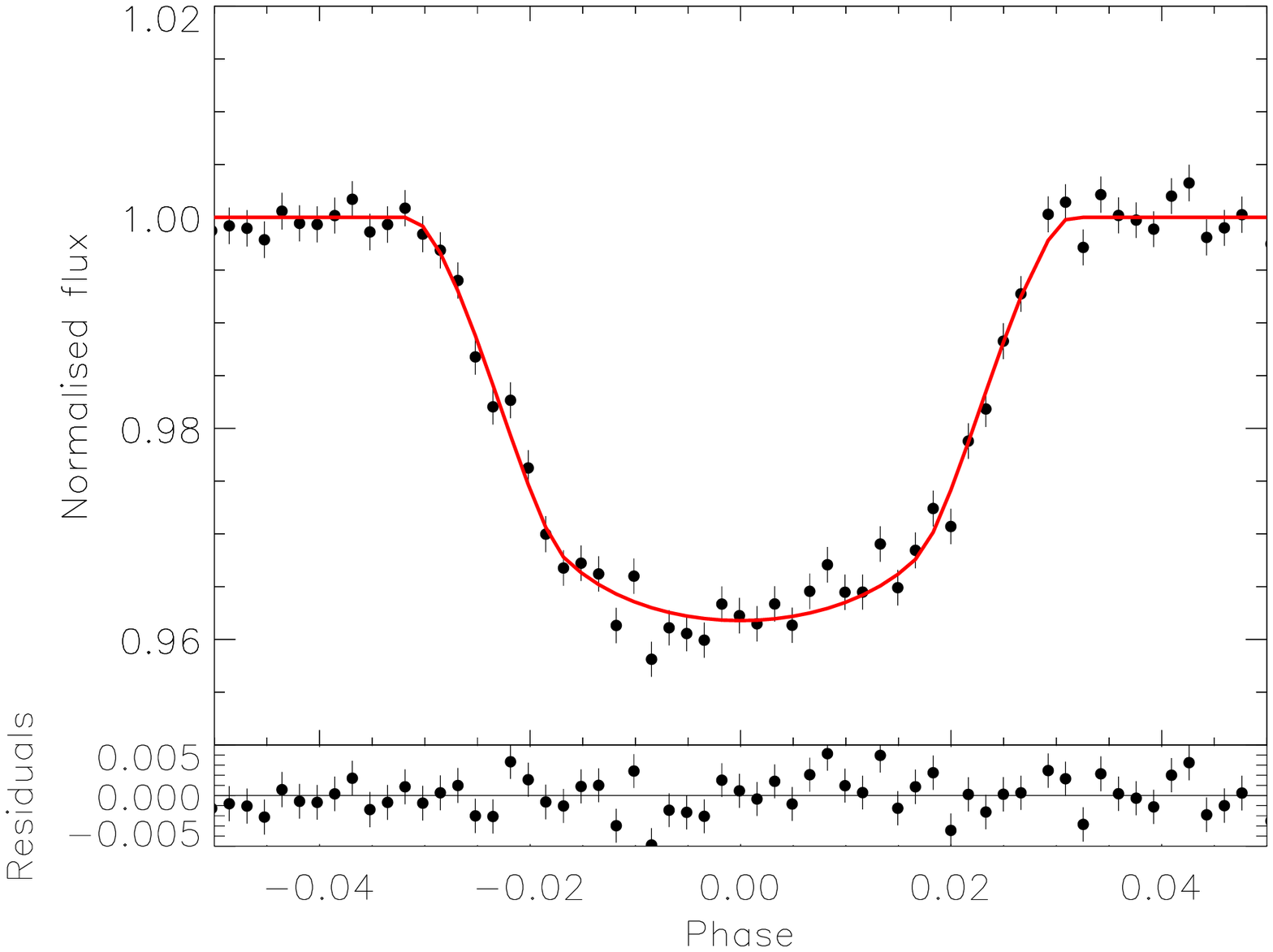}
\caption{{\bf Top:} Phase-folded WTS $J$-band light curve for WTS-2 b
  corrected for dilution by a faint red source and zoomed around the
  transit. The adopted best model from a simultaneous fit to the
  $J$-band and $i$-band dilution-corrected light curves is shown by
  the solid red line (see Section~\ref{sec:transitfit}). The errors
  have been scaled such that the out-of-transit data has
  $\chi_{\nu}^{2}=1$ when compared to a flat line. The lower panel
  shows the residuals to the model. {\bf Bottom:} Same as above but
  for the dilution-corrected INT $i$-band light curve. All the
  available data is shown. Note the change in y-axis scale for the
  residuals.}
\label{fig:lc_i}
\end{figure}

The CASU INT/WFC data reduction pipeline \citep{Irw01,Irw07} was used
to reduce the $i$-band images. The pipeline follows a standard CCD
reduction of de-biasing, correcting for nonlinearity, flat-fielding
and defringing. A master source catalogue was extracted from a stacked
image of the 20 best frames and variable aperture photometry was
performed for all sources in all images to generate light curves. The
out-of-transit RMS in the WTS-2 $i$-band light curve is $1.5$
mmag. The light curve is used simultaneously with the $J$-band light
curve to find the best-fitting model transit to WTS-2 b in
Section~\ref{sec:transitfit}. The WTS-2 b $i'$-band light curve is
shown in Figure~\ref{fig:lc_i} and the data are given in
Table~\ref{tab:lc_i}.

\begin{table}
  \centering
  \begin{tabular}{@{\extracolsep{\fill}}ccc}
    \hline
    \hline
    HJD&Normalised flux&Error\\
    \hline
    2455396.56449361&1.0047&0.0015\\
    2455396.56754940&0.9988&0.0015\\
    ...&...&...\\
    \hline
  \end{tabular}
  \caption{The observed  INT $i$-band light curve for WTS-2
    b without correction for dilution. The errors were derived in the same manner as for the $J$-band
    but have been scaled so that the out-of-transit baseline has a
    reduced $\chi^{2}=1$ when compared to a flat line to avoid
    underestimating the errors (see Figure~\ref{fig:lc_i}). (This
    table is published in full in the online journal and is
    shown partially here for guidance regarding its form and content.)}
  \label{tab:lc_i}
\end{table}

\subsection{CAHA/TWIN intermediate-resolution spectroscopy}
We carried out intermediate-resolution reconnaissance spectroscopy of
WTS-2 to obtain an estimate of the host star effective temperature and
its surface gravity (see Section~\ref{sec:char_star}), and to measure
preliminary radial velocity (RV) variations to test for the presence
of a blended or grazing eclipsing binary system (see
Section~\ref{sec:falsepos} and Table~\ref{tab:recon}). Spectroscopic
observations of WTS-2 and several RV standards were taken over 6
nights during June-August 2011 as part of a wider follow-up campaign
of the WTS planet candidates and M-dwarf eclipsing binaries (Cruz et
al., \emph{in prep.}). We used the Cassegrain Twin Spectrograph (TWIN)
mounted on the 3.5-m telescope at the Calar Alto Observatory (CAHA) in
southern Spain, with its T10 grism and a $1.2\arcsec$ slit, resulting
in a dispersion of $\sim0.39$\AA/pix ($R\sim8000$) and a wavelength
coverage of $\sim6200-6950$\AA. A total of 18 epochs were observed for
WTS-2 with integration times between $600$ and $900$ seconds.
 
\begin{figure}
\centering
\includegraphics[width=0.5\textwidth]{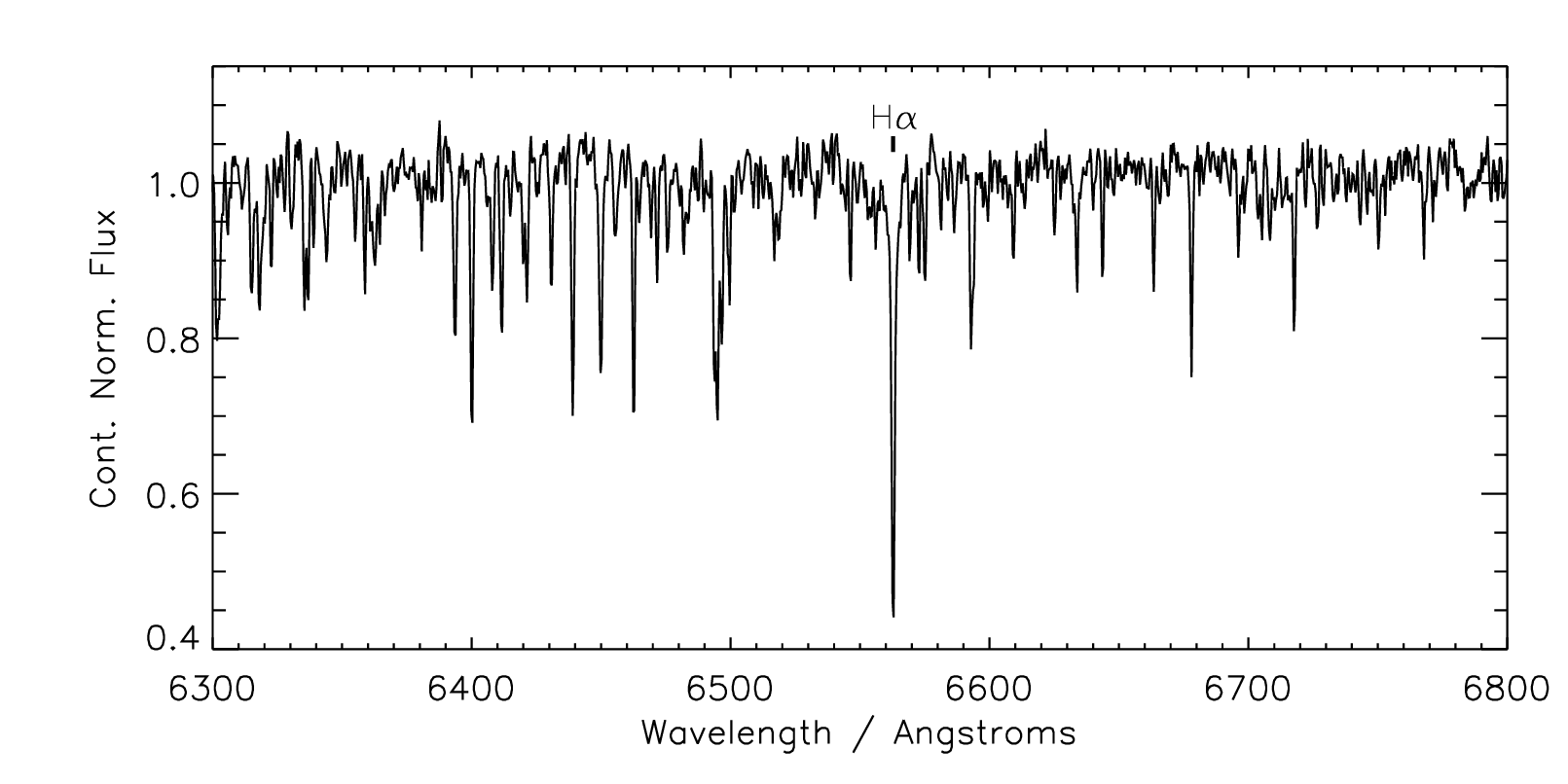}
\caption{A stacked spectrum of WTS-2 using eight of the TWIN
  observations from August 2011. The individual spectra have been
  aligned to the same RV and continuum-normalised. The rest-wavelength
  of H$\alpha$ is labeled.}
\label{fig:TWIN_spec}
\end{figure}

The spectra were reduced in the standard way using {\sc iraf}
packages. To measure the RV variations of WTS-2 and the RV standards,
the {\sc iraf} package {\sc fxcor} was used to perform Fourier
cross-correlation of the observed spectra with synthetic templates
generated from \citet{Mun05}. The effective temperature and surface
gravity of the cross-correlation template was chosen to match the
results of the SED fit in Section~\ref{sec:char_star} but with a solar
metallicity. We also use the TWIN spectra in
Section~\ref{sec:char_star} to confirm the stellar characteristics
found via the SED fit. For this, we used a spectrum created by
aligning and stacking eight of the TWIN spectra obtained in August
2011 into a single spectrum with SNR$\sim25$. The stacked spectrum is
shown in Figure~\ref{fig:TWIN_spec}.

\begin{table}
\centering
\begin{tabular}{ccc}
  \hline
  \hline
  HJD&RV&$\sigma_{\rm RV}$\\
  &(km/s)&(km/s)\\
  \hline
  2455721.417173&   -19.794&   1.540\\
  2455721.501447&   -19.122&   2.015\\
  2455721.586693&   -19.091&   1.820\\
  2455762.651947&   -19.050&   1.888\\
  2455762.659494&   -18.415&   1.670\\
  2455763.590589&   -21.118&   1.560\\
  2455763.658298&   -20.079&   1.614\\
  2455763.665845&   -18.796&   1.656\\
  2455783.377699&   -22.410&   1.599\\
  2455783.567465&   -19.215&   1.374\\
  2455783.645763&   -18.072&   1.888\\
  2455783.656805&   -19.118&   1.851\\
  2455784.508995&   -19.121&   1.132\\
  2455784.661690&   -20.206&   1.710\\
  2455784.672731&   -20.004&   1.313\\
  2455785.444262&   -19.560&   1.265\\
  2455785.508347&   -21.585&   1.863\\
  2455785.668461&   -19.728&   1.971\\
  \hline
\end{tabular}
\caption{Reconnaissance radial velocities from CAHA 3.5m/TWIN.}
\label{tab:recon}
\end{table}

\subsection{HET high-resolution spectroscopy}
High-resolution spectroscopic observations of WTS-2 were obtained
between August-November 2011 at the McDonald Observatory in Austin,
Texas, using the High Resolution Spectrograph (HRS, \citealt{Tul98})
at the Hobby-Eberly Telescope (HET). These spectra were used to
measure the RV variations of the star and hence calculate the
Keplerian parameters of the WTS-2 spectroscopic orbit (see
Section~\ref{sec:precisionRV}), and to measure the bisector variations
to help assess false-positive scenarios (see
Section~\ref{sec:falsepos}). The relative faintness of WTS-2
necessitated a large aperture telescope in order to achieve
high-precision RV measurements. HET has an effective aperture of 9.2
meters \citep{Ram98} and sits at a fixed elevation angle of
$55^{\circ}$, rotating in azimuth to access $81\%$ of the sky visible
from the observatory. HRS is a single-object fibre-coupled
spectrograph with two additional sky fibres that uses a mosaic of two
R-4 echelle gratings with cross-dispersing gratings to separate the
spectral orders. We used an effective slit width of $2\arcsec$ with
the 600g5271 grating to give a resolution of $R=60,000$ and a
wavelength coverage of $\sim4400-6280$\AA, separated into 40 echelle
orders across the two CCD detectors (18 on the red CCD, 22 on the blue
CCD). Each science image was a 1 hour integration, split into
$2\times30$ minute exposures. Due to the faintness of the star, we did
not use the Iodine gas cell but instead observed several exposures of
the ThAr arc lamp before and after each science frame for wavelength
calibration and to monitor any systematic shifts. A high
signal-to-noise (SNR) exposure of a white dwarf was also obtained as a
telluric standard.

The {\sc iraf.echelle}\footnote{http://iraf.net/irafdocs/ech.pdf}
package was used to reduce the HET spectra. After subtracting the bias
and flat-fielding the images, the science and sky spectra for each 30
minute exposure were extracted order-by-order, and the corresponding
sky spectrum was then subtracted. Wavelength calibration was achieved
using the extracted ThAr arc lamp spectra. The dispersion functions
calculated for the ThAr spectra (RMS$\sim0.003$\AA) taken before and
after the science frames were checked for consistency and then
linearly interpolated to create a final dispersion function to apply
to the stellar spectra. No significant drift or abnormalities were
observed in the wavelength solution during each observing run. Before
combining the two 30-minute exposures at each epoch, the individual
spectra were continuum-normalised, filtered for residual cosmic rays,
and corrected for telluric features at the redder wavelengths (using
the extracted white dwarf spectrum) using a custom set of {\sc matlab}
programs. After combining the exposures at each epoch we obtained a
total of seven spectra with average SNRs of $\sim15$.

We note here that due to the faintness of WTS-2, the cores of the
deepest lines in the HRS spectra are distorted during the calibration
process, particularly after sky subtraction. This means only the
weaker lines in these high-resolution spectra are suitable for any
detailed spectroscopic analysis of the host star, such as abundance
calculations or measuring the projected rotational velocity (see
Section~\ref{sec:char_star}).

\begin{figure}
\centering
\includegraphics[width=0.5\textwidth]{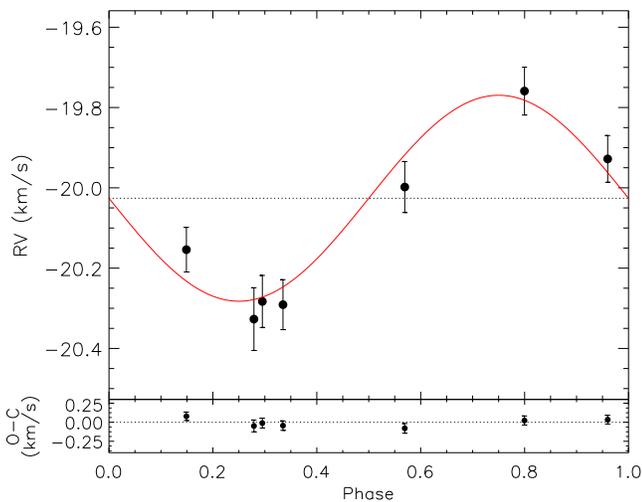}
\caption{{\bf Top:} The radial velocity curve of WTS-2 as a function
  of the orbital phase, measured using high-resolution spectra from
  HET. The red solid curve shows the best-fitting model to the data
  ($K_{\star}=256\pm32$ m/s, see Section~\ref{sec:precisionRV}), while
  the dotted horizontal line marks the measured systemic velocity of
  the system. The best-fitting model parameters are given in
  Table~\ref{tab:parameters}. A circular orbit was assumed in the
  model. The RV error bars in this plot have been scaled such that the
  fit gives $\chi^{2}_{\nu}=1$. {\bf Bottom:} Residuals after
  subtracting the best-fitting model.}
\label{fig:rv_het}
\end{figure}

To measure the RVs, each echelle order in the spectrum was
cross-correlated with a synthetic template using {\sc iraf.fxcor}. The
template was taken from the MAFAGS-OS grid of model atmospheres
\citep{Grup04} with $\teff=4800$K, $\log(g)=4.5$ and solar
metallicity. The template parameters are within the errors of the
final host star properties obtained by the detailed analysis in
Section~\ref{sec:char_star}, and the variation of the RVs for
different templates within these errors is negligible compared to the
errors on the measured RVs. The RVs reported in
Table~\ref{tab:HET_RVs} and shown in Figure~\ref{fig:rv_het} are the
mean RV from all the echelle orders at a given epoch with the
uncertainties equal to the standard deviation on the mean of the RVs.

\begin{table}
\centering
\begin{tabular}{cccccccc}
  \hline
  \hline
  HJD&Phase&RV&$\sigma _{\rm RV}$&BS&$\sigma_{\rm BS}$\\
  &&(km/s)&(km/s)&(km/s)&(km/s)\\
  \hline
  2455790.83253&0.9599&-19.922&0.043&1.73&0.90\\
  2455822.73790&0.2790&-20.332&0.061&0.83&0.93\\
  2455845.68006&0.7997&-19.761&0.046&0.72&0.95\\
  2455856.65080&0.5693&-20.021&0.047&-0.44&0.81\\
  2455867.61787&0.3349&-20.295&0.051&0.04&0.90\\
  2455869.61445&0.2952&-20.282&0.048&-0.48&0.75\\
  2455876.59697&0.1490&-20.115&0.040&0.08&0.86\\
  \hline
\end{tabular}
\caption{Radial velocity data for WTS-2 derived from the HET/HRS
  spectra and their associated bisector span (BS) variations (see
  Section~\ref{sec:falsepos}). The error on each
  RV data point ($sigma_{\rm RV}$) in this table is the standard
  deviation on the mean of measured
  RVs across all echelle orders for that epoch. The phases are
  calculated using the best-fitting period from the simultaneous light
  curve analysis in Section~\ref{sec:transitfit}.}
\label{tab:HET_RVs}
\end{table}

\subsection{High-resolution $i/z$-band AstraLux/CAHA lucky
  imaging}\label{sec:lucky}
\begin{figure}
  \centering
{\setlength{\fboxsep}{0pt}
\setlength{\fboxrule}{0.5pt}
  \fbox{\includegraphics[width=0.4\textwidth]{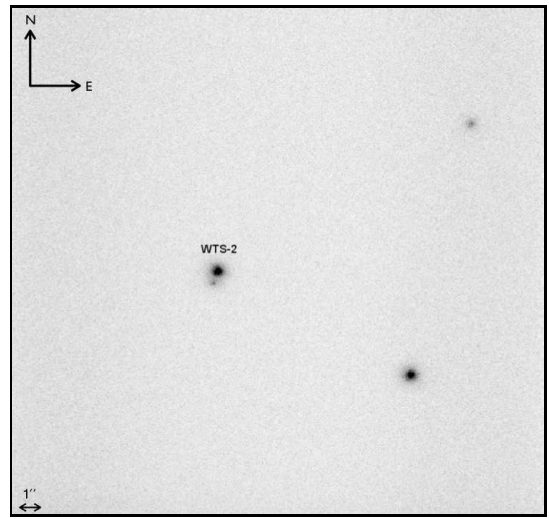}}}
\caption{AstraLux $z$-band image of WTS-2 and two other bright stars
  in the field. The field of view is $24\times24$ arcsec. A faint
  source is clearly visible $0.567\pm0.005$ arcsec South of WTS-2.}
  \label{fig:AstraLux_z}
\end{figure}

Although the WFCAM $J$-band survey images are of relatively high
spatial resolution ($0.4$ arcsec/pixel) compared to most ground-based
transit surveys, in order to adequately address false positive
scenarios and search for unresolved stellar companions, we obtained
high-resolution images of WTS-2 with the lucky imaging camera AstraLux
\citep{Horm08} mounted on the CAHA 2.2m telescope. The observations
were carried out on the night of June 14th 2013, with a mean seeing of
0.6 arcsec. We obtained $30000$ frames in the $i$- and $z$-bands with
single frame exposure times of $0.08$ and $0.06$ s, respectively. The
basic reduction, frame selection and image combination were carried
out with the AstraLux pipeline\footnote{www.mpia.mpg.de/ASTRALUX}
\citep{horm07}. During the reduction process, the images are resampled
to half their pixel size. The calculated plate solution is then
$23.61\pm0.20$ mas/pixel (Lillo-Box et al. 2014, submitted). The plate
scale was measured with the {\sc ccmap} package of {\sc iraf} by
matching the $XY$ positions of $66$ stars identified in an AstraLux
image with their counterparts in the \citet{Yan94} catalog of the
Hubble Space Telescope (see \citealt{Lil12} for a more detailed
explanation of this method which was used to study $\sim100$ Kepler
planet host candidates). For our analysis, we used the best $1\%$ of
exposures in the $i$- and $z$-bands, which have PSFs with FWHMs of
$0.24$ and $0.18$ arcsec, respectively. Figure \ref{fig:AstraLux_z}
shows the $z$-band stack in which a faint source is visible
$0.567\pm0.005$ arcsec South of WTS-2. We performed an iterative PSF
fitting of WTS-2 and this nearby source to estimate the flux of each
of them. The PSF was constructed using the two brighter sources in the
field of view. We find that the nearby faint source is contributing
$10.4\pm1.0\%$ and $13.1\pm1.0\%$ of the total light in the $i$- and
$z$-bands, respectively. We can exclude any other companions beyond a
projected separation of 0.4 arcsec from WTS-2 down to a magnitude
difference of $\Delta m=3$ mag at the $3\sigma$ level.

Motivated by this result, we extended our analysis to $ZYJHK$-band
images taken with WFCAM (see Section~\ref{sec:broadband}). Although
these data have a significantly larger pixel scale ($0.4$
arcsec/pixel) and considerably larger PSFs (FWHM $\sim1.2-1.7$
arcsec), we were able to perform a simultaneous fit of two PSFs and
estimated in this way the blending light coming from the faint source
that is located South of WTS-2. For the PSF fitting we made use of the
position information we obtained from the high resolution AstraLux
images by restricting the separation and position angle to the one
measured on the $i$- and $z$-band images. For the five WFCAM bands we
find that the faint source is contributing $9.5\pm3.0\%$,
$10.7\pm4.0\%$, $19.0\pm4.0\%$, $19.8\pm4.0\%$ and $22.5\pm3.0\%$ in
the $Z$-, $Y$-, $J$-, $H$- and $K$-bands respectively.

The resulting magnitudes for the contaminant are thus as follows:
SDSS-$i=17.66$ mag, WFCAM-$Z=17.12$ mag, SDSS-$z=17.14$ mag,
WFCAM-$Y=16.97$ mag, WFCAM-$J=15.77$ mag, WFCAM-$H=15.23$ mag and
WFCAM-$K=14.98$ mag. The $H-K$ and $J-H$ colours of the contaminant
correspond to a spectral type of $\sim$M$1$V and $\teff\sim3600K$ when
compared with the \citet{Bar98} models. This situation is similar to
the hot Jupiter WASP-12 b which was also recently shown to be diluted
by a faint M-dwarf at $1$ arcsec separation \citep{Cross12}. If the
M-dwarf in our aperture is gravitationally bound to the K-dwarf, the
projected separation would correspond to an orbital separation of
$\sim600$ AU, and an orbital period of $\sim12~500$ years, assuming a
face-on, circular orbit. To assess the likelihood of physical
association between the two sources, we estimated the probability that
the faint source is a chance alignment star using the Besan\c{c}on
stellar population synthesis models \citep{Robin03}. We extracted the
predicted number of stars in a 1.6 sq. degree region centred on the
coordinates of WTS-2 in the magnitude range of the faint source
($I=16.95-17.1$ mag, where the SDSS $i$ magnitude was converted to $I$
using the relations on the SDSS
website\footnote{http://www.sdss.org/dr5/algorithms/sdssUBVRITransform.html\#Lupton2005}).
In this range, we find $2\times10^{-4}$ stars per sq.
arcsec. Multiplying this by our aperture area
($\pi(2^{\prime\prime})^{2}$), we find a priori probability of
$<0.26\%$ of finding a suitably faint red star in our aperture. This
value is a factor of ten lower if one only considers stars within the
projected separation of the WTS-2 and the faint source. We therefore
conclude that the faint source is likely to be a wide-orbit companion
to WTS-2, however the small proper motion means confirmation of
physical association could take many years.

\section{Characterization of the host star}\label{sec:char_star}
The properties of the planet WTS-2 b depend directly on the
characterisation of its host star. Due to the faintness of the host
star, the usual method of deriving the stellar parameters from very
high-resolution spectra (see e.g. \citealt{Tor12}) is not appropriate
because the SNR in our high-resolution HET spectra is too low, and
furthermore the previously mentioned issue of distorted features in
the cores of the deepest lines could bias the results. Instead, we use
two datasets of lower resolution and complementary analyses to arrive
at consistent estimates of the stellar properties, albeit with
comparatively larger uncertainties. Table~\ref{tab:parameters} gives
the final adopted parameters and their errors based on the results of
this section.

We note here that we have not corrected the following analysis for
contamination by the faint red source within the aperture or slit of
the observations. The majority of this analysis is based on data at
optical wavelengths where the contribution from the faint red source
is low ($<10\%$), hence our derived host star properties are unlikely
to deviate outside the presented uncertainties when accounting for the
faint red source.

\subsection{Effective temperature, surface gravity, metallicity,
  lithium abundance, and rotation}\label{sec:SED}
\subsubsection{Photometric analysis}
To begin, we refined the initial SED fit to the WFCAM photometry using
{\sc vosa} to add more bandpasses and to explore a wider range of
$\teff$, surface gravities, metallicities, and to fit for reddening.
{\sc vosa} calculates synthetic photometry by convolving theoretical
atmospheric models with the filter transmission curves of the observed
bandpasses, then performs a $\chi^{2}$ minimisation to find the
best-fitting model to the data \citep{Bay08,Bay13}. We used a grid of
Kurucz ATLAS9 model spectra \citep{Cast97} in the range
$3500\leq\teff\leq6000$ K in steps of $250$ K, with $\log(g)=4.0-5.0$
in steps of $0.5$ dex (to be consistent with the light curve stellar
density estimate), [Fe/H]=$[-0.5,0.0,+0.2,+0.5]$, and $0\leq
A_{V}\leq0.5$ in steps of $0.025$. The upper boundary on the
extinction range was chosen to approximately match the total
integrated line-of-sight extinction for the $2$ degree region around
the centre of the 19hr field ($A_{V}=0.439$ mag, $E(B-V)=0.132$ mag),
calculated using the infrared dust maps of
\citet{Sch98}. Figure~\ref{fig:SED} shows the best-fitting SED for
WTS-2, which has a reduced $\chi_{\nu}^{2}=3.9$.  Note that a $2\%$
error was added in quadrature to the SDSS and WFCAM photometric errors
given in Table~\ref{tab:broadband}, to allow for calibration between
the surveys, but that the magnitudes have not been corrected for the
presence of the faint red source within 0.6 arcsec of the brighter
star (see Sec~\ref{sec:lucky}). In addition to the $\chi^{2}$ model
fit, {\sc vosa} performs a Bayesian analysis of the model fit,
resulting in a posterior probability density function covering the
range of fitted values for each parameter. A Gaussian-fit to the
$\teff$ and $A_{V}$ distributions gives approximate errors as follows:
$\teff=5000\pm140$ K and $A_{V}=0.27\pm0.07$. For $\log(g)$ the
distribution is essentially flat due to the intrinsic insensitivity of
the available broadband photometry to gravity sensitive features, so
we adopt $\log(g)=4.5\pm0.5$. For [Fe/H], higher metallicity is
preferred, with the most probable solutions being [Fe/H]$=+0.2$ and
$+0.5$ ($37\%$ and $42\%$, respectively). The $\teff$ from the refined
SED fit is higher than our original estimate, which is not surprising
given that the initial estimate was made using models known to
underestimate $\teff$ for stars earlier than M-type.

\subsubsection{Spectroscopic analysis}
We checked the results of the SED fitting in two ways; firstly by
fitting synthetic spectra to the stacked TWIN spectrum, and secondly
through a standard spectroscopic abundance analysis of the same
spectrum. From the latter, we also derived estimates of the rotational
and microturbulence velocities, and an upper limit on the lithium
abundance. Firstly, we compared the stacked TWIN spectrum of WTS-2 to
synthetic spectra in the \citet{Coe05} library. The spectral library
was generated by the {\sc pfant} code \citep{Barb03}, which computes
the synthetic spectra using the updated ATLAS9 model atmospheres of
\citet{Cast04} (with a mixing length equal to twice the scale height)
and a list of atomic and molecular lines, under the assumption of
local thermodynamic equilibrium. Before we performed a $\chi^{2}$
minimisation to find the best-fitting model, the synthetic spectra
were degraded to the resolution of the TWIN spectra, then normalised
to their continuum along with the observed stacked spectrum. Our model
grid covered $4250\leq\teff\leq5500$ K in steps of $250$ K,
$4.0\leq\log(g)\leq5.0$ in steps of $0.5$, and
[Fe/H]$=[-1.0,-0.5,0.0,+0.2,+0.5]$. We noted that the synthetic
spectra systematically under-predicted the depth of some absorption
features in the Solar spectrum (most likely due to neglect of non-LTE
effects and/or errors in the continuum normalisation) such that our
$\chi^{2}$ minimisation would preferentially select metal-rich spectra
(see \citealt{Cap12} for a more detailed explanation). We therefore
only performed the $\chi^{2}$ analysis on those lines that were
well-reproduced for the Solar spectrum. The best-fitting model was
consistent with the {\sc vosa} result, giving $\teff=5000\pm250$ K,
$\log(g)=4.5\pm 0.5$, and [Fe/H]=$+0.2^{+0.3}_{-0.2}$, where the
errors correspond simply to the step-size in the models. This
corresponds to a spectral type of K$2$V$\pm2$, according to Table B1
of \citet{Gray08}.

For the standard spectroscopic abundance analysis, we measured the
excitation potential of neutral Fe I and ionised Fe II lines in the
TWIN stacked spectrum and compared them to synthetic spectra. All
synthetic spectra were calculated using 1D LTE model atmospheres
computed with SAM12 and WITA6 routines \citep{Pav03} and constants
taken from the VALD2 \citep{Kup99}. For a complete description of our
procedure, see \citet{Pav12}. For a range of synthetic models with
microturbulence velocity $\xi=0.0-2.5$ km/s, in steps of $0.25$ km/s,
and $\teff=4900-5100$ K in steps of $100$ K, we found that the
ionisation equilibrium condition was met at $\xi=0.75\pm0.50$ for
$\log(g)=4.45\pm0.25$. The corresponding iron abundance was
[Fe/H]=$+0.095\pm0.021$, again consistent with the SED-fitting
results.

\begin{figure}
  \centering
  \includegraphics[width=0.5\textwidth]{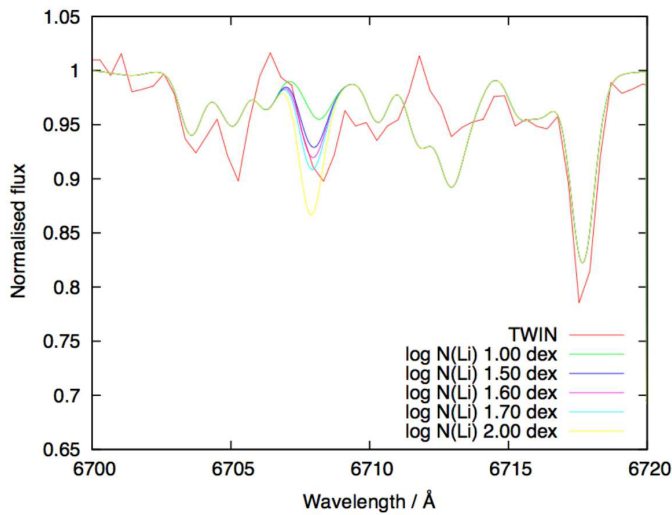}
  \caption{A section of the stacked TWIN spectrum covering the lithium
    feature at $6707.55-6709.21$\AA~with models overlaid for different
    Li abundances. The apparent redshift of the observed Li feature is
    most likely due to noise contamination, so we place an upper limit
    at $\log$N(Li)$_{\rm LTE}=1.8$ dex.}
  \label{fig:lithium}
\end{figure}

To measure the rotational velocity of the star, the $v\sin(i)$ value
was calculated independently for each of the 20 Fe II lines in the
TWIN stacked spectrum, by convolving the model line profile with a set
of rotation profiles \citep{Gray08}, ranging from $0-6$ km/s in steps
of $0.2$ km/s.  The average and standard deviation of all the lines
was $v\sin(i)=2.2\pm1.0$ km/s. Finally, we placed an upper limit on
the lithium abundance of $\log N$(Li) $< 1.8$ ($=12+\log N$(Li/H)),
with an equivalent width upper limit of EW(Li)$\sim 0.089$\AA. Only
upper limits are possible due to noise contamination and relatively
low resolution of the TWIN spectrum (see Figure~\ref{fig:lithium}).

\subsection{Mass and age constraints}\label{sec:mass-age}
\begin{figure}
\centering
\includegraphics[width=0.49\textwidth]{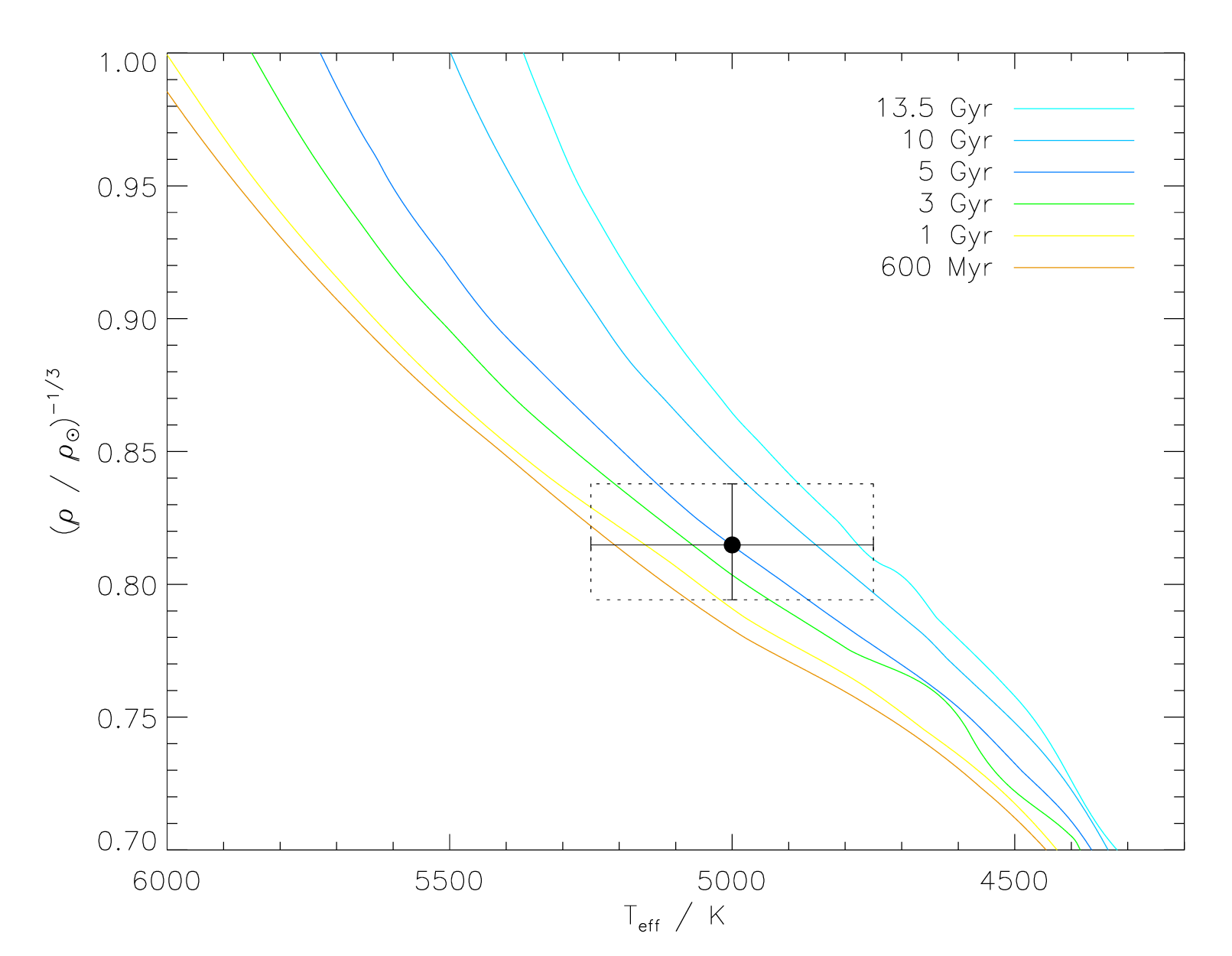}
\caption{Modified Hertzsprung-Russell diagram comparing the effective
  temperature and light curve stellar density for WTS-2, with the
  PARSEC v1.0 stellar evolution isochrones \citep{Bres12} for
  Z=0.019. The data point and its error box marks the allowed values
  for WTS-2. Assuming the star is not on the pre-main sequence, the
  isochrones give an age constraint of $>600$ Myr.}
\label{fig:evo_tracks}
\end{figure}

The mass of WTS-2 was derived using a modified Hertzsprung-Russell
diagram, as shown in Figure~\ref{fig:evo_tracks}, comparing the
spectroscopically measured $\teff$ to the stellar density measured
from the light curves. Model isochrones were generated using the
PARSEC (PAdova and TRieste Stellar Evolution Code) v1.0 code, which
includes the pre-main sequence phase \citep{Bres12}, for $Z=0.019$
(i.e. Solar). The observational errors on $\teff$ allow solutions in
the pre-main sequence phase; however, we rule out young ages using
other indicators. For example, comparing the upper limit on the
lithium abundance of WTS-2 to that observed in open clusters of a
known age \citep{Sest05}, constrains the age to $>250$ Myr. This
already places the system beyond the pre-main sequence phase of the
isochrones. For K-dwarfs, one can also obtain age constraints via
gyrochronology \citep{Barne07,Mam08}, which depends on the stellar
rotation period, usually measured from star spot modulation in the
light curve. However, all significant peaks in the WTS-2 periodogram
are consistent with aliases of the observing window function or of
long-term systematic trends present in all the WTS light curves. This
is not surprising, given that we are using infrared light curves,
which have less contrast between the spot and stellar temperatures,
resulting in lower amplitude rotational modulation signals
\citep{Goul12}. However, the maximum possible rotation period is set
by the upper limit on $v\sin(i)$ ($3.2$ km/s for WTS-2). Using
Equation 7 of \citet{Mald10}, we find that the upper limit on
$v\sin(i)$ is consistent with a gyrochronology lower age limit of
$\gtrsim600$ Myr. Our spectra do not cover sufficient activity
sensitive spectral features so we cannot use the age-activity
relationship, although the lack of emission in the H$\alpha$ line
rules out a very young star. However, the age constraint is in
agreement with association to the young and young-old Galactic disk
\citep{Leg92}, determined from the space velocities given in
Table~\ref{tab:parameters}, which were derived using proper motions
from SDSS DR7 \citep{Munn04,Munn08} and the systemic velocity derived
in Section~\ref{sec:precisionRV}. The model isochrones between
$0.6-13.5$ Gyrs allow a mass range of $M_{\star}=0.820\pm0.082\msun$,
which we adopt as the mass of WTS-2. As a K-dwarf, WTS-2 has a deeper
outer convective envelope than the Sun. According to the models of
\citet{Sad12}, the physical depth of the convective envelope is at
$R_{\rm cz}\sim0.54\rsun$ ($\sim30\%$ of the stellar radius), and it
has a mass of $M_{\rm cz}\sim0.05\msun$ according to the \citet{Pin01}
models (compared to $M_{\rm cz}\sim0.001\msun$ for a late F star such
as WASP-18).

\section{System parameters}\label{sec:char_planet}
The orbital elements and physical properties of WTS-2 b are derived
from a simultaneous fitting of the $J$-band and $i$-band light curves,
then combining the results with a separate analysis of the RVs
measured with HET. Given that we have an estimate of the blended light
contribution in both the $i$ and $J$-band filters, we present an
analysis of both the diluted and dilution-corrected light curves for
completeness. We adopt the dilution-corrected solution for the
remainder of this paper; however, many of the derived parameters are
consistent within the $1\sigma$ error bars from both analyses due to
the relatively large errors on the fractions of blended light. We also
address the limits we can place on the $J$-band secondary eclipse of
WTS-2 b.

\subsection{Light curve analysis}\label{sec:transitfit}
In both the diluted and dilution-corrected cases, the $J$-band and
$i$-band light curves were modelled jointly using the analytic
formulae presented by \citet{Mand02}. A Markov-Chain Monte Carlo
analysis (MCMC) was used to derive the uncertainties on the fitted
parameters and their correlations. We fixed the limb-darkening
coefficients in the fit by adopting values from the tables of
\citet{Cla11}. We used the ATLAS atmospheric models and the flux
conversion method (FCM) to obtain the quadratic law limb-darkening
coefficients in the $i$- and $J$-bands ($\gamma_{1i}, \gamma_{2i},
\gamma_{1J}, \gamma_{2J}$) corresponding to $\teff=5000$ K,
$\log(g)=4.5$, [Fe/H]$=+0.2$, and $\xi=2$ km/s. This gave
$\gamma_{1i}=0.4622$, $\gamma_{2i}=0.1784$, $\gamma_{1J}=0.2609$, and
$\gamma_{2J}=0.2469$. Before fitting the light curves, we applied a
scaling factor to the per data point errors in the $J$- and $i$-band
light curves, such that the out-of-transit data when compared to a
flat line gave a $\chi^{2}_{\nu}$ of unity. This was to account for
any under-estimation of the errors. The following parameters were
allowed to vary in the MCMC analysis: the period ($P$), the epoch of
mid-transit ($T_{0}$), the planet/star radius ratio
($R_{P}/R_{\star}$), the impact parameter ($b=a\cos(i)/R_{\star}$),
where $i$ in the inclination of the system to our line-of-sight, and
the semi-major axis in units of the stellar radius
($a/R_{\star}$). Note that the radius ratio was assumed to be the same
in both the $J$-band and $i$-band transit models. The orbit was
assumed to be circular, hence the eccentricity ($e$) was fixed to
zero. Three chains of $1\times10^{6}$ steps were run each time to
check convergence, then combined after discarding the first $10\%$ of
each chain (the burn-in length).

\begin{figure*}
\centering
\includegraphics[width=\textwidth]{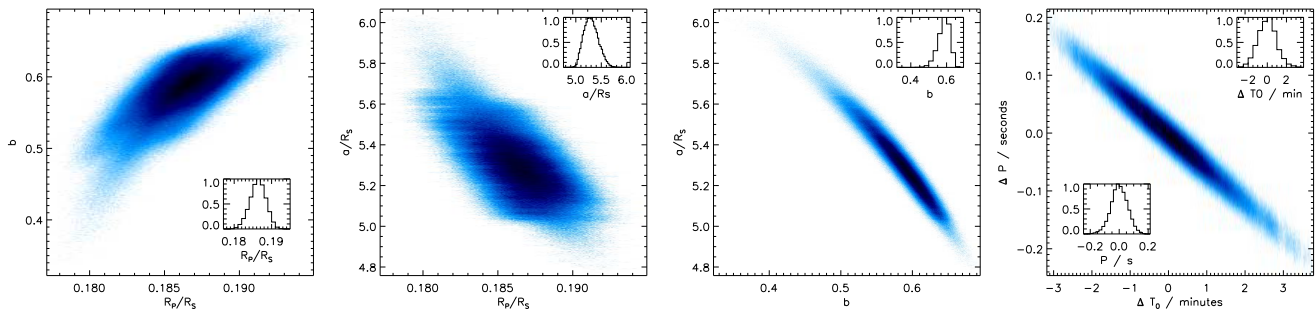}
\caption{Distributions of correlated parameters in the MCMC analysis
  of the dilution-corrected light curves. The insets show individual
  parameter (normalised) histograms to highlight the skew in the
  distributions.}
\label{fig:mcmc_correlations}
\end{figure*}

The dilution-corrected light curves and their combined best-fitting
model are shown in Figure~\ref{fig:lc_i}, and the resulting
best-fitting model parameters are listed in
Table~\ref{tab:parameters}.  Figure~\ref{fig:mcmc_correlations} shows
the extent of correlation between some of the more correlated model
parameters in this analysis. The distributions are not perfectly
Gaussian and result in slightly asymmetric errors for the $68.3\%$
confidence interval about the median. In order to propagate these
errors into the calculation of absolute dimensions, we have
symmetrized the errors by adopting the mean of the $68.3\%$ boundaries
(the $15.85\%$ and $84.15\%$ confidence limits) as the parameter value
(rather than the median), and we then quote the $68.3\%$ confidence
interval as the $\pm1\sigma$ errors.  The full extent of the of
relatively large errors on the blending fractions was explored by
running the MCMC analysis on light curves corrected with the
$\pm1\sigma$ limits of the estimated blending fractions. The quoted
errors and parameter values were derived using the distributions from
all of these runs. The results of fitting the original, diluted light
curves are also given in Table~\ref{tab:parameters} for completeness.

\begin{table*}
  \centering
  \begin{tabular}{lll}
    \hline
    \hline
    \multicolumn{1}{l}{Stellar properties:}
    &Diluted&Dilution-corrected\\
    \hline
    Names& WTS-2&---\\
    & 2MASS 19345587+3648557&---\\
    &SDSS J193455.87+364855.6&---\\
    &WISE J193455.86+364855.6&---\\
    &KIC 1173581&---\\
    RA&19h34m55.87s (293.732792 deg)&---\\
    Dec&+36d48m55.79s (36.815497 deg)&---\\
    $\teff$& $5000\pm250$ K&---\\
    Spectral Type&K$2(\pm2)$V&---\\
    $\log(g)^{a}$& $4.5\pm0.5$&---\\
    $\log(g)^{b}$&$4.589\pm0.023$&$4.600\pm0.023$\\
    $\rm [Fe/H]$&$+0.2^{+0.3}_{-0.2}$&---\\
    $v\sin(i)$& $2.2\pm1.0$ km/s&---\\
    $\xi$& $0.75\pm0.5$ km/s&---\\
    $\log N(Li)_{\rm LTE}$& $<1.8$ dex&---\\
    $M_{\star}^{b}$& $0.820\pm0.082$ $\msun$& $0.820\pm0.082\msun$\\
    $R_{\star}^{b}$&$0.761\pm0.033\rsun$&$0.752\pm0.032\rsun$\\
    $\rho_{\star}^{c}$&$1.86\pm0.15\rho_{\rm{\odot}}$&$1.93\pm0.16\rho_{\rm{\odot}}$\\ 
    $R_{\rm cz}$&$\sim0.54\rsun$&$\sim0.54\rsun$\\
    $M_{\rm cz}$&$\sim0.05\msun$&$\sim0.05\msun$\\
    Age& $>600$ Myr&---\\
    $A_{V}$&$0.27\pm0.07$ mag&---\\
    Distance& $\sim1$ kpc&---\\
    $\mu_{\alpha}\cos\delta$&$2.3\pm2.3$ mas/yr&---\\
    $\mu_{delta}$&$-1.9\pm2.3$ mas/yr&---\\
    $U$&$-13.3\pm5.6$ km/s&---\\
    $V$&$-0.3\pm7.7$ km/s&---\\
    $W$&$-15.1\pm5.3$ km/s&---\\
    \hline
    \multicolumn{1}{l}{System properties:}
    &Diluted&Dilution-corrected\\
    \hline
    $P$&$1.0187074\pm7.1\times10^{-7}$ days&$1.0187068\pm6.5\times10^{-7}$ days\\
    $T_{0}-2454317$&$0.81264\pm6.4\times10^{-4}$ HJD &$0.81333\pm6.5\times10^{-4}$ HJS\\ 
    $R{\rm p}/R_{\star}$&$0.1755\pm0.0018$&$0.1863\pm0.0021$\\
    $b$&$0.597\pm0.032$&$0.584\pm0.033$\\
    $i$&$83.43\pm0.53$ $^{\circ}$&$83.55\pm0.53$ $^{\circ}$\\
    $a$&$0.01855\pm0.00062$ AU&$0.01855\pm0.00062$ AU\\
    &$1.51\pm$$0.11$$a_{Roche}$&$1.44\pm$$0.12$$a_{Roche}$\\
    $\bar{\chi}_{\rm LC}^{2}$&$457.8$&$387.2$\\
    $K$&$256\pm$$32$ m/s&---\\
    $V_{\rm sys}$&$-20.026\pm0.019$ km/s&---\\
    $e$&$0$ (fixed)&---\\
    \hline
    \multicolumn{1}{l}{Planet properties:}
    &Diluted&Dilution-corrected\\
    \hline
    $M_{\rm P}$&$1.12\pm$$0.16$$\mjup$&$1.12\pm0.16\mjup$\\
    $R_{\rm p}$&$1.300\pm0.058\rjup$&$1.363\pm0.061\rjup$\\
    $\rho_{\rm P}$&$0.63\pm$$0.12$ gcm$^{-3}$ ($0.477\pm$$0.093$$\rhojup$)&$0.54\pm$$0.11$ gcm$^{-3}$ ($0.413\pm$$0.080$$\rhojup$)\\
    $g_{b}$&$16.4\pm$$2.7$  ms$^{-2}$&$14.9\pm2.5$  ms$^{-2}$\\    
    $F_{inc}$&$1.29\times10^{9}\pm0.29\times10^{-9}$ erg/s/cm$^{2}$&$1.26\times10^{9}\pm0.29\times10^{-9}$ erg/s/cm$^{2}$\\
    $\rm T_{\rm eq}^{d}$&$2000\pm100$ K&$2000\pm100$ K\\
    $\Theta$&$0.0389\pm$$0.0070$&$0.0371\pm0.0068$\\
    \hline
  \end{tabular}
  \caption{Characterisation of the WTS-2 system. The
    `Dilution-corrected' column is based on an analysis in which the
    light curves were corrected for contamination by a faint red
    source contributing $10.4\pm1\%$ and $19.0\pm4\%$ of the flux in
    the aperture in the $i$- and $J$-bands, respectively (see
    Section~\ref{sec:lucky}). The `Diluted' column gives the results
    without correction for the faint red source.  $^{a}$From
    spectroscopic analysis. $^{b}$From light curve mean stellar
    density and stellar evolution isochrones. $^{c}$From light curve
    analysis. The proper motions $\mu_{\alpha}\cos\delta$ and
    $\mu_{delta}$ are from the SDSS DR7 database. The space velocities
    $U,V,W$ are with respect to the Sun (heliocentric) but for a
    left-handed coordinate system, i.e. U is positive away from the
    Galactic centre. $\bar{\chi}_{\rm LC}^{2}$ is the mean of the
    $\chi^{2}$ values in the MCMC runs. $^{d}$ Equilibrium temperature
    assuming $A_{B}=0$ and $f=2/3$. Note that we used the equatorial
    radius of Jupiter ($\rjup=7.1492\times10^7$m). $a_{Roche}$ is the
    Roche limit separation i.e. the critical distance inside which the
    planet would lose mass via Roche lobe overflow. $g_{b}$ is the
    planet surface gravity according to equation 7 in
    \citet{Sou08}. $\Theta$ is the Safronov number
    ($\Theta=\frac{1}{2}(V_{esc}/V_{orb})^{2}=(a/R_{P})(M_{P}/M_{\star})$)
    \citep{Han07}. The errors on the light curve parameters are the
    $68.3\%$ confidence interval while the parameter value is the mean
    of the $68.3\%$ confidence level boundaries, such that the errors
    are symmetric. Despite its KIC name, WTS-2 b is unfortunately not
    in the Kepler field-of-view.}
  \label{tab:parameters}
\end{table*}

\subsubsection{$J$-band secondary eclipse limits}\label{sec:secondary}
\begin{figure}
\centering
\includegraphics[width=0.49\textwidth]{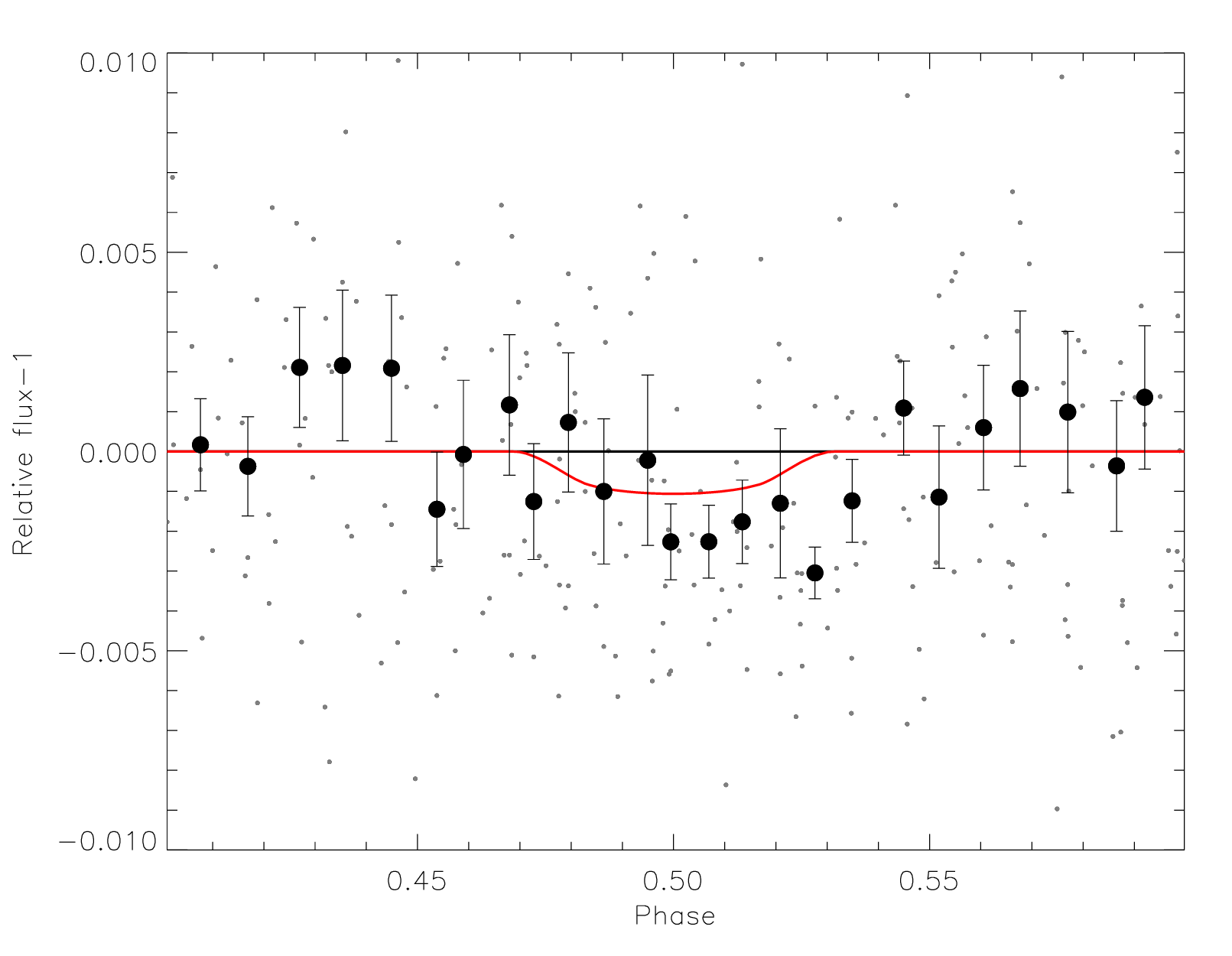}
\caption{The observed WTS-2 b J-band light curve i.e. without
  correction for dilution, zoomed around the expected secondary
  eclipse phase. The observed data is shown in grey dots, while the
  black filled circles are the data binned in phase with uncertainties
  equal to the standard error on the mean. The black solid horizontal
  line shows a flat model i.e. no secondary eclipse detection. The
  solid red line is the best-fitting model light curve from a basic
  linear regression analysis, which gave a depth of
  $(F_{P}/F_{\star})=0.93\times10^{-3}\pm0.69\times10^{-3}$, i.e. a
  non-detection. The model was fitted to data between phases
  $0.4<\phi<0.6$.}
\label{fig:sec_eclip}
\end{figure}

WTS-2 b orbits very close to its host star and receives a high level
of incident radiation ($F_{inc}\sim1.3\times10^{9}$ erg/s/cm$^{2}$).
Following the prescription of \citet{Lop07b} and assuming that the
atmosphere has a zero-albedo ($A_{B}=0$) and instantaneously
re-radiates the incident stellar flux (i.e. no advection, $f=2/3$),
the expected equilibrium temperature of the planet is $\rm T_{\rm
  eq}=2000\pm100$ K. Although this is not as high as the hottest
hot-Jupiters (e.g. KOI-13 b has $\rm T_{\rm eq}\sim2900$ K,
\citealt{Mis12}), WTS-2 b is one of the hottest planets orbiting a
K-dwarf. Adopting this value as the maximum day-side temperature of
the planet and approximating the spectra of the planet and star as
black-bodies, we expect the observed secondary eclipse depth in the
WTS $J$-band light curve to be $\sim0.63$ mmag. This value is
calculated using the dilution-corrected light curve analysis and then
adding back in the $\sim19\%$ contamination of the WTS $J$-band light
curve by the additional red source in the aperture. The out-of-eclipse
data in the WTS-2 b $J$-band light curve has an RMS of $\sim4$ mmag,
and is the typical RMS of the WTS $J$-band light curves at $J=13.9$
mag. There are $88$ data points in the expected secondary eclipse of
the WTS-2 b $J$-band light curve (according to the best-fitting
model). Assuming white-noise only, this would result in a precision of
$\sim0.43$ mmag on the secondary eclipse depth. We performed a basic
linear regression fit to the WTS-2 $J$-band light curve with a model
from the \citet{Mand02} routines to attempt to detect the secondary
eclipse of WTS-2 b. The best-fitting model is shown in
Figure~\ref{fig:sec_eclip} and corresponds to a flux ratio of
$(F_{P}/F_{\star})=0.93\times10^{-3}\pm0.69\times10^{-3}$. The large
uncertainty is unsurprising and means that the WTS survey light curve
is not capable of detecting the secondary eclipse, hence we are unable
to constrain the properties of the planet's day-side from the WTS
data. The sparse sampling of the eclipse in the WTS light curve and
the randomised observing pattern of the survey over many nights make
it difficult to monitor the systematic effects during a single
eclipse, hindering a robust measurement of the flux ratio. However, we
do note that we find no evidence for an anomalously deep event, which
supports the planetary nature of WTS-2 b. A single dedicated night of
observation would in principle be able to measure the eclipse depth to
a sufficient precision. The potential for follow-up studies of the
planet's atmosphere is discussed in Section~\ref{sec:followup}.

\subsection{Radial velocity analysis}\label{sec:precisionRV}
The RV curve has been modelled with constraints from the light curve
fit, rather than being fitted simultaneously with the light curve
data, due to the limited amount of RV data. To fit the RV curve, we
adopted the well-defined period and transit ephemeris from the light
curves and fixed these parameters in the RV curve model. We also fixed
the orbit to be circular as we do not have enough data to model an
eccentric orbit. Furthermore, a circular orbit is arguably the most
reasonable approximation for a planet so close to its host star (see
e.g. \citealt{Ande12}). The model takes the form of:

\begin{equation}
  \rm RV=V_{sys}+K_{\star}\sin(2\pi\phi)
\label{eqn:sinefit}
\end{equation}

\noindent where $\phi$ is the phase, $K_{\star}$ is the RV
semi-amplitude, and $V_{sys}$ is the systemic velocity of the WTS-2
system. The phase-folded radial velocities and the best-fitting model
are plotted in Figure~\ref{fig:rv_het}, while
Table~\ref{tab:parameters} gives the resulting model parameter
values. In the fit, the RV error bars have been scaled by
$\sqrt(\chi^{2}_{\nu})=1.32$ such that $\chi^{2}_{\nu}=1$. This
accounts for possible under-estimation of the RV errors, or
conversely, reflects the quality of the fit, and acts to enlarge the
uncertainties on the model parameters, which are the $1\sigma$ errors
from the $\chi^{2}$-fit. The best-fitting model gives a planet mass of
$1.12\pm0.16 M_{J}$, where the error is calculated by propagating the
errors of the relevant observables ($K_{\star}$, $M_{\star}$, $P$, and
$i$).

\section{Eliminating false positives}\label{sec:falsepos}
Wide-field transit surveys invariably suffer from transit mimics,
usually caused by eclipsing binaries, either as grazing systems, or by
eclipsing binaries contaminated by a source of third light. Given that
WTS-2 b is a relatively unusual planetary system, and the presence of
a faint third light source in our aperture, it is important to
investigate viable false positive scenarios.

\subsection{Non-blended false positives}
Due to the faintness of our target, before proceeding to precision RV
measurements with the 9.2-m HET, we carried out reconnaissance
intermediate-resolution spectroscopy with the 3.5-m telescope at CAHA
to check for large RV variations indicative of non-blended false
positives such as a grazing binary, or a binary containing two
identical size stars whose light curve has been erroneously
phase-folded on half of the true orbital period. The CAHA spectra were
single-lined, with no evidence for a double-peak in the
cross-correlation functions, indicating that the system was not a
non-blended false positive. Such scenarios would also have been
reflected in the stellar density measured from the transit shape, as
the density depends strongly on $P^{-2}$ \citep{Sea03}. The measured
RVs, given in Table~\ref{tab:recon}, had an RMS of $1.1$ km/s and were
consistent with no significant RV variation within the precision of
the measurements, ruling out companion masses $>5\mjup$ for
non-blended scenarios.

\subsection{Blended false positives}
\begin{figure}
  \centering
  \includegraphics[width=0.5\textwidth]{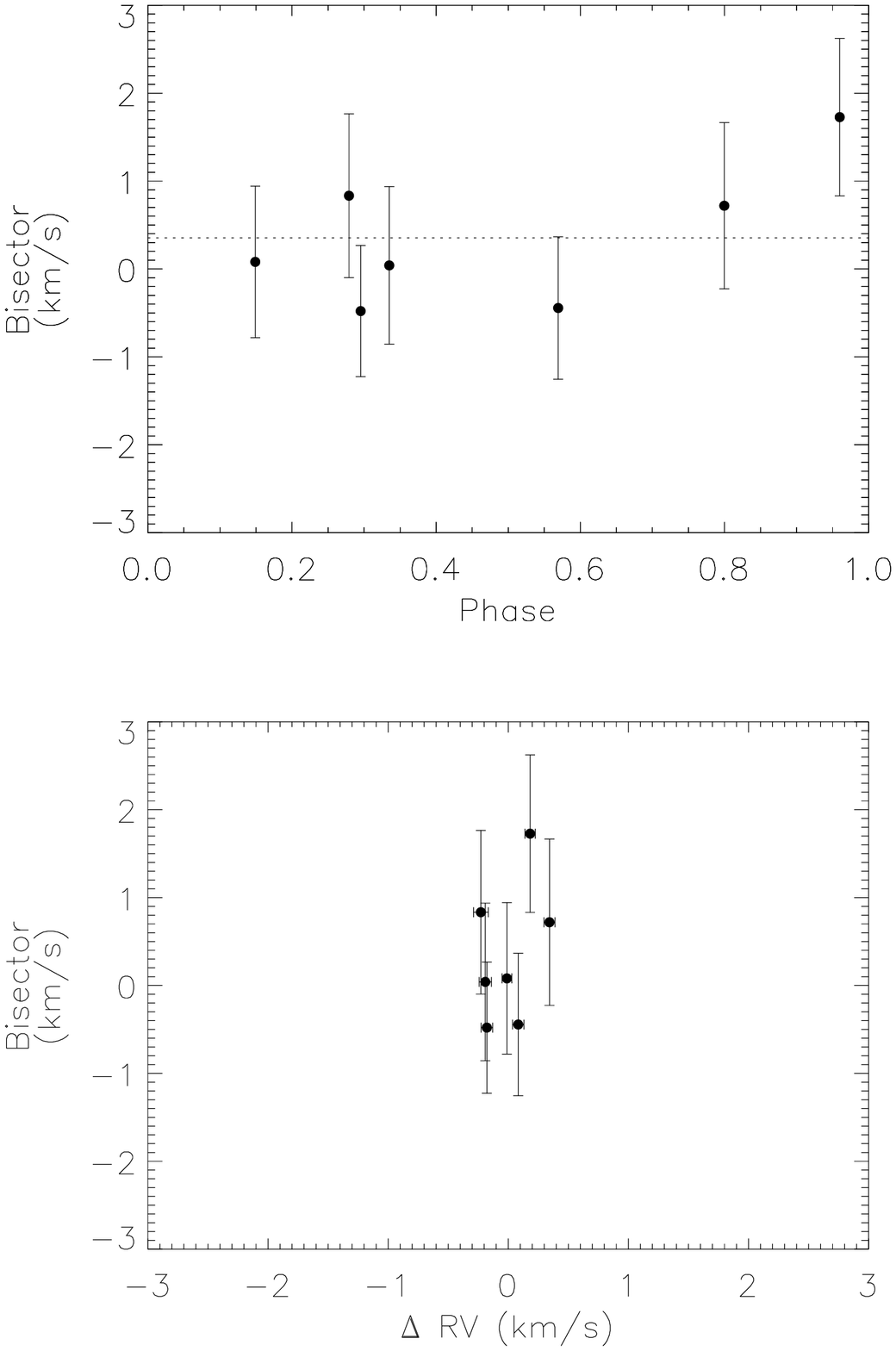}
  \caption{Bisector spans for WTS-2 b measured from high-resolution
    HET spectra, as a function of phase and as a function of the
    change in RV. Due to the large errors on the bisectors, we cannot
    use them to assess the possibility of a blending eclipsing binary
    in the aperture and instead use other methods detailed in this
    section.}
  \label{fig:bisectors}
\end{figure}

Despite the orders of magnitude larger RV variations expected for a
binary system, in the case where the binary spectral lines are blended
with a brighter foreground star, the overall variations in the
cross-correlation profile can have significantly smaller amplitudes,
potentially as small as that expected for a giant planet. Such a
system would produce significant line-profile variations, so we
measured the bisector spans (i.e. the difference between the bisector
values at the top and at the bottom of the correlation function,
\citealt{Tor05}) for each epoch of high-resolution HET spectra. In the
case of contamination from a blended binary, or stellar atmospheric
oscillations, we would expected to measure bisector spans values
consistently different from zero, and as a strong function of the
measured radial velocities \citep{Que01,Man05}.
Figure~\ref{fig:bisectors} shows the measured bisector spans as a
function of phase and RV. Although the bisector span values are
scattered around zero, they have large errors and a RMS scatter
($\sim1$ km/s) that exceeds the measured RV semi-amplitude ($0.256$
km/s). The result is that they are too noisy to conclusively rule out
any blended eclipsing binary scenario.

Instead, to further rule out blended eclipsing binary scenarios, we
consider the following information. Firstly, the transit depths in the
$i$- and $J$-bands are very consistent. Thus, if the light curves were
generated by a background eclipsing binary blended with a bright
foreground K-dwarf, then the colour (i.e. surface temperature) of the
eclipsed star should also be similar to a K-dwarf. Secondly, the mean
stellar density derived from the best-fitting transit model is in
excellent agreement with the stellar density inferred from
spectroscopic observations of the brightest source in the aperture,
i.e. a K-dwarf. Again, this implies the eclipsed star should be
similar in nature to the spectroscopically observed K-dwarf. Now, if
we assume a significant fraction of the light in the observed light
curves originates from the foreground K-dwarf and subtract it, we find
that the transit can no longer be fitted by a K-dwarf star, instead
requiring a cooler, denser star to fit the transit shape, which is in
contradiction to our first two statements. This already indicates that
a blended eclipsing binary scenario can be rejected but it is
important to robustly rule out the detected red object within the
aperture as the source of the occultations.

To further explore the role of additional light in the observed light
curves, we use constraints provided by a simultaneous modelling of the
$i$- and $J$-band light curves. Following a similar method outlined by
\citet{Sne09} and \citet{Kopp13} for assessing the blend scenarios for
OGLE-TR-L9 b and POTS-1 b, we simulate background eclipsing binary
systems blended by different amounts of light from a third star using
the \citet{Mand02} algorithms. This analysis was carried out on the
observed light curves, i.e. the light curves that have not been
corrected for the known amount of dilution by the faint red source
identified by in AstraLux imaging (see Section~\ref{sec:lucky}), and
can thus be considered an independent test of the blending fraction.

\begin{figure}
  \centering
  \includegraphics[width=0.5\textwidth]{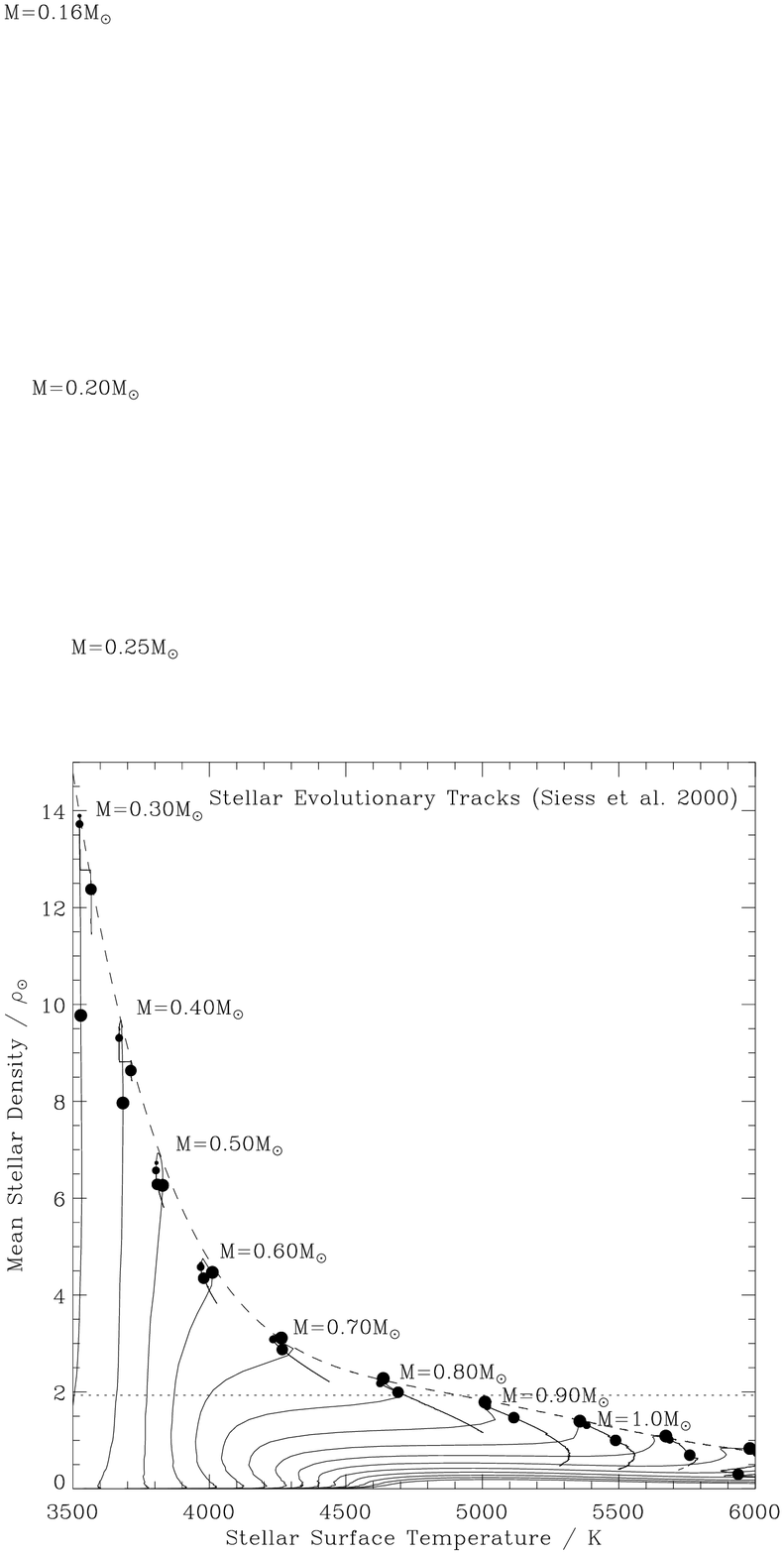}
  \includegraphics[width=0.5\textwidth]{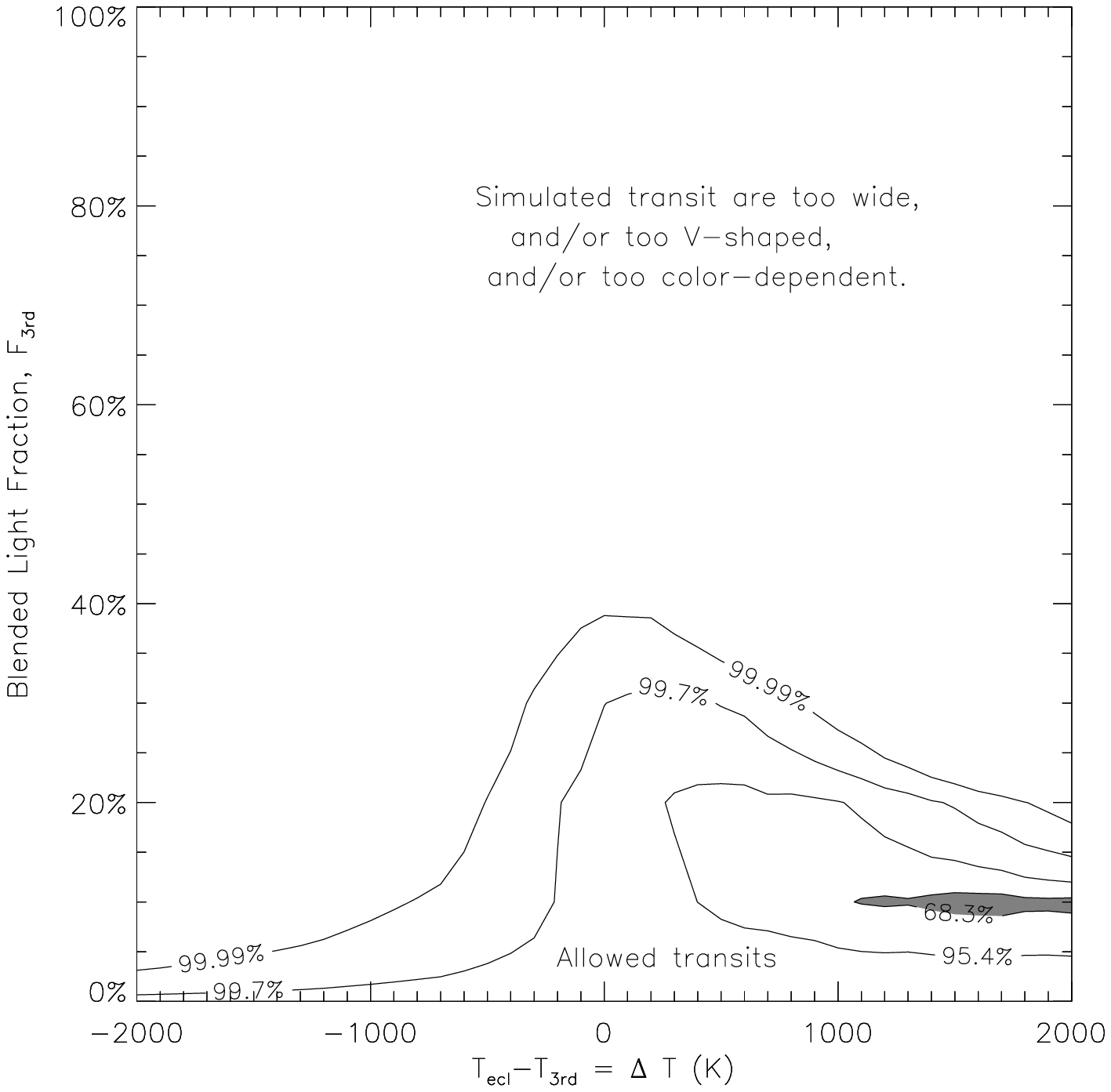}
  \caption{{\bf Top:} Stellar evolutionary tracks showing the
    variation in mean stellar density with stellar surface temperature
    for a range of ages and masses, based on the \citet{Sie00}
    models. The dashed line marks the maximum possible density for a
    given surface temperature. The horizontal dotted line marks the
    stellar density measured from the best-fitting model to the
    dilution-corrected light curves. {\bf Bottom:} Confidence
    intervals from the simultaneous $\chi^{2}$ analysis of all
    possible blended eclipsing binary scenarios fitted to the $i$- and
    $J$-band light curves. The y-axis gives the blended light fraction
    for the $i$-band and the x-axis gives the difference in surface
    temperature between the eclipsed star and the third light
    star. The combined light curves significantly favour a low level
    of blending by a source that is redder than the main contributor
    of light in the aperture. Note that blended background M-dwarf
    eclipsing binaries solutions lie in the upper left of this plot
    and are very poor-fits to the light curves in the two different
    bandpasses.}
  \label{fig:blends}
\end{figure}

In the simulations, we vary two parameters: i) the difference in
surface temperature between the eclipsed star and the blending source,
$\Delta T$, and ii) the fraction of light from the blending source
($0<F_{3^{rd}}<90\%$). The combined light should produce a spectrum
with a temperature that matches the spectroscopic measurement,
i.e. $5000$ K. Any small fraction of light originating from the
eclipsing star is included in $F_{3^{rd}}$. We exclude models which
give stellar densities inconsistent with stellar evolutionary tracks,
although we allow any evolutionary status for the eclipsed star since
we do not insist that the contaminant is bound to the observed
K-dwarf, even though this is quite likely (see
Section~\ref{sec:lucky}). Figure~\ref{fig:blends} shows the upper
limits on the allowed eclipsed star density across a range of masses
based on the \citet{Sie00} evolutionary tracks. For each combination
of $F_{3^{rd}}$ and $\Delta T$, we allow the binary radius ratio and
the impact parameter to vary freely in the simulation, while the
density of the eclipsed star is limited to be below the maximum
density allowed based on the temperature of the eclipsed star
($T_{ecl}$). The fractional contribution of the light from the third
star is also adjusted in each waveband based on blackbody
spectra. Given that we are most concerned about blends with a
background eclipsing M-dwarf, we set the limb darkening coefficients
to be appropriate for a $3500K$ eclipsed star throughout. Although
this is not strictly valid for hotter models, the effect of the limb
darkening is marginal compared to the large chromatic variations
caused by observing in different filters.

The bottom panel of Figure~\ref{fig:blends} shows the $\chi^{2}$
confidence contours of fitting blended transit models to the two light
curves simultaneously. $\Delta T=0$ K corresponds to
$T_{ecl}=T_{3^{rd}}=5000$ K, while $F_{3^{rd}}$ refers to the fraction
of blending light in the $i$-band. It shows that the data can only be
fitted well by a low level of blending light from a source that is
redder than the occulted star. In fact, the preferred solution is for
a blending fraction of $\sim10\%$ by an object of $T_{3^{rd}}\sim3600$
K. This matches extremely well with the independent measurement of the
blended light fraction from the AstraLux imaging (see
Section~\ref{sec:lucky}). If the observed light curve had been
generated by a foreground K-dwarf diluting eclipses from a background
M-dwarf eclipsing binary, the simulations would have congregated in
the upper left corner of the plot. However, these models produce
transit shapes that are too wide, too V-shaped or too color-dependent
to match the data.

Finally, we note that it is unlikely that star spots are responsible
for the RV variations as a $\sim1$-day period would correspond to a
rotational velocity of $\sim40$ km/s for a K2V star, which is
inconsistent with our measured $vsini$ unless there is a high degree
of spin-orbit axis misalignment, which seems to be unlikely for cool
dwarfs \citep{Win10}.

All of these factors combined lead us to conclude that the planetary
nature of the detected system is robust, despite that lack of a
conclusive bisector span analysis.

\section{Discussion}\label{sec:discussion}
\begin{figure}
\centering
\includegraphics[width=0.5\textwidth]{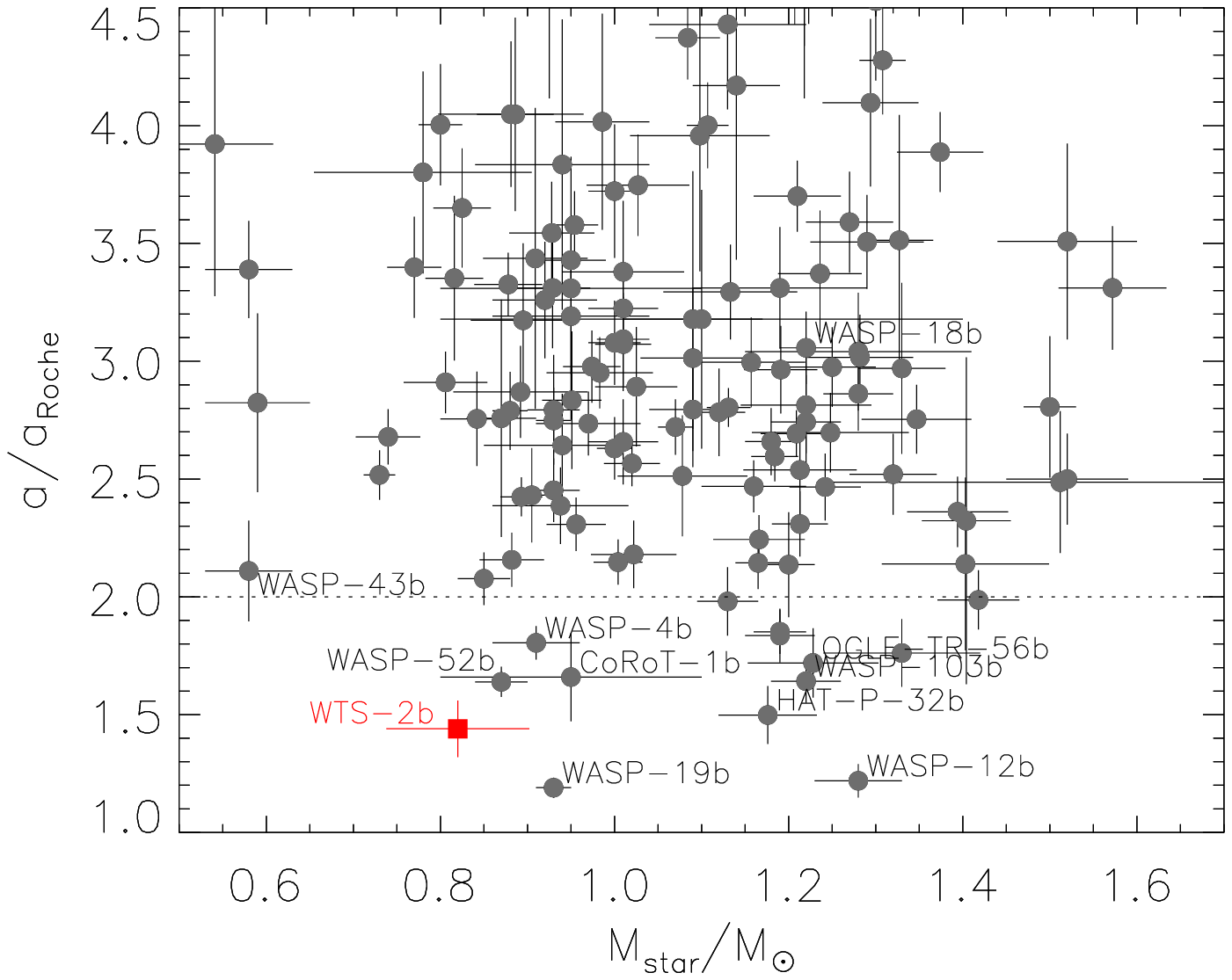}
\caption{The distribution of $a/a_{Roche}$ as a function of stellar
  mass for all known transiting exoplanets. WTS-2 b is marked by the
  red filled square and is one of the closest exoplanets to the
  separation at which it would begin to lose mass. A few other systems
  of note are labelled. The horizontal dashed line marks the ideal
  circularisation radius \citep{Ford06}, i.e. where highly eccentric
  orbits caused by e.g. planet-planet scattering are
  circularised. Data from exoplanets.org.}
\label{fig:roche}
\end{figure}

We have presented WTS-2 b, the second planet to have been discovered
in the infrared light curves of the WTS. The notable property of this
otherwise typical hot Jupiter is its orbital separation of just
$a=0.01855$ AU, which places the planet in the small but growing
sample of extreme giant planets in sub-$0.02$ AU orbits. The planet's
orbit is just 1.5 times the tidal destruction radius i.e. the critical
separation inside which the planet would to lose mass via Roche lobe
overflow, $a_{Roche}\approx2.16R_{P}(M_{\star}/M_{P})^{1/3}$
\citep{Fab05,Ford06}. Figure~\ref{fig:roche} shows the distribution of
$a/a_{Roche}$ as a function of stellar mass for transiting exoplanets,
marking WTS-2 b as one of the closest systems to tidal destruction,
particularly for low-mass host stars. Throughout this discussion, we
use parameter values for WTS-2 b derived from the analysis of the
dilution-corrected light curve (see Section~\ref{sec:transitfit}).

\subsection{Remaining lifetime}
The close proximity of WTS-2b to its host star suggests that its
orbital evolution is dominated by tidal forces
(e.g. \citealt{Ras96,Paz02}). The tide raised on the star by the
planet exerts a strong torque that transfers the angular momentum of
the planetary orbit to the stellar spin
(e.g. \citealt{Gold66,Zah77,Hut81,Egg98}), causing the planet to
spiral inwards and the star to spin up. In our case, the tide raised
on the planet by the star is ignored as we have (reasonably) assumed
that the planet is on a circular orbit and synchronised. Following
\citet{Mats10}, we find that the total angular momentum, $L_{\rm
  tot}$, in the WTS-2 b system compared to the critical angular
momentum, $L_{\rm crit}$, required for the star--planet system to
reach a state of tidal equilibrium, i.e. dual synchronisation, is
$L_{\rm tot}/L_{\rm crit}\sim0.57$ which is $<1$, indicating that
WTS-2 b will never reach a stable orbit and will continue to spiral in
towards the host star under tidal forces until it is inside $a_{\rm
  Roche}$, where it will presumably be destroyed by Roche lobe
overflow \citep{Gu03}. First, let us estimate how long it will it take
before the planet meets its demise and if the orbital decay will be
directly observable on the decade timescale, according the standard
$\qprime=10^{6}$ calibration.

To estimate the remaining lifetime of WTS-2 b, we take a simple model
of tidal interactions, namely the damping of the equilibrium tide by
viscous forces inside the star, i.e. the hydrostatic adjustment of the
star to the imposed gravitational field of the planet, with the tidal
bulge lagging the planet by a constant time \citep{Hut81,Egg98}. Note
that we chose this model as it has been shown by \citet{Soc12} that
the constant time lag model has a better physical motivation than the
constant phase lag model \citep{Gold66}, as it is independent of the
orbital configuration. In this model, the rate of semi-major axis
decay is given by \citep{Mats10}:
\begin{equation}
\frac{\dot{a}}{a}=-6k_2\Delta t \frac{M_{P}}{M_{\star}}\left(\frac{R_{\star}}{a}\right)^5n^2,
\label{eqn:adot}
\end{equation}
under the simplifying assumptions that the relevant tidal frequency is
simply the planet's mean motion ($n=2\pi/P$), the orbit is circular
($e=0$), the planet rotation is synchronised with the orbit, and that
the star is non-rotating. Here, $k_{2}$ is the star's second-order
Love number (related to the star's density profile) and $\Delta t$ is
the constant time lag. In this model, assuming that the planet does
not change the star's spin significantly, integrating
equation~\ref{eqn:adot} gives the future lifetime:
\begin{equation}
t_{\rm life}=-\frac{a^8}{48k_{2}\Delta tGM_{P}R_{\star}^{5}}.
\label{eqn:tlife}
\end{equation}
Note that $t_{\rm life}$ is the time until $a=0$ AU, but that the
difference in time between this and $a=a_{Roche}$ is negligible.  As
mentioned previously, the strength of tidal forces is commonly
parametrised by means of the tidal quality factor $\qprime$, with a
higher $\qprime$ meaning weaker tidal dissipation. While in the highly
simplified constant phase lag model \citep{Gold66} $\qprime$ is a
constant, this is not in general true.  In our adopted constant time
lag model, $\qprime$ is related to the lag time by \citep{Mats10}:
\begin{equation}
\qprime=\frac{3}{4k_2\Delta tn}.
\label{eqn:qprime}
\end{equation}
Adopting $\qprime=10^{6}$ for the current-day WTS-2 b system, based on
previous studies of $\qprime$ \citep{Trill98,Mei05,Jack08}, we find a
remaining lifetime of $\sim40$ Myr, which is just $7\%$ of the
youngest possible age of the system ($\gtrsim600$ Myr), and $<1\%$ for
the more typical older field star ages allowed by the stellar model
isochrones used in Section~\ref{sec:mass-age}. Two situations arise
from this, either i) the system is undergoing a rapid orbital decay
and is genuinely close to destruction, in which case we can measure
the tidal decay directly by monitoring the transit time shift over
tens of years, or ii) $\qprime$ is larger so that the system decays
more slowly, or has a more complicated dependency on other system
parameters.

\subsection{Transit arrival time shift}\label{sec:tshift}
In the case of scenario i), we can calculate how long it would take to
observe a significant shift in the transit arrival time of WTS-2
b. For this, we need to know the current rate of orbital angular
frequency change ($dn/dt$), which can be calculated via the chain rule
using equation~\ref{eqn:adot} and the derivative of Kepler's third law
in terms of $n$ with respect to $a$ (i.e. $dn/da$):
\begin{eqnarray}
  \frac{dn}{dt} = \left(\frac{dn}{da}\right)\dot{a}
  =-\left(\frac{27}{4}\right)n^{2}\left(\frac{M_{P}}{M_{\star}}\right)\left(\frac{R_{\star}}{a}\right)^{5}\left(\frac{1}{\qprime}\right),
\label{eqn:dndt}
\end{eqnarray}
For WTS-2 b, assuming $\qprime=10^{6}$, we find that
{$dn/dt=1.0594599\times10^{-20}$ rad/s$^{2}$. To calculate the
  expected transit time shift, $T_{\rm shift}$, after a time $T$, we
  note that the angle $\theta$ swept out by a planet orbiting with
  angular frequency $n=d\theta/dT$ increasing at a constant rate of
  $dn/dT$ is, via Taylor expansion:
\begin{equation}
\theta=n_{0}T + \frac{1}{2}T^{2}\left(\frac{dn}{dT}\right).
\end{equation}
The angular difference between the linear ephemeris and the quadratic
ephemeris after $T$ years is simply the quadratic term, thus the
transit arrival time shift is:
\begin{equation}
T_{\rm shift}=\frac{1}{2}T^{2}\left(\frac{dn}{dT}\right)\left(\frac{P}{2\pi}\right),
\label{eqn:shift}
\end{equation}
where $P$ is the orbital period. Note that equation~\ref{eqn:dndt} and
equation~\ref{eqn:shift} carry the same assumptions as
equation~\ref{eqn:adot}. Assuming that current instrumentation can
reach a timing accuracy of $5$ seconds (see e.g. \citealt{Gil09}), the
decay of WTS-2 b's orbit would be detectable after $\sim15$ years
($T_{\rm shift}\sim17$ s for $\qprime=10^{6}$), but it remains the
best target to observe this phenomenon for early-to-mid K-dwarf host
stars. If no detectable transit time shift is found in the WTS-2 b
system, it provides a stringent lower limit for the value of $\qprime$
in the sparsely sampled K-dwarf regime, thus helping to constrain
tidal evolution theories that argue $\qprime$ is dependent on the
depth and mass of the convective outer envelope of the host star
\citep{Bark09,Bark10,Pen11}.

We have predicted the $T_{\rm shift}$ values for a sample of known
transiting hot Jupiters ($M_{P}>0.3\mjup$) to determine if direct
observational constraints across the entire mass range of planet host
stars is achievable within a decade. Such constraints could be used to
address the dependence of $\qprime$ on the depth of the stellar
convective envelope. The sample was selected from the exoplanets.org
database, choosing systems with approximately circular orbits
($e<0.01$), and contains $101$ planets (as of $20^{\rm th}$ January
2014). We assume these hot Jupiter systems contain only one planet and
do not have stellar companions, such that additional transit timing
variations can be ignored. Choosing near circular orbit systems also
allows us to neglect issues such as precession of the orbit, the
stellar oblateness, and the value of the planetary tidal quality
factor (although these affects are likely to be small). For these
reasons, the well-known close-in hot Jupiter WASP-12 b is excluded
from our sample, owing to its stellar and planetary companions, and
slightly eccentric orbit \citep{Hus11,Berg13,Maci13}. The left panel
of Figure~\ref{fig:shift} shows our predicted $T_{\rm shift}$ values
for the sample after 10 years as a function of stellar host mass
assuming $\qprime=10^{6}$ in equation~\ref{eqn:dndt}. There are a
number of systems with feasibly observable variations in their transit
arrival time whose host stars span a variety of stellar internal
structures and could provide direct observational constraints on
$\qprime$ within a decade with current instrumentation. The right hand
panel of Figure~\ref{fig:shift} depicts how long one would need to
wait in order to place a lower limit constraint on $\qprime$ in our
adopted model for some of the most perturbed systems, e.g. after $25$
years, if no detectable transit time shift is observed, one could rule
out values of $\qprime<1\times10^{7}$ across a wide range of stellar
masses. Note however that future transit arrival time shift
measurements require similarly accurate measurements of the planets'
current-day periods and ephemerides. Intriguingly, tentative
measurements of period decay rates in WASP-43 b and OGLE-TR-113 b for
example, which orbit M- and K-dwarf host stars, suggest relatively
small values of $\qprime$, on the order of $10^{3}-10^{4}$; however,
further data over several years is required to confirm these results
\citep{Ada10,Ble14,Murg14}.

\begin{figure*}
  \centering
  \includegraphics[width=0.49\textwidth]{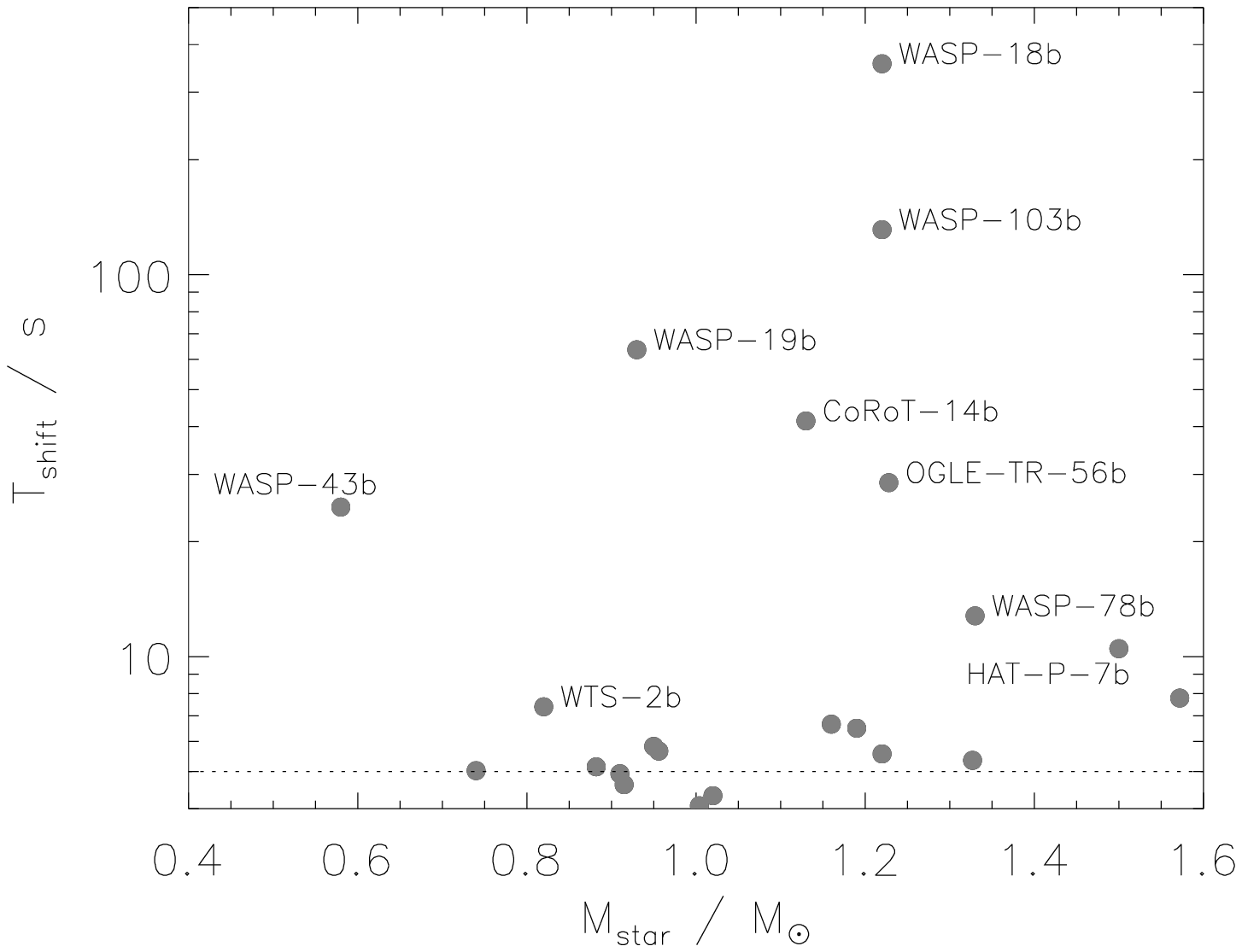}
  \includegraphics[width=0.49\textwidth]{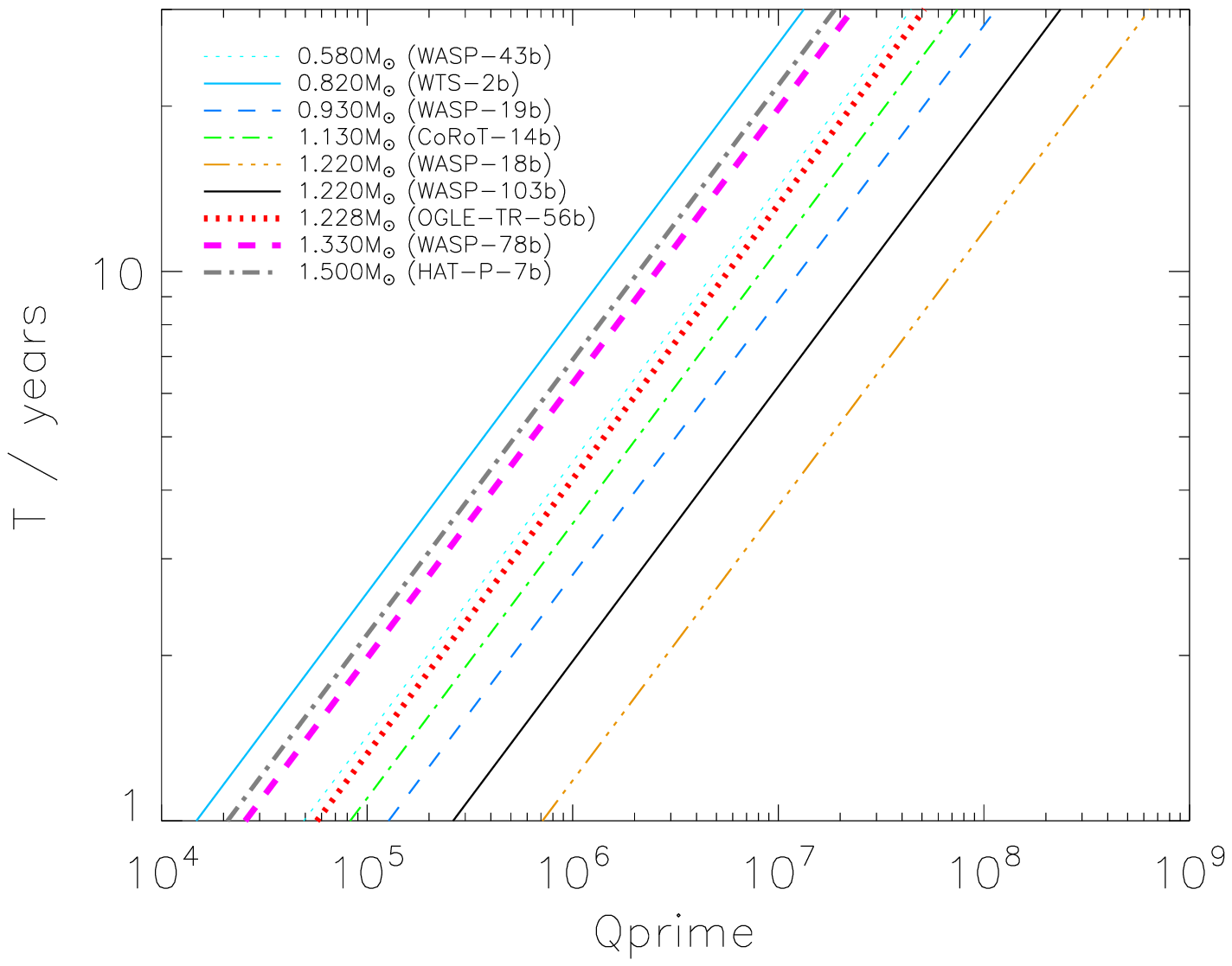}
  \caption{{\bf Left:} Transit time shifts after 10 years for known
    transiting hot Jupiters assuming $\qprime=10^{6}$ in
    equation~\ref{eqn:dndt}. The more significantly shifted planets
    are labeled. The horizontal dotted line marks the 5 second timing
    accuracy possible with current instrumentation. After 10 years,
    strong observational constraints on $\qprime$ would be available
    across the full stellar mass range of exoplanets host stars. {\bf
      Right:} The amount of time after discovery one would need to
    wait to detect $T_{\rm shift}=5$ seconds for a given $\qprime$,
    e.g. after $T\sim25$ years, one could rule out $\qprime\leq10^{7}$
    across a range of stellar masses if no detectable shift was
    observed.}
  \label{fig:shift}
\end{figure*}

\subsubsection{WASP-18 b}\label{sec:wasp-18}
Importantly, we note here that for the most extreme planet, WASP-18 b,
equations~\ref{eqn:dndt} and~\ref{eqn:shift} give
$dn/dt=5.53\times10^{-19}$ rad/s$^{2}$ (which corresponds to a rate of
change of period of $dP/dt=-0.018$ s/yr) and a corresponding $T_{\rm
  shift}\sim356$ seconds after $10$ years for $\qprime=10^{6}$, which
is significantly more than the predicted $T_{\rm shift}=28$ seconds
reported in \citet{Hel09}. There are several differences between our
methods of calculating $T_{\rm shift}$ and the WASP-18 b discovery
paper, for example, \citet{Hel09} used the tidal evolution formalism
of \citet{Dobb04}, which defines $\qprime$ to be a factor of 2
different to our adopted formalism, and they also included the effects
of stellar rotation and the stellar wind which we have neglected here.
However, none of these factors are sufficient to explain the order of
magnitude difference between the predicted $T_{\rm shift}$ values for
WASP-18 b. We also note that a $T_{\rm shift}$ of order 100s of
seconds for WASP-18 b is consistent with scaling the theoretical
calculations of \citet{Pen11} for $\qprime=10^{6}$. Given that our
equations give a similar remaining lifetime for WASP-18 b ($\sim0.72$
Myr) to that reported by \citeauthor{Hel09} ($0.65$ Myr), our orbital
evolution tracks and in-spiral times appear to agree. We have
therefore concluded that a simple numerical error occurred in the
WASP-18 b discovery paper at the final stage of converting the orbital
evolution into $dP/dt$ and a corresponding transit arrival time shift
(Collier Cameron, priv. comm), and that under the assumption of
$\qprime=10^{6}$, observable shifts in the transit timing of WASP-18
will arrive much earlier than previously thought. In fact, we
calculate that a shift of $28$ seconds for the WASP-18 b transit would
only take $\sim3$ years, which is a positive outcome. \citet{Max13}
found no evidence for variations in the times of transit from a linear
ephemeris for WASP-18 b greater than 100 seconds after 3 years, but if
$\qprime$ is genuinely close to $10^{6}$, we expect to see evidence of
this much sooner than a decade. We also note that our predicted timing
variation for WASP-18 b over 10 years is now much larger than that
predicted to be caused by the Applegate effect on similar timescales
\citep{Wat10}.

\subsection{Current observational constraints on
  $\qprime$}\label{sec:pop} Rather than waiting to observe a
decaying orbital period by measuring transit arrival time shifts, can
we already rule out low values of $\qprime$ ($\lesssim10^{6}$)?  For
example, in the individual case of WASP-19 b, \citet{Hel09} suggest
$\qprime\sim10^{7}$, else the probability of observing the planet in
its current evolutionary state is unlikely given the known population
of hot Jupiters. However, the growing number of very close in hot
Jupiters suggests that the population should be treated as
whole. \citet{Pen12} performed a population study of transiting
exoplanets in circular orbits around stars with surface convective
zones, to find a $\qprime$ that would give a statistically likely
distribution of remaining planet lifetimes. They assumed that the
orbits of the planets initially evolved only under gas disc migration
and then by tidal forces alone since the zero-age main-sequence. They
integrated the orbital evolution from $5$ Myrs based on the given ages
of its host star, and argued that $\qprime\gtrsim10^7$ in order to fit
the observed population at the $99\%$ confidence level. Their largest
source of uncertainty was the error on the stellar ages, but even
accounting for this they still found inconsistency with a low values
of $\qprime$. However, \citet{Pen12} point out that their result may
not be valid for other giant planet migration mechanisms, such as
dynamical scattering, and that their model is not valid for stars
without surface convective layers so they excluded any host star with
$M_{\star}>1.25\msun$, which could be subject to a different mode of
tidal dissipation. We also note that high values of $\qprime$ for
those planets deposited close to the host star before the dispersal of
the gas disk ($\lesssim10$ Myr, \citealt{Her07,Wya08}) are perhaps
expected as the tidal migration would need to be slow over the host
star's main sequence lifetime.

Here, we attempt a complementary study to that of \citet{Pen12}, in
that we assume the population of hot Jupiters instead migrated by
scattering onto eccentric orbits (it is interesting to note here that
the likely bound M-dwarf at $0.6$ arcsec separation from WTS-2 is a
potential source of Kozai perturbations which could also trigger the
migration of the gas giant).  Planets scattered such that their
eccentric orbit just grazes $a=a_{Roche}$ are tidally circularised to
$2a_{Roche}$ \citep{Ford06,Nag08}, and we assume that any inside
$2a_{Roche}$ at present-day are assumed to have migrated under tidal
forces alone from there (see Figure~\ref{fig:roche}). The key
difference is that we assume the scattering event can occur at any
point during the planet's total lifetime so the tidal forces have not
necessarily been dominant during the majority of the planet's
lifetime. This assumption means that the pile-up of planets near
$2a_{Roche}$ is constantly replenished. If planets are continuously
falling in from the pile-up at a constant rate in time due to tidal
forces, then our model given in equation 2 will give a distribution of
remaining lifetimes that is uniformly distributed in time. For
example, for every one planet we see with a remaining lifetime of
$0.1-1$ Myr, we expect to see $10$ with remaining lifetimes of $1-10$
Myr, $100$ with remaining lifetimes of $10-100$ Myr, and so on. If the
calculated remaining lifetime distribution for the observed population
diverges from this, our model and adopted value of $\qprime=10^{6}$
are not observationally supported. However, if the distribution
matches, planets such as WASP-18 b and WASP-19 b are consistent with
being genuinely close to destruction and their detection is not so
unlikely. For simplicity, we have used the sample of hot Jupiters that
we created in Section~\ref{sec:tshift}. Due to the dependence of
$\qprime$ on the orbital period, we assign a current-day $\qprime$ to
each system by assuming that it had $\qprime=10^{6}$ at its $3$-day
orbital separation. This ensures that $\Delta t$ in
equation~\ref{eqn:tlife} is constant for all systems, allowing a
physically meaningful comparison between planets.

\begin{figure}
\centering
\includegraphics[width=0.49\textwidth]{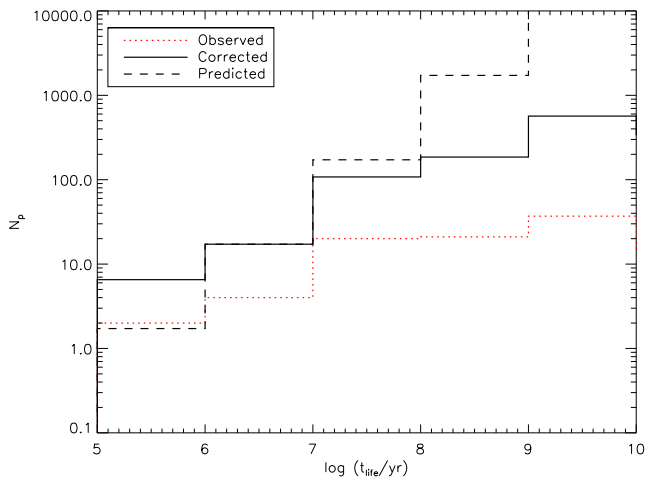}
\caption{Histogram of the calculated remaining lifetimes of observed
  systems (red dotted line) using $\qprime=10^{6}$. The black solid
  line histogram shows the observed distribution after a correction
  for transit probability and survey completeness. The dashed black
  line shows the distribution we expect to observe if planets are
  falling in under tidal forces at a constant rate in time, using
  $\qprime=10^{6}$. The predicted distribution is scaled using the
  1-10 Myr bin which we have assumed is the least affected by
  incompleteness. The corrected and predicted distributions are
  discrepant at long remaining lifetimes, implying that our model of
  the tidal forces is incorrect, but the survey completeness
  correction makes these bins uncertain (see Section~\ref{sec:pop}).}
\label{fig:remaining}
\end{figure}

The resulting distribution of observed remaining lifetimes is shown as
a red dotted-line histogram in Figure~\ref{fig:remaining}. We correct
the observed distribution of remaining lifetimes to account for the
geometrical alignment bias in the transit detection probability, such
that each planet observed is representative of a population of
$\sim(a/R_\star)$ planets. We also correct for survey incompleteness
using the detection probability function described by \citet{Pen12},
which is $100\%$ complete out to $2$-days and tails off at longer
periods. Applying these corrections yields the histogram shown by the
solid black line in Figure~\ref{fig:remaining}. The predicted
distribution (dashed line), i.e. that which increases by a factor of
10 for each bin, is created by scaling to the $1-10$ Myr bin. This bin
was chosen for the scaling as it has the best combination of sample
size and completeness. In the longest remaining lifetime bins, the
bias-corrected distribution is highly discrepant with the predicted
one, suggesting that either $\qprime$ is indeed higher, so that
planets do not typically spiral into their hosts within the age of the
system, or that $\qprime$ may have a complicated frequency dependence
making the future lifetime of the system hard to predict. The latter
possibility is predicted by various dynamical tide mechanisms i.e. the
excitation of normal modes in the star by the imposed gravitational
field (see e.g. \citealt{Ogi07}), with the tidal quality factor
varying by orders of magnitude with small changes in the planet's
orbital frequency as different modes are excited in the star. In this
case, the planets with supposedly short lifetimes could be temporarily
stuck in a region of high $\qprime$ after migrating rapidly from a
feeding region where $\qprime$ is lower. A third possibility is that
mass loss as the planet's size approaches its Roche lobe causes orbit
expansion that retards the tidal decay
\citep{Li10,Fos10,Has12}. However, using the equations of
\citet{Li10}, we estimate that at least in the case of WTS-2b such
mass loss is negligible, around $10^{4}$ times less than that for
WASP-12b.

However, the simplifications in our population study bias us against
longer remaining lifetimes, i.e. we have excluded eccentric systems
which tend to have longer periods, and we do not have a detailed
treatment of the long-period sensitivity of the transit surveys
contributing to the sample. While RV surveys suggest that it is
unlikely the number of longer lifetime (longer period) systems will
increase dramatically, our bias-corrected distribution of remaining
lifetimes is still uncertain, and it is not straight-forward to
reconstruct it.

Although more detailed population studies, such as that by
\citep{Pen12}, strongly advocate $\qprime\gtrsim10^{7}$ for the
general population of exoplanets, this is under one specific set of
initial conditions (e.g. gas disk migration) with some idealised
assumptions about the chances of a planet candidate being confirmed by
follow-up considering the human element involved in its assessment and
the availability of resources. Such studies will always be hampered by
these uncertainties and while they provide some generalised
constraints on $\qprime$, we conclude that the most informative and
straight-forward constraints on $\qprime$ are best obtained through
the monitoring of orbital periods in individual close-in giant planet
systems. Even in the case of no detectable period decay, this places a
constraint on the rate of change of orbital period, and hence
definitive limits on the value of $\qprime$. Importantly, each system
acts as a probe of different parameters that $\qprime$ may be
dependent on e.g. the internal structure of the host star, such that
even a relatively small sample of planets can lead to strong
observational constraints $\qprime$ (see right panel of
Figure~\ref{fig:shift}). To achieve the same results with population
studies, i.e. studying $\qprime$ as a function of host star mass,
would require many more well-characterised systems per host star mass
bin, and although future space-based and ground-based planet discovery
missions may provide this, it is likely to be on a similar timescale
to the technological advancements in precision timing
measurements. Consequently, we find that monitoring changes in orbital
periods of close-in giant planets will be the most informative and
least assumption-prone method for observationally constraining
$\qprime$.

\subsection{Follow-up potential}\label{sec:followup}
In terms of planetary mass and host star, WTS-2 b is very similar to
the well-known hot Jupiter HD 189733 b \citep{Bou05b}. However, WTS-2
b receives almost three times as much incident stellar radiation on
account of its closer orbit, resulting in an expected maximum day-side
temperature that is $\sim500$ K hotter than HD 189733 b. Stellar
irradiation is expected to be a dominant factor in determining the
atmospheric properties of a hot Jupiter (e.g. Fortney et al. 2008).
From this, WTS-2 b is expected to have an inversion layer
(stratosphere) in its atmosphere, caused by gaseous absorbing
compounds \citep{Bur08b}. In cooler atmospheres, these absorbers
condense out, as may be the case for HD189733 b, which does not
exhibit an inversion layer \citep{Char08,Bir13}. Multi-wavelength
measurements of the WTS-2 b secondary eclipse depth will allow the
temperature structure of its atmosphere to be determined. The fact
that WTS-2 b is very hot and orbits a relatively small star means that
its secondary eclipse depths will be deeper compared to other hot
Jupiters of similar $T_{\rm eq}$ orbiting more luminous stars. To
assess the potential of ground-based follow-up studies of WTS-2 b's
atmosphere, we have calculated the expected secondary eclipse depths
for WTS-2 b at optical and infrared wavelengths again following the
equations of \citet{Lop07b}. We approximate the stellar and planetary
spectra as black-bodies, and assume the maximum day-side temperature
for the planet $T_{\rm eff,p}=2000$ K i.e. zero-albedo (no reflection)
and no advection of incident energy from the day-side to the
night-side. The expected planet/star flux ratios in the $I$, $Z$, $J$,
$H$, and $K_{s}$ bandpasses, based on the dilution-corrected light
curve analysis, are $\sim0.14\times10^{-3}$, $\sim0.19\times10^{-3}$,
$\sim0.75\times10^{-3}$, $\sim1.5\times10^{-3}$, and
$\sim2.6\times10^{-3}$, respectively. Note that any potential
follow-up observations would need to add the expected contamination
from the M-dwarf companion to these values. For example, if the
M-dwarf is entirely contained within the photometric aperture, the
expected observed depths would be $\sim0.12\times10^{-3}$,
$\sim0.17\times10^{-3}$, $\sim0.63\times10^{-3}$,
$\sim1.25\times10^{-3}$, and $\sim2.12\times10^{-3}$, in the $I$, $Z$,
$J$, $H$, and $K_{s}$ bandpasses, respectively.

\begin{figure}
\centering
\includegraphics[width=0.49\textwidth]{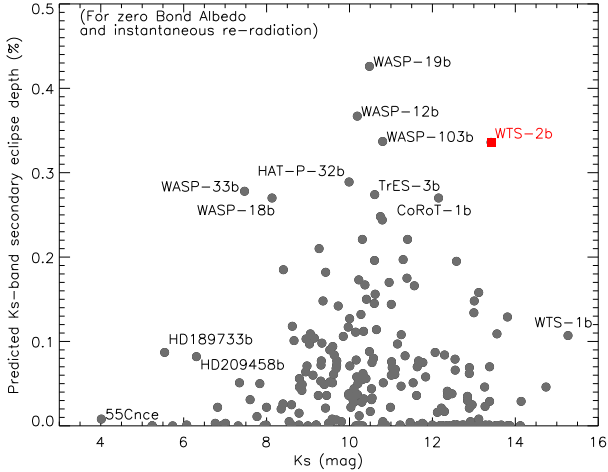}
\caption{Expected $K_{s}$-band secondary eclipse depths for the
  currently known exoplanets as a function of the system's
  $K_{s}$-band magnitude. Eclipse depths are calculated assuming the
  maximum day-side temperature ($A_{B}=0$, $f=2/3$ i.e. no advection
  of energy from the day-side to the night-side) for all planets.
  WTS-2 is marked by the red square and is one of the deepest
  predicted $Ks$-band secondary eclipse depths, making it more
  favourable for ground-based atmospheric characterisation
  studies. Note that these are not measurements and that the true
  depth can deviate e.g. HD189733b ($0.039\%$, \citealt{Swa09}. Also
  note that WTS-2 b is the expected depth without adding in
  contamination from the red companion, and will depend on the spatial
  resolution of the observations. Planet and stellar data from
  exoplanets.org 20/01/2014.}
\label{fig:all_secs}
\end{figure}

Figure~\ref{fig:all_secs} shows the expected non-contaminated
$K_{s}$-band secondary eclipse depth of WTS-2 b in the context of
other transiting exoplanets, again assuming that each planet has
$A_{B}=0$ and $f=2/3$.  We find that WTS-2 b has one of the deepest
predicted $K_{s}$-band secondary eclipses amongst the known exoplanet
population. Although the host star is relatively faint, such a deep
secondary eclipse could potentially be detected with ground-based
infrared facilities. For example, both the Long-slit Intermediate
Resolution Infrared Spectrograph (LIRIS) at the 4-m William Herschel
Telescope in La Palma, and WFCAM on UKIRT in Hawaii have a proven
record for detecting such events (see
e.g. \citealt{Sne07,Moo09,Moo11}). It has also been shown that the
presence of an inversion layer may depend on the activity of the host
star, whereby UV flux from an active host star causes
photodissociation of the absorbing compounds in the planet's upper
atmosphere preventing the temperature inversion \citep{Knu10}. A
measurement of the activity level in the WTS-2 host star is not only a
useful ageing diagnostic, but key to understanding the planet's
atmospheric properties. Measurements of the WTS-2 b secondary eclipse
would also help constrain the eccentricity of the system, and improve
the ephemeris of the orbit, aiding future studies of orbital decay in
the system.

\section{Conclusions}\label{sec:conclusion}
We have reported the discovery of WTS-2 b, a typical transiting hot
Jupiter in an unusually close orbit around a K2V star, which has a
likely gravitationally-bound M-dwarf companion at a projected
separation of $0.6$ arcsec. The proximity of the planet to its host
star places it at just 1.5 times the separation at which it would be
destroyed by Roche lobe overflow. The system provides a calibration
point for theories describing the effect of tidal forces on the
orbital evolution of giant planets, which are poorly constrained by
observations. In particular, the system is useful for constraining
theories that predict host stars with deeper convective envelopes lead
to more efficient tidal dissipation. Using a simple model of tidal
orbital evolution with a tidal dissipation quality factor
$\qprime=10^{6}$, we calculated a remaining lifetime for WTS-2 b of
just $40$ Myr. The decaying orbit corresponds to a shift in the
transit arrival time of WTS-2 b of $\sim17$ seconds after $15$
years. We have also reported a correction to the previously published
predicted shift in the transit arrival time of WASP-18 b, which used a
very similar model for the stellar tides. We have calculated that the
WASP-18 b transit time shift is $356$ seconds after 10 years for
$\qprime=10^{6}$, which is much larger than the previously reported
$28$ seconds. We found that transit arrival time measurements in
individual systems could place stringent observational constraints on
$\qprime$ across the full mass spectrum of exoplanet host stars within
the next decade.  Our attempt to constraint $\qprime$ via a study of
the observed population of currently known transiting hot Jupiters was
inconclusive, requiring a more detailed and precise determination of
transit survey sensitivities at long periods. We conclude that the
most informative and straight-forward constraints on $\qprime$ and the
theory of tidal orbital evolution for exoplanets will be provided by
transit arrival time shifts in individual systems. Finally, WTS-2 b is
one of the most highly irradiated gas giants orbiting a K-dwarf and is
therefore expected to have an inversion layer in its atmosphere. This
is in contrast to the non-inverted atmosphere of HD 189733 b, which
has a very similar planet mass and host star to WTS-2 b, but receives
$\sim3$ times less incident radiation. Despite the relatively faint
magnitude of the host star, the system size ratio and hot day-side
temperature result in predicted infrared secondary eclipses that are
within the reach of current ground-based instrumentation.

\section*{Acknowledgements}
The authors would like to thank A. Collier Cameron and C. Hellier for
their time and help in addressing the WASP-18 b transit arrival time
shift discrepancy.  JLB would also like to thank Doug Lin for some
engaging and very helpful discussions, and to thank our anonymous
referee for asking some very pertinent questions that improved this
manuscript. We also thank the excellent TOs and support staff at
UKIRT, and all those observers who clicked on U/CMP/2. All authors of
this paper have received support from the RoPACS network during this
research, a Marie Curie Initial Training Network funded by the
European Commission’s Seventh Framework Programme.

The United Kingdom Infrared Telescope is operated by the Joint
Astronomy Centre on behalf of the Science and Technology Facilities
Council of the U.K. This article is based on observations made with
the INT operated on the island of La Palma by the ING in the Spanish
Observatorio del Roque de los Muchachos. The Hobby-Eberly Telescope
(HET) is a joint project of the University of Texas at Austin, the
Pennsylvania State University, Stanford University,
Ludwig-Maximilians-Universit\"at M\"unchen, and
Georg-August-Universit\"at G\"ottingen. The HET is named in honor of
its principal benefactors, William P. Hobby and Robert E. Eberly. This
article is based on Calar Alto Observatory, the German-Spanish
Astronomical Center, Calar Alto, jointly operated by the
Max-Planck-Institut f\"ur Astronomie Heidelberg and the Instituto de
Astrof\'isica de Andaluc\'ia (CSIC).

This research has been funded by the Spanish National Plan of R\&D
grants AYA2010-20630, AYA2010-19136, AYA2010-21161-C02-02,
AYA2011-30147-C03-03, AYA2012-38897-C02-01, CONSOLIDER-INGENIO GTC
CSD2006-00070 and PRICIT-S2009/ESP-1496. This work was partly funded
by the Funda\c{c}\~ao para a Ci\^encia e a Tecnologia (FCT)-Portugal
through the project PEst-OE/EEI/UI0066/2011. NL was funded by the
Ram\'on y Cajal fellowship number 08-303-01-02 by the Spanish ministry
of science and innovation. Lillo-Box thanks the CSIC JAE-predoc
program for the PhD fellowship.

This publication makes use of VOSA, developed under the Spanish
Virtual Observatory project supported from the Spanish MICINN through
grant AyA2008-02156.This research has made use of the Exoplanet Orbit
Database and the Exoplanet Data Explorer at exoplanets.org
\citep{Wri10} and the Extrasolar Planets Encyclopaedia exoplanet.eu
\citep{Sch11}. This research uses products from SDSS DR7. Funding for
the SDSS and SDSS-II has been provided by the Alfred P. Sloan
Foundation, the Participating Institutions, the National Science
Foundation, the U.S. Department of Energy, the National Aeronautics
and Space Administration, the Japanese Monbukagakusho, the Max Planck
Society, and the Higher Education Funding Council for England. The
SDSS Web Site is http://www.sdss.org/. This publication makes use of
data products from the Two Micron All Sky Survey, which is a joint
project of the University of Massachusetts and the Infrared Processing
and Analysis Center/California Institute of Technology, funded by the
National Aeronautics and Space Administration and the National Science
Foundation. This work also makes use of NASA’s Astrophysics Data
System (ADS) bibliographic services, and the SIMBAD database, operated
at CDS, Strasbourg, France. {\sc iraf} is distributed by the National
Optical Astronomy Observatory, which is operated by the Association of
Universities for Research in Astronomy (AURA) under cooperative
agreement with the National Science Foundation.

\bibliographystyle{mn2e}
\bibliography{referencesjlb}{}

\begin{thebibliography}{}

\bibitem[\protect\citeauthoryear{{Adams}, {L{\'o}pez-Morales}, {Elliot},
  {Seager} \& {Osip}}{{Adams} et~al.}{2010}]{Ada10}
{Adams} E.~R.,  {L{\'o}pez-Morales} M.,  {Elliot} J.~L.,  {Seager} S.,
  {Osip} D.~J.,  2010, \apj, 721, 1829

\bibitem[\protect\citeauthoryear{{Aigrain} \& {Irwin}}{{Aigrain} \&
  {Irwin}}{2004}]{Aig04}
{Aigrain} S.,  {Irwin} M.,  2004, \mnras, 350, 331

\bibitem[\protect\citeauthoryear{{Anderson}, {Collier Cameron}, {Gillon},
  {Hellier}, {Jehin}, {Lendl}, {Maxted}, {Queloz}, {Smalley}, {Smith},
  {Triaud}, {West}, {Pepe}, {Pollacco}, {S{\'e}gransan}, {Todd} \&
  {Udry}}{{Anderson} et~al.}{2012}]{Ande12}
{Anderson} D.~R.,  {Collier Cameron} A.,  {Gillon} M.,  {Hellier} C.,  {Jehin}
  E.,  {Lendl} M.,  {Maxted} P.~F.~L.,  {Queloz} D.,  {Smalley} B.,  {Smith}
  A.~M.~S.,  {Triaud} A.~H.~M.~J.,  {West} R.~G.,  {Pepe} F.,  {Pollacco} D.,
  {S{\'e}gransan} D.,  {Todd} I.,    {Udry} S.,  2012, \mnras, 422, 1988

\bibitem[\protect\citeauthoryear{{Baraffe}, {Chabrier}, {Allard} \&
  {Hauschildt}}{{Baraffe} et~al.}{1998}]{Bar98}
{Baraffe} I.,  {Chabrier} G.,  {Allard} F.,    {Hauschildt} P.~H.,  1998, \aap,
  337, 403

\bibitem[\protect\citeauthoryear{{Barbuy}, {Perrin}, {Katz}, {Coelho},
  {Cayrel}, {Spite} \& {Van't Veer-Menneret}}{{Barbuy} et~al.}{2003}]{Barb03}
{Barbuy} B.,  {Perrin} M.-N.,  {Katz} D.,  {Coelho} P.,  {Cayrel} R.,  {Spite}
  M.,    {Van't Veer-Menneret} C.,  2003, \aap, 404, 661

\bibitem[\protect\citeauthoryear{{Barker} \& {Ogilvie}}{{Barker} \&
  {Ogilvie}}{2009}]{Bark09}
{Barker} A.~J.,  {Ogilvie} G.~I.,  2009, \mnras, 395, 2268

\bibitem[\protect\citeauthoryear{{Barker} \& {Ogilvie}}{{Barker} \&
  {Ogilvie}}{2010}]{Bark10}
{Barker} A.~J.,  {Ogilvie} G.~I.,  2010, \mnras, 404, 1849

\bibitem[\protect\citeauthoryear{{Barnes}}{{Barnes}}{2007}]{Barne07}
{Barnes} S.~A.,  2007, \apj, 669, 1167

\bibitem[\protect\citeauthoryear{{Batalha}, {Borucki}, {Caldwell},
  {Chandrasekaran}, {Gautier}, {Jenkins} \& {Koch}}{{Batalha}
  et~al.}{2006}]{Bat06}
{Batalha} N.~M.,  {Borucki} W.,  {Caldwell} D.~A.,  {Chandrasekaran} H.,
  {Gautier} T.~N.,  {Jenkins} J.,    {Koch} D.~G.,  2006, in American
  Astronomical Society Meeting Abstracts Vol.~38 of Bulletin of the American
  Astronomical Society, {Optimization of the Kepler Field of View}.
p. 210.08

\bibitem[\protect\citeauthoryear{{Bayo}, {Rodrigo}, {Barrado}, {Solano},
  {Allard} \& {Joergens}}{{Bayo} et~al.}{2013}]{Bay13}
{Bayo} A.,  {Rodrigo} C.,  {Barrado} D.,  {Solano} E.,  {Allard} F.,
  {Joergens} V.,  2013, International Workshop on Stellar Spectral Libraries
  2013, ASICS

\bibitem[\protect\citeauthoryear{{Bayo}, {Rodrigo}, {Barrado Y Navascu{\'e}s},
  {Solano}, {Guti{\'e}rrez}, {Morales-Calder{\'o}n} \& {Allard}}{{Bayo}
  et~al.}{2008}]{Bay08}
{Bayo} A.,  {Rodrigo} C.,  {Barrado Y Navascu{\'e}s} D.,  {Solano} E.,
  {Guti{\'e}rrez} R.,  {Morales-Calder{\'o}n} M.,    {Allard} F.,  2008, \aap,
  492, 277

\bibitem[\protect\citeauthoryear{{Bergfors}, {Brandner}, {Daemgen}, {Biller},
  {Hippler}, {Janson}, {Kudryavtseva}, {Gei{\ss}ler}, {Henning} \&
  {K{\"o}hler}}{{Bergfors} et~al.}{2013}]{Berg13}
{Bergfors} C.,  {Brandner} W.,  {Daemgen} S.,  {Biller} B.,  {Hippler} S.,
  {Janson} M.,  {Kudryavtseva} N.,  {Gei{\ss}ler} K.,  {Henning} T.,
  {K{\"o}hler} R.,  2013, \mnras, 428, 182

\bibitem[\protect\citeauthoryear{{Birkby}, {Nefs}, {Hodgkin}, {Kov{\'a}cs},
  {Sip{\H o}cz}, {Pinfield}, {Snellen} \& {Mislis}}{{Birkby}
  et~al.}{2012}]{Bir12}
{Birkby} J.,  {Nefs} B.,  {Hodgkin} S.,  {Kov{\'a}cs} G.,  {Sip{\H o}cz} B.,
  {Pinfield} D.,  {Snellen} I.,    {Mislis} D. e.~a.,  2012, \mnras, 426, 1507

\bibitem[\protect\citeauthoryear{{Birkby}, {de Kok}, {Brogi}, {de Mooij},
  {Schwarz}, {Albrecht} \& {Snellen}}{{Birkby} et~al.}{2013}]{Bir13}
{Birkby} J.~L.,  {de Kok} R.~J.,  {Brogi} M.,  {de Mooij} E.~J.~W.,  {Schwarz}
  H.,  {Albrecht} S.,    {Snellen} I.~A.~G.,  2013, \mnras, 436, L35

\bibitem[\protect\citeauthoryear{{Blecic}, {Harrington}, {Madhusudhan},
  {Stevenson}, {Hardy}, {Cubillos}, {Hardin}, {Bowman}, {Nymeyer}, {Anderson},
  {Hellier}, {Smith} \& {Collier Cameron}}{{Blecic} et~al.}{2014}]{Ble14}
{Blecic} J.,  {Harrington} J.,  {Madhusudhan} N.,  {Stevenson} K.~B.,  {Hardy}
  R.~A.,  {Cubillos} P.~E.,  {Hardin} M.,  {Bowman} O.,  {Nymeyer} S.,
  {Anderson} D.~R.,  {Hellier} C.,  {Smith} A.~M.~S.,    {Collier Cameron} A.,
  2014, \apj, 781, 116

\bibitem[\protect\citeauthoryear{{Bouchy}, {Udry}, {Mayor}, {Moutou}, {Pont},
  {Iribarne}, {da Silva}, {Ilovaisky}, {Queloz}, {Santos}, {S{\'e}gransan} \&
  {Zucker}}{{Bouchy} et~al.}{2005}]{Bou05b}
{Bouchy} F.,  {Udry} S.,  {Mayor} M.,  {Moutou} C.,  {Pont} F.,  {Iribarne} N.,
   {da Silva} R.,  {Ilovaisky} S.,  {Queloz} D.,  {Santos} N.~C.,
  {S{\'e}gransan} D.,    {Zucker} S.,  2005, \aap, 444, L15

\bibitem[\protect\citeauthoryear{{Bressan}, {Marigo}, {Girardi}, {Salasnich},
  {Dal Cero}, {Rubele} \& {Nanni}}{{Bressan} et~al.}{2012}]{Bres12}
{Bressan} A.,  {Marigo} P.,  {Girardi} L.,  {Salasnich} B.,  {Dal Cero} C.,
  {Rubele} S.,    {Nanni} A.,  2012, \mnras, 427, 127

\bibitem[\protect\citeauthoryear{{Burrows}, {Budaj} \& {Hubeny}}{{Burrows}
  et~al.}{2008}]{Bur08b}
{Burrows} A.,  {Budaj} J.,    {Hubeny} I.,  2008, \apj, 678, 1436

\bibitem[\protect\citeauthoryear{{Cappetta}, {Saglia}, {Birkby},
  {Koppenhoefer}, {Pinfield}, {Hodgkin}, {Cruz} \& {Kov{\'a}cs}}{{Cappetta}
  et~al.}{2012}]{Cap12}
{Cappetta} M.,  {Saglia} R.~P.,  {Birkby} J.~L.,  {Koppenhoefer} J.,
  {Pinfield} D.~J.,  {Hodgkin} S.~T.,  {Cruz} P.,    {Kov{\'a}cs} G. e.~a.,
  2012, \mnras, 427, 1877

\bibitem[\protect\citeauthoryear{{Casali}, {Adamson}, {Alves de Oliveira},
  {Almaini}, {Burch}, {Chuter}, {Elliot}, {Folger}, {Foucaud}, {Hambly},
  {Hastie}, {Henry}, {Hirst}, {Irwin} \& et al.}{{Casali} et~al.}{2007}]{Cas07}
{Casali} M.,  {Adamson} A.,  {Alves de Oliveira} C.,  {Almaini} O.,  {Burch}
  K.,  {Chuter} T.,  {Elliot} J.,  {Folger} M.,  {Foucaud} S.,  {Hambly} N.,
  {Hastie} M.,  {Henry} D.,  {Hirst} P.,  {Irwin} M.,    et al. 2007, \aap,
  467, 777

\bibitem[\protect\citeauthoryear{{Castelli}, {Gratton} \& {Kurucz}}{{Castelli}
  et~al.}{1997}]{Cast97}
{Castelli} F.,  {Gratton} R.~G.,    {Kurucz} R.~L.,  1997, \aap, 318, 841

\bibitem[\protect\citeauthoryear{{Castelli} \& {Kurucz}}{{Castelli} \&
  {Kurucz}}{2004}]{Cast04}
{Castelli} F.,  {Kurucz} R.~L.,  2004, IAU Symp. No 210, Modelling of Stellar
  Atmospheres

\bibitem[\protect\citeauthoryear{{Charbonneau}, {Knutson}, {Barman}, {Allen},
  {Mayor}, {Megeath}, {Queloz} \& {Udry}}{{Charbonneau} et~al.}{2008}]{Char08}
{Charbonneau} D.,  {Knutson} H.~A.,  {Barman} T.,  {Allen} L.~E.,  {Mayor} M.,
  {Megeath} S.~T.,  {Queloz} D.,    {Udry} S.,  2008, \apj, 686, 1341

\bibitem[\protect\citeauthoryear{{Claret} \& {Bloemen}}{{Claret} \&
  {Bloemen}}{2011}]{Cla11}
{Claret} A.,  {Bloemen} S.,  2011, \aap, 529, A75

\bibitem[\protect\citeauthoryear{{Coelho}, {Barbuy}, {Mel{\'e}ndez}, {Schiavon}
  \& {Castilho}}{{Coelho} et~al.}{2005}]{Coe05}
{Coelho} P.,  {Barbuy} B.,  {Mel{\'e}ndez} J.,  {Schiavon} R.~P.,    {Castilho}
  B.~V.,  2005, \aap, 443, 735

\bibitem[\protect\citeauthoryear{{Crossfield}, {Barman}, {Hansen}, {Tanaka} \&
  {Kodama}}{{Crossfield} et~al.}{2012}]{Cross12}
{Crossfield} I.~J.~M.,  {Barman} T.,  {Hansen} B.~M.~S.,  {Tanaka} I.,
  {Kodama} T.,  2012, \apj, 760, 140

\bibitem[\protect\citeauthoryear{{de Mooij}, {de Kok}, {Nefs} \& {Snellen}}{{de
  Mooij} et~al.}{2011}]{Moo11}
{de Mooij} E.~J.~W.,  {de Kok} R.~J.,  {Nefs} S.~V.,    {Snellen} I.~A.~G.,
  2011, \aap, 528, A49+

\bibitem[\protect\citeauthoryear{{de Mooij} \& {Snellen}}{{de Mooij} \&
  {Snellen}}{2009}]{Moo09}
{de Mooij} E.~J.~W.,  {Snellen} I.~A.~G.,  2009, \aap, 493, L35

\bibitem[\protect\citeauthoryear{{Dobbs-Dixon}, {Lin} \&
  {Mardling}}{{Dobbs-Dixon} et~al.}{2004}]{Dobb04}
{Dobbs-Dixon} I.,  {Lin} D.~N.~C.,    {Mardling} R.~A.,  2004, \apj, 610, 464

\bibitem[\protect\citeauthoryear{{Eggleton}, {Kiseleva} \& {Hut}}{{Eggleton}
  et~al.}{1998}]{Egg98}
{Eggleton} P.~P.,  {Kiseleva} L.~G.,    {Hut} P.,  1998, \apj, 499, 853

\bibitem[\protect\citeauthoryear{{Faber}, {Rasio} \& {Willems}}{{Faber}
  et~al.}{2005}]{Fab05}
{Faber} J.~A.,  {Rasio} F.~A.,    {Willems} B.,  2005, \icarus, 175, 248

\bibitem[\protect\citeauthoryear{{Ford} \& {Rasio}}{{Ford} \&
  {Rasio}}{2006}]{Ford06}
{Ford} E.~B.,  {Rasio} F.~A.,  2006, \apjl, 638, L45

\bibitem[\protect\citeauthoryear{{Fossati}, {Haswell}, {Froning}, {Hebb},
  {Holmes}, {Kolb}, {Helling} \& {Carter}}{{Fossati} et~al.}{2010}]{Fos10}
{Fossati} L.,  {Haswell} C.~A.,  {Froning} C.~S.,  {Hebb} L.,  {Holmes} S.,
  {Kolb} U.,  {Helling} C.,    {Carter} A. e.~a.,  2010, \apjl, 714, L222

\bibitem[\protect\citeauthoryear{{Gillon}, {Anderson}, {Collier-Cameron},
  {Delrez}, {Hellier}, {Jehin}, {Lendl}, {Maxted} \& et al.}{{Gillon}
  et~al.}{2014}]{Gil14}
{Gillon} M.,  {Anderson} D.~R.,  {Collier-Cameron} A.,  {Delrez} L.,  {Hellier}
  C.,  {Jehin} E.,  {Lendl} M.,  {Maxted} P.~F.~L.,    et al. 2014, \aap, 562,
  L3

\bibitem[\protect\citeauthoryear{{Gillon}, {Smalley}, {Hebb}, {Anderson},
  {Triaud}, {Hellier}, {Maxted}, {Queloz} \& {Wilson}}{{Gillon}
  et~al.}{2009}]{Gil09}
{Gillon} M.,  {Smalley} B.,  {Hebb} L.,  {Anderson} D.~R.,  {Triaud}
  A.~H.~M.~J.,  {Hellier} C.,  {Maxted} P.~F.~L.,  {Queloz} D.,    {Wilson}
  D.~M.,  2009, \aap, 496, 259

\bibitem[\protect\citeauthoryear{{Goldreich} \& {Soter}}{{Goldreich} \&
  {Soter}}{1966}]{Gold66}
{Goldreich} P.,  {Soter} S.,  1966, \icarus, 5, 375

\bibitem[\protect\citeauthoryear{{Goulding}, {Barnes}, {Pinfield},
  {Kov{\'a}cs}, {Birkby}, {Hodgkin}, {Catal{\'a}n}, {Sip{\H o}cz}, {Jones},
  {Del Burgo}, {Jeffers}, {Nefs}, {G{\'a}lvez-Ortiz} \& {Martin}}{{Goulding}
  et~al.}{2012}]{Goul12}
{Goulding} N.~T.,  {Barnes} J.~R.,  {Pinfield} D.~J.,  {Kov{\'a}cs} G.,
  {Birkby} J.,  {Hodgkin} S.,  {Catal{\'a}n} S.,  {Sip{\H o}cz} B.,  {Jones}
  H.~R.~A.,  {Del Burgo} C.,  {Jeffers} S.~V.,  {Nefs} S.,  {G{\'a}lvez-Ortiz}
  M.~C.,    {Martin} E.~L.,  2012, \mnras, 427, 3358

\bibitem[\protect\citeauthoryear{{Gray}}{{Gray}}{2008}]{Gray08}
{Gray} D.~F.,  2008, {The Observation and Analysis of Stellar Photospheres}.
Cambridge University Press

\bibitem[\protect\citeauthoryear{{Grupp}}{{Grupp}}{2004}]{Grup04}
{Grupp} F.,  2004, \aap, 420, 289

\bibitem[\protect\citeauthoryear{{Gu}, {Lin} \& {Bodenheimer}}{{Gu}
  et~al.}{2003}]{Gu03}
{Gu} P.-G.,  {Lin} D.~N.~C.,    {Bodenheimer} P.~H.,  2003, \apj, 588, 509

\bibitem[\protect\citeauthoryear{{Hansen} \& {Barman}}{{Hansen} \&
  {Barman}}{2007}]{Han07}
{Hansen} B.~M.~S.,  {Barman} T.,  2007, \apj, 671, 861

\bibitem[\protect\citeauthoryear{{Haswell}, {Fossati}, {Ayres}, {France},
  {Froning}, {Holmes}, {Kolb}, {Busuttil}, {Street}, {Hebb}, {Collier Cameron},
  {Enoch}, {Burwitz}, {Rodriguez}, {West}, {Pollacco}, {Wheatley} \&
  {Carter}}{{Haswell} et~al.}{2012}]{Has12}
{Haswell} C.~A.,  {Fossati} L.,  {Ayres} T.,  {France} K.,  {Froning} C.~S.,
  {Holmes} S.,  {Kolb} U.~C.,  {Busuttil} R.,  {Street} R.~A.,  {Hebb} L.,
  {Collier Cameron} A.,  {Enoch} B.,  {Burwitz} V.,  {Rodriguez} J.,  {West}
  R.~G.,  {Pollacco} D.,  {Wheatley} P.~J.,    {Carter} A.,  2012, \apj, 760,
  79

\bibitem[\protect\citeauthoryear{{Hebb}, {Collier-Cameron}, {Triaud}, {Lister},
  {Smalley}, {Maxted}, {Hellier} \& {Anderson}}{{Hebb} et~al.}{2010}]{Heb10}
{Hebb} L.,  {Collier-Cameron} A.,  {Triaud} A.~H.~M.~J.,  {Lister} T.~A.,
  {Smalley} B.,  {Maxted} P.~F.~L.,  {Hellier} C.,    {Anderson} e.~a.,  2010,
  \apj, 708, 224

\bibitem[\protect\citeauthoryear{{Hellier}, {Anderson}, {Cameron}, {Gillon},
  {Hebb}, {Maxted} et~al.,}{{Hellier} et~al.}{2009}]{Hel09}
{Hellier} C.,  {Anderson} D.~R.,  {Cameron} A.~C.,  {Gillon} M.,  {Hebb} L.,
  {Maxted} P.~F.~L.,    et~al., 2009, \nat, 460, 1098

\bibitem[\protect\citeauthoryear{{Hellier}, {Anderson}, {Collier Cameron},
  {Doyle}, {Fumel}, {Gillon}, {Jehin} \& {Lendl}}{{Hellier}
  et~al.}{2012}]{Hel12}
{Hellier} C.,  {Anderson} D.~R.,  {Collier Cameron} A.,  {Doyle} A.~P.,
  {Fumel} A.,  {Gillon} M.,  {Jehin} E.,    {Lendl} M. e.~a.,  2012, \mnras,
  426, 739

\bibitem[\protect\citeauthoryear{{Hellier}, {Anderson}, {Collier-Cameron},
  {Miller}, {Queloz}, {Smalley} \& {Southworth}}{{Hellier}
  et~al.}{2011}]{Hel11}
{Hellier} C.,  {Anderson} D.~R.,  {Collier-Cameron} A.,  {Miller} G.~R.~M.,
  {Queloz} D.,  {Smalley} B.,    {Southworth} J. e.~a.,  2011, \apjl, 730, L31

\bibitem[\protect\citeauthoryear{{Hern{\'a}ndez}, {Hartmann}, {Megeath},
  {Gutermuth}, {Muzerolle}, {Calvet}, {Vivas}, {Brice{\~n}o}, {Allen},
  {Stauffer}, {Young} \& {Fazio}}{{Hern{\'a}ndez} et~al.}{2007}]{Her07}
{Hern{\'a}ndez} J.,  {Hartmann} L.,  {Megeath} T.,  {Gutermuth} R.,
  {Muzerolle} J.,  {Calvet} N.,  {Vivas} A.~K.,  {Brice{\~n}o} C.,  {Allen} L.,
   {Stauffer} J.,  {Young} E.,    {Fazio} G.,  2007, \apj, 662, 1067

\bibitem[\protect\citeauthoryear{{Hodgkin}, {Irwin}, {Hewett} \&
  {Warren}}{{Hodgkin} et~al.}{2009}]{Hod09}
{Hodgkin} S.~T.,  {Irwin} M.~J.,  {Hewett} P.~C.,    {Warren} S.~J.,  2009,
  \mnras, 394, 675

\bibitem[\protect\citeauthoryear{{Hormuth}}{{Hormuth}}{2007}]{horm07}
{Hormuth} F.,  2007, Master's thesis, University of Heidelberg

\bibitem[\protect\citeauthoryear{{Hormuth}, {Hippler}, {Brandner}, {Wagner} \&
  {Henning}}{{Hormuth} et~al.}{2008}]{Horm08}
{Hormuth} F.,  {Hippler} S.,  {Brandner} W.,  {Wagner} K.,    {Henning} T.,
  2008, in Society of Photo-Optical Instrumentation Engineers (SPIE) Conference
  Series Vol.~7014 of Society of Photo-Optical Instrumentation Engineers (SPIE)
  Conference Series, {AstraLux: the Calar Alto lucky imaging camera}

\bibitem[\protect\citeauthoryear{{Howard}, {Marcy}, {Bryson}, {Jenkins},
  {Rowe}, {Batalha}, {Borucki} \& {Koch}}{{Howard} et~al.}{2012}]{How11}
{Howard} A.~W.,  {Marcy} G.~W.,  {Bryson} S.~T.,  {Jenkins} J.~M.,  {Rowe}
  J.~F.,  {Batalha} N.~M.,  {Borucki} W.~J.,    {Koch} D.~G. e.~a.,  2012,
  \apjs, 201, 15

\bibitem[\protect\citeauthoryear{{Husnoo}, {Pont}, {H{\'e}brard}, {Simpson},
  {Mazeh}, {Bouchy}, {Moutou}, {Arnold}, {Boisse}, {D{\'{\i}}az}, {Eggenberger}
  \& {Shporer}}{{Husnoo} et~al.}{2011}]{Hus11}
{Husnoo} N.,  {Pont} F.,  {H{\'e}brard} G.,  {Simpson} E.,  {Mazeh} T.,
  {Bouchy} F.,  {Moutou} C.,  {Arnold} L.,  {Boisse} I.,  {D{\'{\i}}az} R.~F.,
  {Eggenberger} A.,    {Shporer} A.,  2011, \mnras, 413, 2500

\bibitem[\protect\citeauthoryear{{Hut}}{{Hut}}{1981}]{Hut81}
{Hut} P.,  1981, \aap, 99, 126

\bibitem[\protect\citeauthoryear{{Irwin}, {Irwin}, {Aigrain}, {Hodgkin}, {Hebb}
  \& {Moraux}}{{Irwin} et~al.}{2007}]{Irw07}
{Irwin} J.,  {Irwin} M.,  {Aigrain} S.,  {Hodgkin} S.,  {Hebb} L.,    {Moraux}
  E.,  2007, \mnras, 375, 1449

\bibitem[\protect\citeauthoryear{{Irwin} \& {Lewis}}{{Irwin} \&
  {Lewis}}{2001}]{Irw01}
{Irwin} M.,  {Lewis} J.,  2001, New Astronomy Review, 45, 105

\bibitem[\protect\citeauthoryear{{Jackson}, {Greenberg} \& {Barnes}}{{Jackson}
  et~al.}{2008}]{Jack08}
{Jackson} B.,  {Greenberg} R.,    {Barnes} R.,  2008, \apj, 678, 1396

\bibitem[\protect\citeauthoryear{{Knutson}, {Howard} \& {Isaacson}}{{Knutson}
  et~al.}{2010}]{Knu10}
{Knutson} H.~A.,  {Howard} A.~W.,    {Isaacson} H.,  2010, \apj, 720, 1569

\bibitem[\protect\citeauthoryear{{Koppenhoefer}, {Saglia}, {Fossati},
  {Lyubchik}, {Mugrauer}, {Bender}, {Lee}, {Riffeser}, {Afonso}, {Greiner},
  {Henning}, {Neuh{\"a}user}, {Snellen}, {Pavlenko}, {Verdugo} \&
  {Vogt}}{{Koppenhoefer} et~al.}{2013}]{Kopp13}
{Koppenhoefer} J.,  {Saglia} R.~P.,  {Fossati} L.,  {Lyubchik} Y.,  {Mugrauer}
  M.,  {Bender} R.,  {Lee} C.-H.,  {Riffeser} A.,  {Afonso} P.,  {Greiner} J.,
  {Henning} T.,  {Neuh{\"a}user} R.,  {Snellen} I.~A.~G.,  {Pavlenko} Y.,
  {Verdugo} M.,    {Vogt} N.,  2013, \mnras

\bibitem[\protect\citeauthoryear{{Kov{\'a}cs}, {Hodgkin}, {Sip{\H o}cz},
  {Pinfield}, {Barrado}, {Birkby}, {Cappetta}, {Cruz}, {Koppenhoefer},
  {Mart{\'{\i}}n}, {Murgas}, {Nefs}, {Saglia} \& {Zendejas}}{{Kov{\'a}cs}
  et~al.}{2013}]{Kov12}
{Kov{\'a}cs} G.,  {Hodgkin} S.,  {Sip{\H o}cz} B.,  {Pinfield} D.,  {Barrado}
  D.,  {Birkby} J.,  {Cappetta} M.,  {Cruz} P.,  {Koppenhoefer} J.,
  {Mart{\'{\i}}n} E.~L.,  {Murgas} F.,  {Nefs} B.,  {Saglia} R.,    {Zendejas}
  J.,  2013, \mnras, 433, 889

\bibitem[\protect\citeauthoryear{{Kupka}, {Piskunov}, {Ryabchikova}, {Stempels}
  \& {Weiss}}{{Kupka} et~al.}{1999}]{Kup99}
{Kupka} F.,  {Piskunov} N.,  {Ryabchikova} T.~A.,  {Stempels} H.~C.,    {Weiss}
  W.~W.,  1999, \aaps, 138, 119

\bibitem[\protect\citeauthoryear{{Lawrence}, {Warren}, {Almaini}, {Edge},
  {Hambly}, {Jameson}, {Lucas}, {Casali}, {Adamson}, {Dye}, {Emerson},
  {Foucaud}, {Hewett}, {Hirst}, {Hodgkin}, {Irwin}, {Lodieu}, {McMahon} \& et
  al.}{{Lawrence} et~al.}{2007}]{Lawr07}
{Lawrence} A.,  {Warren} S.~J.,  {Almaini} O.,  {Edge} A.~C.,  {Hambly} N.~C.,
  {Jameson} R.~F.,  {Lucas} P.,  {Casali} M.,  {Adamson} A.,  {Dye} S.,
  {Emerson} J.~P.,  {Foucaud} S.,  {Hewett} P.,  {Hirst} P.,  {Hodgkin} S.~T.,
  {Irwin} M.~J.,  {Lodieu} N.,  {McMahon} R.~G.,    et al. 2007, \mnras, 379,
  1599

\bibitem[\protect\citeauthoryear{{Leggett}}{{Leggett}}{1992}]{Leg92}
{Leggett} S.~K.,  1992, \apjs, 82, 351

\bibitem[\protect\citeauthoryear{{Li}, {Miller}, {Lin} \& {Fortney}}{{Li}
  et~al.}{2010}]{Li10}
{Li} S.-L.,  {Miller} N.,  {Lin} D.~N.~C.,    {Fortney} J.~J.,  2010, \nat,
  463, 1054

\bibitem[\protect\citeauthoryear{{Lillo-Box}, {Barrado} \& {Bouy}}{{Lillo-Box}
  et~al.}{2012}]{Lil12}
{Lillo-Box} J.,  {Barrado} D.,    {Bouy} H.,  2012, \aap, 546, A10

\bibitem[\protect\citeauthoryear{{L{\'o}pez-Morales} \&
  {Seager}}{{L{\'o}pez-Morales} \& {Seager}}{2007}]{Lop07b}
{L{\'o}pez-Morales} M.,  {Seager} S.,  2007, \apjl, 667, L191

\bibitem[\protect\citeauthoryear{{Maciejewski}, {Dimitrov}, {Seeliger},
  {Raetz}, {Bukowiecki}, {Kitze}, {Errmann}, {Nowak} \& et al.}{{Maciejewski}
  et~al.}{2013}]{Maci13}
{Maciejewski} G.,  {Dimitrov} D.,  {Seeliger} M.,  {Raetz} S.,  {Bukowiecki}
  {\L}.,  {Kitze} M.,  {Errmann} R.,  {Nowak} G.,    et al. 2013, \aap, 551,
  A108

\bibitem[\protect\citeauthoryear{{Maldonado}, {Mart{\'{\i}}nez-Arn{\'a}iz},
  {Eiroa}, {Montes} \& {Montesinos}}{{Maldonado} et~al.}{2010}]{Mald10}
{Maldonado} J.,  {Mart{\'{\i}}nez-Arn{\'a}iz} R.~M.,  {Eiroa} C.,  {Montes} D.,
     {Montesinos} B.,  2010, \aap, 521, A12

\bibitem[\protect\citeauthoryear{{Mamajek} \& {Hillenbrand}}{{Mamajek} \&
  {Hillenbrand}}{2008}]{Mam08}
{Mamajek} E.~E.,  {Hillenbrand} L.~A.,  2008, \apj, 687, 1264

\bibitem[\protect\citeauthoryear{{Mandel} \& {Agol}}{{Mandel} \&
  {Agol}}{2002}]{Mand02}
{Mandel} K.,  {Agol} E.,  2002, \apjl, 580, L171

\bibitem[\protect\citeauthoryear{{Mandushev}, {Torres}, {Latham},
  {Charbonneau}, {Alonso}, {White}, {Stefanik}, {Dunham}, {Brown} \&
  {O'Donovan}}{{Mandushev} et~al.}{2005}]{Man05}
{Mandushev} G.,  {Torres} G.,  {Latham} D.~W.,  {Charbonneau} D.,  {Alonso} R.,
   {White} R.~J.,  {Stefanik} R.~P.,  {Dunham} E.~W.,  {Brown} T.~M.,
  {O'Donovan} F.~T.,  2005, \apj, 621, 1061

\bibitem[\protect\citeauthoryear{{Mann}, {Gaidos}, {L{\'e}pine} \&
  {Hilton}}{{Mann} et~al.}{2012}]{Mann12}
{Mann} A.~W.,  {Gaidos} E.,  {L{\'e}pine} S.,    {Hilton} E.~J.,  2012, \apj,
  753, 90

\bibitem[\protect\citeauthoryear{{Matsumura}, {Peale} \& {Rasio}}{{Matsumura}
  et~al.}{2010}]{Mats10}
{Matsumura} S.,  {Peale} S.~J.,    {Rasio} F.~A.,  2010, \apj, 725, 1995

\bibitem[\protect\citeauthoryear{{Maxted}, {Anderson}, {Doyle}, {Gillon},
  {Harrington}, {Iro}, {Jehin}, {Lafreni{\`e}re}, {Smalley} \&
  {Southworth}}{{Maxted} et~al.}{2013}]{Max13}
{Maxted} P.~F.~L.,  {Anderson} D.~R.,  {Doyle} A.~P.,  {Gillon} M.,
  {Harrington} J.,  {Iro} N.,  {Jehin} E.,  {Lafreni{\`e}re} D.,  {Smalley} B.,
     {Southworth} J.,  2013, \mnras, 428, 2645

\bibitem[\protect\citeauthoryear{{Meibom} \& {Mathieu}}{{Meibom} \&
  {Mathieu}}{2005}]{Mei05}
{Meibom} S.,  {Mathieu} R.~D.,  2005, \apj, 620, 970

\bibitem[\protect\citeauthoryear{{Mislis} \& {Hodgkin}}{{Mislis} \&
  {Hodgkin}}{2012}]{Mis12}
{Mislis} D.,  {Hodgkin} S.,  2012, \mnras, 422, 1512

\bibitem[\protect\citeauthoryear{{Munari}, {Sordo}, {Castelli} \&
  {Zwitter}}{{Munari} et~al.}{2005}]{Mun05}
{Munari} U.,  {Sordo} R.,  {Castelli} F.,    {Zwitter} T.,  2005, \aap, 442,
  1127

\bibitem[\protect\citeauthoryear{{Munn}, {Monet}, {Levine}, {Canzian}, {Pier},
  {Harris}, {Lupton}, {Ivezi{\'c}}, {Hindsley}, {Hennessy}, {Schneider} \&
  {Brinkmann}}{{Munn} et~al.}{2004}]{Munn04}
{Munn} J.~A.,  {Monet} D.~G.,  {Levine} S.~E.,  {Canzian} B.,  {Pier} J.~R.,
  {Harris} H.~C.,  {Lupton} R.~H.,  {Ivezi{\'c}} {\v Z}.,  {Hindsley} R.~B.,
  {Hennessy} G.~S.,  {Schneider} D.~P.,    {Brinkmann} J.,  2004, \aj, 127,
  3034

\bibitem[\protect\citeauthoryear{{Munn}, {Monet}, {Levine}, {Canzian}, {Pier},
  {Harris}, {Lupton}, {Ivezi{\'c}}, {Hindsley}, {Hennessy}, {Schneider} \&
  {Brinkmann}}{{Munn} et~al.}{2008}]{Munn08}
{Munn} J.~A.,  {Monet} D.~G.,  {Levine} S.~E.,  {Canzian} B.,  {Pier} J.~R.,
  {Harris} H.~C.,  {Lupton} R.~H.,  {Ivezi{\'c}} {\v Z}.,  {Hindsley} R.~B.,
  {Hennessy} G.~S.,  {Schneider} D.~P.,    {Brinkmann} J.,  2008, \aj, 136, 895

\bibitem[\protect\citeauthoryear{{Murgas}, {Palle}, {Zapatero Osorio},
  {Nortmann}, {Hoyer} \& {Cabrera-Lavers}}{{Murgas} et~al.}{2014}]{Murg14}
{Murgas} F.,  {Palle} E.,  {Zapatero Osorio} M.~R.,  {Nortmann} L.,  {Hoyer}
  S.,    {Cabrera-Lavers} A.,  2014, ArXiv e-prints

\bibitem[\protect\citeauthoryear{{Nagasawa}, {Ida} \& {Bessho}}{{Nagasawa}
  et~al.}{2008}]{Nag08}
{Nagasawa} M.,  {Ida} S.,    {Bessho} T.,  2008, \apj, 678, 498

\bibitem[\protect\citeauthoryear{{Ogilvie} \& {Lin}}{{Ogilvie} \&
  {Lin}}{2007}]{Ogi07}
{Ogilvie} G.~I.,  {Lin} D.~N.~C.,  2007, \apj, 661, 1180

\bibitem[\protect\citeauthoryear{{P{\"a}tzold} \& {Rauer}}{{P{\"a}tzold} \&
  {Rauer}}{2002}]{Paz02}
{P{\"a}tzold} M.,  {Rauer} H.,  2002, \apjl, 568, L117

\bibitem[\protect\citeauthoryear{{Pavlenko}}{{Pavlenko}}{2003}]{Pav03}
{Pavlenko} Y.~V.,  2003, Astronomy Reports, 47, 59

\bibitem[\protect\citeauthoryear{{Pavlenko}, {Jenkins}, {Jones}, {Ivanyuk} \&
  {Pinfield}}{{Pavlenko} et~al.}{2012}]{Pav12}
{Pavlenko} Y.~V.,  {Jenkins} J.~S.,  {Jones} H.~R.~A.,  {Ivanyuk} O.,
  {Pinfield} D.~J.,  2012, \mnras, 422, 542

\bibitem[\protect\citeauthoryear{{Penev}, {Jackson}, {Spada} \& {Thom}}{{Penev}
  et~al.}{2012}]{Pen12}
{Penev} K.,  {Jackson} B.,  {Spada} F.,    {Thom} N.,  2012, \apj, 751, 96

\bibitem[\protect\citeauthoryear{{Penev} \& {Sasselov}}{{Penev} \&
  {Sasselov}}{2011}]{Pen11}
{Penev} K.,  {Sasselov} D.,  2011, \apj, 731, 67

\bibitem[\protect\citeauthoryear{{Pinsonneault}, {DePoy} \&
  {Coffee}}{{Pinsonneault} et~al.}{2001}]{Pin01}
{Pinsonneault} M.~H.,  {DePoy} D.~L.,    {Coffee} M.,  2001, \apjl, 556, L59

\bibitem[\protect\citeauthoryear{{Pont}, {Zucker} \& {Queloz}}{{Pont}
  et~al.}{2006}]{Pon06b}
{Pont} F.,  {Zucker} S.,    {Queloz} D.,  2006, \mnras, 373, 231

\bibitem[\protect\citeauthoryear{{Queloz}, {Henry}, {Sivan}, {Baliunas},
  {Beuzit}, {Donahue}, {Mayor}, {Naef}, {Perrier} \& {Udry}}{{Queloz}
  et~al.}{2001}]{Que01}
{Queloz} D.,  {Henry} G.~W.,  {Sivan} J.~P.,  {Baliunas} S.~L.,  {Beuzit}
  J.~L.,  {Donahue} R.~A.,  {Mayor} M.,  {Naef} D.,  {Perrier} C.,    {Udry}
  S.,  2001, \aap, 379, 279

\bibitem[\protect\citeauthoryear{{Ramsey}, {Adams}, {Barnes}, {Booth},
  {Cornell}, {Fowler}, {Gaffney}, {Glaspey}, {Good}, {Hill}, {Kelton},
  {Krabbendam}, {Long}, {MacQueen}, {Ray}, {Ricklefs}, {Sage}, {Sebring},
  {Spiesman} \& {Steiner}}{{Ramsey} et~al.}{1998}]{Ram98}
{Ramsey} L.~W.,  {Adams} M.~T.,  {Barnes} T.~G.,  {Booth} J.~A.,  {Cornell}
  M.~E.,  {Fowler} J.~R.,  {Gaffney} N.~I.,  {Glaspey} J.~W.,  {Good} J.~M.,
  {Hill} G.~J.,  {Kelton} P.~W.,  {Krabbendam} V.~L.,  {Long} L.,  {MacQueen}
  P.~J.,  {Ray} F.~B.,  {Ricklefs} R.~L.,  {Sage} J.,  {Sebring} T.~A.,
  {Spiesman} W.~J.,    {Steiner} M.,  1998, in {Stepp} L.~M.,  ed., Society of
  Photo-Optical Instrumentation Engineers (SPIE) Conference Series Vol.~3352 of
  Society of Photo-Optical Instrumentation Engineers (SPIE) Conference Series,
  {Early performance and present status of the Hobby-Eberly Telescope}.
pp 34--42

\bibitem[\protect\citeauthoryear{{Rasio} \& {Ford}}{{Rasio} \&
  {Ford}}{1996}]{Ras96}
{Rasio} F.~A.,  {Ford} E.~B.,  1996, Science, 274, 954

\bibitem[\protect\citeauthoryear{{Robin}, {Reyl{\'e}}, {Derri{\`e}re} \&
  {Picaud}}{{Robin} et~al.}{2003}]{Robin03}
{Robin} A.~C.,  {Reyl{\'e}} C.,  {Derri{\`e}re} S.,    {Picaud} S.,  2003,
  \aap, 409, 523

\bibitem[\protect\citeauthoryear{{Schlegel}, {Finkbeiner} \&
  {Davis}}{{Schlegel} et~al.}{1998}]{Sch98}
{Schlegel} D.~J.,  {Finkbeiner} D.~P.,    {Davis} M.,  1998, \apj, 500, 525

\bibitem[\protect\citeauthoryear{{Schneider}, {Dedieu}, {Le Sidaner}, {Savalle}
  \& {Zolotukhin}}{{Schneider} et~al.}{2011}]{Sch11}
{Schneider} J.,  {Dedieu} C.,  {Le Sidaner} P.,  {Savalle} R.,    {Zolotukhin}
  I.,  2011, \aap, 532, A79

\bibitem[\protect\citeauthoryear{{Seager} \& {Mall{\'e}n-Ornelas}}{{Seager} \&
  {Mall{\'e}n-Ornelas}}{2003}]{Sea03}
{Seager} S.,  {Mall{\'e}n-Ornelas} G.,  2003, \apj, 585, 1038

\bibitem[\protect\citeauthoryear{{Sestito} \& {Randich}}{{Sestito} \&
  {Randich}}{2005}]{Sest05}
{Sestito} P.,  {Randich} S.,  2005, \aap, 442, 615

\bibitem[\protect\citeauthoryear{{Siess}, {Dufour} \& {Forestini}}{{Siess}
  et~al.}{2000}]{Sie00}
{Siess} L.,  {Dufour} E.,    {Forestini} M.,  2000, \aap, 358, 593

\bibitem[\protect\citeauthoryear{{Skrutskie}, {Cutri}, {Stiening}, {Weinberg},
  {Schneider}, {Carpenter}, {Beichman} \& {Capps}}{{Skrutskie}
  et~al.}{2006}]{Skr06}
{Skrutskie} M.~F.,  {Cutri} R.~M.,  {Stiening} R.,  {Weinberg} M.~D.,
  {Schneider} S.,  {Carpenter} J.~M.,  {Beichman} C.,    {Capps} R. e.~a.,
  2006, \aj, 131, 1163

\bibitem[\protect\citeauthoryear{{Snellen} \& {Covino}}{{Snellen} \&
  {Covino}}{2007}]{Sne07}
{Snellen} I.~A.~G.,  {Covino} E.,  2007, \mnras, 375, 307

\bibitem[\protect\citeauthoryear{{Snellen}, {Koppenhoefer}, {van der Burg},
  {Dreizler}, {Greiner}, {de Hoon}, {Husser}, {Kr{\"u}hler}, {Saglia} \&
  {Vuijsje}}{{Snellen} et~al.}{2009}]{Sne09}
{Snellen} I.~A.~G.,  {Koppenhoefer} J.,  {van der Burg} R.~F.~J.,  {Dreizler}
  S.,  {Greiner} J.,  {de Hoon} M.~D.~J.,  {Husser} T.~O.,  {Kr{\"u}hler} T.,
  {Saglia} R.~P.,    {Vuijsje} F.~N.,  2009, \aap, 497, 545

\bibitem[\protect\citeauthoryear{{Socrates} \& {Katz}}{{Socrates} \&
  {Katz}}{2012}]{Soc12}
{Socrates} A.,  {Katz} B.,  2012, ArXiv e-prints

\bibitem[\protect\citeauthoryear{{Southworth}}{{Southworth}}{2008}]{Sou08}
{Southworth} J.,  2008, \mnras, 386, 1644

\bibitem[\protect\citeauthoryear{{Swain}, {Vasisht}, {Tinetti}, {Bouwman},
  {Chen}, {Yung}, {Deming} \& {Deroo}}{{Swain} et~al.}{2009}]{Swa09}
{Swain} M.~R.,  {Vasisht} G.,  {Tinetti} G.,  {Bouwman} J.,  {Chen} P.,  {Yung}
  Y.,  {Deming} D.,    {Deroo} P.,  2009, \apjl, 690, L114

\bibitem[\protect\citeauthoryear{{Torres}, {Fischer}, {Sozzetti}, {Buchhave},
  {Winn}, {Holman} \& {Carter}}{{Torres} et~al.}{2012}]{Tor12}
{Torres} G.,  {Fischer} D.~A.,  {Sozzetti} A.,  {Buchhave} L.~A.,  {Winn}
  J.~N.,  {Holman} M.~J.,    {Carter} J.~A.,  2012, \apj, 757, 161

\bibitem[\protect\citeauthoryear{{Torres}, {Konacki}, {Sasselov} \&
  {Jha}}{{Torres} et~al.}{2005}]{Tor05}
{Torres} G.,  {Konacki} M.,  {Sasselov} D.~D.,    {Jha} S.,  2005, \apj, 619,
  558

\bibitem[\protect\citeauthoryear{{Trilling}, {Benz}, {Guillot}, {Lunine},
  {Hubbard} \& {Burrows}}{{Trilling} et~al.}{1998}]{Trill98}
{Trilling} D.~E.,  {Benz} W.,  {Guillot} T.,  {Lunine} J.~I.,  {Hubbard} W.~B.,
     {Burrows} A.,  1998, \apj, 500, 428

\bibitem[\protect\citeauthoryear{{Tull}}{{Tull}}{1998}]{Tul98}
{Tull} R.~G.,  1998, in {D'Odorico} S.,  ed., Society of Photo-Optical
  Instrumentation Engineers (SPIE) Conference Series Vol.~3355 of Society of
  Photo-Optical Instrumentation Engineers (SPIE) Conference Series,
  {High-resolution fiber-coupled spectrograph of the Hobby-Eberly Telescope}.
pp 387--398

\bibitem[\protect\citeauthoryear{{van Saders} \& {Pinsonneault}}{{van Saders}
  \& {Pinsonneault}}{2012}]{Sad12}
{van Saders} J.~L.,  {Pinsonneault} M.~H.,  2012, \apj, 746, 16

\bibitem[\protect\citeauthoryear{{Watson} \& {Marsh}}{{Watson} \&
  {Marsh}}{2010}]{Wat10}
{Watson} C.~A.,  {Marsh} T.~R.,  2010, \mnras, 405, 2037

\bibitem[\protect\citeauthoryear{{Winn}, {Fabrycky}, {Albrecht} \&
  {Johnson}}{{Winn} et~al.}{2010}]{Win10}
{Winn} J.~N.,  {Fabrycky} D.,  {Albrecht} S.,    {Johnson} J.~A.,  2010, \apjl,
  718, L145

\bibitem[\protect\citeauthoryear{{Wright}, {Eisenhardt}, {Mainzer}, {Ressler},
  {Cutri}, {Jarrett}, {Kirkpatrick} \& {Padgett}}{{Wright}
  et~al.}{2010}]{Wrig10}
{Wright} E.~L.,  {Eisenhardt} P.~R.~M.,  {Mainzer} A.~K.,  {Ressler} M.~E.,
  {Cutri} R.~M.,  {Jarrett} T.,  {Kirkpatrick} J.~D.,    {Padgett} D. e.~a.,
  2010, \aj, 140, 1868

\bibitem[\protect\citeauthoryear{{Wright}, {Fakhouri}, {Marcy}, {Han}, {Feng},
  {Johnson}, {Howard}, {Fischer}, {Valenti}, {Anderson} \& {Piskunov}}{{Wright}
  et~al.}{2011}]{Wri10}
{Wright} J.~T.,  {Fakhouri} O.,  {Marcy} G.~W.,  {Han} E.,  {Feng} Y.,
  {Johnson} J.~A.,  {Howard} A.~W.,  {Fischer} D.~A.,  {Valenti} J.~A.,
  {Anderson} J.,    {Piskunov} N.,  2011, \pasp, 123, 412

\bibitem[\protect\citeauthoryear{{Wyatt}}{{Wyatt}}{2008}]{Wya08}
{Wyatt} M.~C.,  2008, \araa, 46, 339

\bibitem[\protect\citeauthoryear{{Yanny}, {Guhathakurta}, {Bahcall} \&
  {Schneider}}{{Yanny} et~al.}{1994}]{Yan94}
{Yanny} B.,  {Guhathakurta} P.,  {Bahcall} J.~N.,    {Schneider} D.~P.,  1994,
  \aj, 107, 1745

\bibitem[\protect\citeauthoryear{{York}, {Adelman}, {Anderson} Jr., {Anderson},
  {Annis}, {Bahcall}, {Bakken}, {Barkhouser} \& {SDSS Collaboration}}{{York}
  et~al.}{2000}]{Yor00}
{York} D.~G.,  {Adelman} J.,  {Anderson} Jr. J.~E.,  {Anderson} S.~F.,  {Annis}
  J.,  {Bahcall} N.~A.,  {Bakken} J.~A.,  {Barkhouser} R.~e.~a.,    {SDSS
  Collaboration} 2000, \aj, 120, 1579

\bibitem[\protect\citeauthoryear{{Zahn}}{{Zahn}}{1977}]{Zah77}
{Zahn} J.,  1977, \aap, 57, 383

\bibitem[\protect\citeauthoryear{{Zendejas}, {Koppenhoefer}, {Saglia},
  {Birkby}, {Hodgkin}, {Kovacs}, {Pinfield}, {Sipocz}, {Barrado}, {Bender},
  {del Burgo}, {Cappetta}, {Martin}, {Nefs}, {Riffeser} \& {Steele}}{{Zendejas}
  et~al.}{2013}]{Zen13}
{Zendejas} J.,  {Koppenhoefer} J.,  {Saglia} R.~P.,  {Birkby} J.~L.,  {Hodgkin}
  S.~T.,  {Kovacs} G.,  {Pinfield} D.~J.,  {Sipocz} B.,  {Barrado} D.,
  {Bender} R.,  {del Burgo} C.,  {Cappetta} M.,  {Martin} E.~L.,  {Nefs} S.~V.,
   {Riffeser} A.,    {Steele} P.,  2013, \aap, 560, A92

\end{thebibliography}

\label{lastpage}
\end{document}